\documentclass[11pt,a4paper]{article}
\usepackage[english]{babel}
\usepackage[utf8]{inputenc}
\usepackage{epsf}
\usepackage{cite}
\usepackage{graphicx}
\usepackage{subcaption} 

\usepackage{verbatim} 

\usepackage{amsmath}
\usepackage{amssymb}
\usepackage{mathrsfs}

\usepackage{mathbbol}

\usepackage{amsfonts}
\usepackage{bbding}
\usepackage{array}
\usepackage[
colorlinks=true
,urlcolor=black
,anchorcolor=black
,citecolor=black
,filecolor=black
,linkcolor=black
,menucolor=black
,linktocpage=true
,pdfproducer=medialab
,pdfa=true
]{hyperref}

\usepackage{color} 

\usepackage[left=2cm,right=2cm,top=2cm,bottom=2cm]{geometry}
\usepackage{tikz}
\usetikzlibrary{shapes,arrows,positioning,automata,backgrounds,calc,er,patterns}
\usepackage[compat=1.1.0]{tikz-feynman}

\vspace*{1cm}
\allowdisplaybreaks
\begin{document}

\begin{center}
\vspace*{1mm}
\vspace{1.3cm}

\bigskip 
\mathversion{bold}
{\Large\bf
 LFV Higgs and $Z$-boson decays: 
leptonic CPV phases \\ \vspace*{3mm} 
and CP asymmetries
}
\mathversion{normal}

\vspace*{1.2cm}

{\bf A.~Abada $^{a}$, J.~Kriewald $^{b}$, E. Pinsard $^{b}$, 
S.~Rosauro-Alcaraz $^{a}$ and A.~M.~Teixeira $^{b}$}

\vspace*{.5cm}
$^{a}$ P\^ole Th\'eorie, Laboratoire de Physique des 2 Infinis Irène Joliot Curie (UMR 9012), \\
CNRS/IN2P3,
15 Rue Georges Clemenceau, 91400 Orsay, France

\vspace*{.2cm}
$^{b}$ Laboratoire de Physique de Clermont (UMR 6533), CNRS/IN2P3,\\
Univ. Clermont Auvergne, 4 Av. Blaise Pascal, 63178 Aubi\`ere Cedex,
France

\end{center}

\vspace*{5mm}
\begin{abstract}

\noindent
Heavy neutral leptons are motivated by several extensions of the Standard Model and
their presence induces modifications in the lepton mixing matrix, including new Dirac and Majorana CP violating phases. 
It has been recently shown that these phases play an important role in lepton number and lepton flavour violating decays and transitions, with a striking impact for the predicted rates of certain observables. In this work, we now
consider the potential role of the heavy neutral fermions and of the new CP violating phases in Higgs and $Z$-boson lepton flavour violating decays $H,Z\to \ell_\alpha^\pm\ell_\beta^\mp$, as well as in the corresponding CP asymmetries. In order to allow for a thorough  analysis, we derive the full analytical expressions of the latter observables, taking into account the effect of the (final state) charged lepton masses.
A comprehensive exploration of
lepton flavour violating $Z$ and Higgs decays confirms that these are very sensitive to the presence of the additional heavy neutral leptons. Moreover, the new CPV phases have a clear impact on the decay rates, 
leading to both destructive and constructive interferences.
Interestingly, in the $\mu\tau$ sector, 
the $Z\to \ell_\alpha^\pm\ell_\beta^\mp$ rates are
within reach of FCC-ee, with associated CP asymmetries that can potentially reach up to 20\% -- 30\%.

\end{abstract}
\vspace*{10mm}

\newpage
\section{Introduction}
Numerous New Physics (NP) models have been explored in recent years in order to explain and understand the phenomenology of neutrino oscillations.
Extensions of the Standard Model (SM) accounting for neutrino data open the door to new leptonic mixings - possibly going beyond the minimal Pontecorvo-Maki-Nagakawa-Sakata (PMNS) left-handed leptonic
mixing matrix $U_\text{PMNS}$ - as well as to new sources of CP violation, through the presence of Dirac (and possibly Majorana) phases. Notice that unless CP conservation is enforced, CP violating (CPV) phases are in general associated with multi-generational fermion mixings. Leptonic sources of CP violation are a key ingredient to constructions aiming at explaining the observed baryon asymmetry of the Universe via leptogenesis; moreover, they have been found to have a significant impact on a number of observables, which are currently being investigated at high-energy colliders or in dedicated high-intensity experiments. 

Among the many NP constructions successfully addressing the problem of neutrino mass generation and the observed lepton mixing pattern, one of the most minimal (yet very appealing) possibilities is that of 
SM extensions via heavy neutral 
leptons (HNL), in general of Majorana nature as is the case of right-handed neutrinos. Present in numerous scenarios (such as the type I seesaw and its variants, or arising from complete constructions as $\mathrm{SO}(10)$ grand unified models, among other examples), sterile fermions can be at the source of extensive contributions to numerous phenomena beyond neutrino oscillations: depending on the mass of these new states (and their interactions with the active neutrinos), such phenomena range from flavour observables, to high-energy collider signals, with a non-negligible impact on astroparticle physics or even cosmology. 

Be it in the framework of a specific mechanism of neutrino mass generation, or in simple minimal ``ad-hoc" constructions, the role of heavy neutral leptons regarding lepton flavour observables has been the object of numerous dedicated studies, both concerning lepton number violating (LNV) observables  
(see for instance~\cite{Ali:2001gsa,Atre:2005eb,Atre:2009rg,Chrzaszcz:2013uz,Deppisch:2015qwa,Cai:2017mow,Abada:2017jjx,Drewes:2019byd,Maiezza:2015lza,Helo:2013dla,Blaksley:2011ey,Ibarra:2011wi}) and charged lepton flavour violating (cLFV) transitions and decays (for example, see~\cite{Riemann:1982rq,Riemann:1999ab,Illana:1999ww,Mann:1983dv,Illana:2000ic,Alonso:2012ji,Ilakovac:1994kj,Ma:1979px,Gronau:1984ct,Deppisch:2004fa,Deppisch:2005zm,Dinh:2012bp,Hambye:2013jsa,Abada:2014kba,Abada:2015oba,Abada:2015zea,Abada:2016vzu,Calibbi:2017uvl,Abada:2018nio,Arganda:2014dta}). 

The new CPV phases (Dirac and/or Majorana) are also expected to lead to significant (and detectable) effects in several searches. In addition to CP-violating observables (as is the case of electric dipole moments of  charged leptons~\cite{Abada:2015trh,Abada:2016awd,Novales-Sanchez:2016sng,deGouvea:2005jj}), the new phases are also at the source of interference effects in LNV (and cLFV) semileptonic meson and tau decays~\cite{Abada:2019bac} - especially in the case in which the SM is extended by two (or more) heavy neutrino states. 
In particular, the role of a second HNL has been explored in studies concerning the effects of CP violation, be it in high-scale seesaw scenarios, in the context of renormalisation group running and thermal  leptogenesis~\cite{Petcov:2005yh,Petcov:2006pc}, in assessing the possibility of resonant CP violation~\cite{Bray:2007ru}, or upon considering possible effects in forward-backward asymmetries at an electron-positron collider~\cite{Dev:2019rxh}. Other studies have compared the expected number of events associated with same-sign and opposite-sign dileptons at colliders in the framework of left-right symmetric models~\cite{Anamiati:2016uxp,Das:2017hmg,Dev:2019rxh}.

Recently, a study was carried out, dedicated to investigating the impact of leptonic CPV phases on numerous cLFV transitions and decays~\cite{Abada:2021zcm}; working in the framework of a minimal low-energy
extension of the SM via 2 Majorana heavy sterile states, the analysis revealed that the new phases could have a significant impact for the predicted cLFV rates, leading to a possible loss of correlation between cLFV observables (which would be otherwise manifest in the CP conserving case). 
Building upon this first thorough analysis of the effects of CPV phases regarding leptonic cLFV transitions and decays, we now address in detail the prospects in what concerns SM neutral boson decays. The present work is thus dedicated to a detailed investigation of the role of the CPV phases in cLFV Higgs and $Z$-boson decays ($H,Z\to\ell_\alpha^\pm\ell_\beta^\mp$). 
A first part of the analysis concerns an in-depth study of the decay rates, revisiting the full computation of the amplitudes. In particular, we derive the full effective cLFV $Z$ and Higgs vertices in the presence of heavy neutral leptons without any simplifying assumptions, finding good agreement with previous calculations~\cite{Ilakovac:1994kj,Illana:2000ic,DeRomeri:2016gum,Hernandez-Tome:2019lkb,Herrero:2018luu,Abada:2015zea,Arganda:2014dta,Arganda:2004bz,Hernandez-Tome:2020lmh,Arganda:2016zvc,Thao:2017qtn,Abada:2014cca}.

\bigskip
As argued in~\cite{Abada:2021zcm}, the possible effects of CPV phases on the cLFV observables can lead to a strong loss of correlation between them, and thus smear out
otherwise clean flavour patterns.
A similar situation can be encountered in leptoquark models (albeit not necessarily due to CP violating interactions), such that an identification of specific leptoquark states would be often unfeasible just by taking into account future data on cLFV (meson) decays. As suggested in the literature, 
these difficulties can be overcome (or at least reduced) by comparing different CP final states, rather than taking the average of the observables (see e.g.~\cite{Becirevic:2016zri,Becirevic:2016oho,Hati:2020cyn}).
In specific cLFV meson decays (e.g. $B\to K\mu^\pm\tau^\mp$), measurements of both possible final state charge assignments have been carried out~\cite{BaBar:2012azg,LHCb:2020khb}, thus allowing to constrain different flavour structures (as for example in the case of certain leptoquark models).

Similarly, in heavy neutral lepton models, it can be difficult to identify the presence of new CPV phases at the origin of peculiar flavour patterns.
Henceforth, it could be interesting to compare the different CP final states of cLFV transitions: as 
a (more) direct manifestation of leptonic CP violation, one can then consider CP asymmetries in cLFV $Z$ (and Higgs) decays. 
The latter observables could offer further hints not only on the underlying source of flavour violation, but also on its CP structure (possibly in connection with leptogenesis~\cite{daSilva:2022mrx}).
We thus carefully examine the impact of Dirac and Majorana CPV phases on such observables, and the results of our phenomenological analysis suggest that the CP asymmetries can reach up to $20-30$\%, in particular for cLFV $Z\to \mu \tau$ decay rates, which can be within the reach of FCC-ee~\cite{Abada:2019lih}.

As we argue throughout this study, CP violating leptonic phases clearly have extensive effects on cLFV transitions and decays, and their presence must be systematically taken into account to fully explore HNL extensions of the SM.

\bigskip
The manuscript is organised as follows: we begin by briefly describing the features of this simple SM extension in Section~\ref{sec:model}, after which we revisit cLFV $Z$ and Higgs boson decays, presenting full analytical expressions in Section~\ref{sec:cLFV_HZ:exp}. Our results concerning the role of the Dirac and Majorana CPV phases are collected 
in Section~\ref{sec:results}, which is followed by a dedicated discussion of CP asymmetries in cLFV neutral boson decays (Section~\ref{sec:asymmetries}).
Finally, we summarise our main findings in the Conclusions. Further information on the analytical derivation and analysis is collected in the Appendices. 

\section{A minimal HNL extension of the SM}\label{sec:model}
As done in a previous analysis~\cite{Abada:2021zcm}, we follow a phenomenological approach, relying on a minimal SM extension which allows studying the effects of CPV phase-induced interferences in the lepton sector. We thus consider the SM extended by 2 sterile Majorana states; the neutral lepton sector is thus composed of 5 massive states (with masses $m_{i}$, $i = 1,...,5)$, which include 3 light (mostly active) neutrinos and two heavier states.
From a low-energy perspective - and independently of the mechanism at the origin of neutrino mass generation -, the leptonic mixings are impacted by the presence of the new heavy states. Instead of the
$3\times 3$ $U_\text{PMNS}$
mixing matrix, a $5\times 5$ unitary matrix $\mathcal{U}$ now parametrises mixings in the lepton sector; notice that the left-handed leptonic mixings are now encoded in its $3\times 3$ upper left block (the 
would-be PMNS matrix, $\tilde U_\text{PMNS}$); 
as a consequence, the mixing in charged current interactions will be parametrised via a rectangular $3 \times 5$%
mixing matrix.
The deviations of $\tilde U_\text{PMNS}$ from unitarity~\cite{Schechter:1980gr,Gronau:1984ct} can be conveniently cast in terms of a matrix $\eta$ as follows~\cite{FernandezMartinez:2007ms}
\begin{equation}
\label{eq:defPMNSeta}
U_\text{PMNS} \, \to \, \tilde U_\text{PMNS} \, = \,(\mathbb{1} - \eta)\, 
U_\text{PMNS}\,.
\end{equation}

In agreement with what was done in~\cite{Abada:2021zcm}, $\mathcal{U}$
can be parametrised through a series of ten (complex) rotations $R_{ij}$ (with $i\neq j$), and a diagonal matrix including the four physical Majorana phases, $\varphi_i$~\cite{Abada:2015trh}
\begin{eqnarray}
    \mathcal{U} \,= \,R_{45}\,R_{35}\,R_{25}\,R_{15}\,
    R_{34}\,R_{24}\,R_{14}\,R_{23}\,R_{13}\,R_{12}\times\mathrm{diag}(1, e^{i\varphi_2}, e^{i\varphi_3}, e^{i\varphi_4}, e^{i\varphi_5})\,.
    \label{eqn:allrot}
\end{eqnarray}
Each of the above rotations $R_{ij}$ is given by a $5\times 5$ unitary matrix, parametrised in terms of a real angle $\theta_{ij}$ and a (Dirac) phase, $\delta_{ij}$, as illustrated below for $R_{45}$:
\begin{equation}\label{eq:R45}
    R_{45} = \begin{pmatrix}
    \mathbb{1}_{3\times3} & \mathbb{0}_{2\times3} \\
    \mathbb{0}_{3\times2} & \Theta_{45}
            \end{pmatrix}\,,
\quad \text{with} \quad
\Theta_{45}= \begin{pmatrix}
\cos\theta_{45} & \sin \theta_{45} e^{-i\delta_{45}}\\
    -\sin\theta_{45} e^{i\delta_{45}} & \cos\theta_{45}
\end{pmatrix}\,.
\end{equation}
Concerning the mixing parameters of the (mostly) active light neutrinos ($\theta_{12}, \theta_{23}, \theta_{13}$ and the Dirac CPV phase $\delta_{13}$), in our analysis we take the central values of the NuFIT 5.1 results~\cite{Esteban:2020cvm} (for a normal ordering of the light neutrino spectrum; notice that the predictions for cLFV transitions are mostly independent of the ordering of the spectrum). 
These are collected 
in Appendix~\ref{app:constraints}, where we also discuss
other relevant constraints on SM extensions with sterile fermions.

As already stated, due to the presence of the additional sterile fermion states, the would-be PMNS matrix is 
no longer unitary, thus leading to modified charged and neutral lepton currents.
In the physical basis, and for the generic case of a number $n_S$ of sterile fermions, the relevant terms of the interaction Lagrangian can be cast as
\begin{align}\label{eq:lagrangian:WGHZ}
& \mathcal{L}_{W^\pm}\, =\, -\frac{g_w}{\sqrt{2}} \, W^-_\mu \,
\sum_{\alpha=1}^{3} \sum_{j=1}^{3 + n_S} \mathcal{U}_{\alpha j} \bar \ell_\alpha 
\gamma^\mu P_L \nu_j \, + \, \text{H.c.}\,, \nonumber \\
& \mathcal{L}_{Z^0}^{\nu}\, = \,-\frac{g_w}{4 \cos \theta_w} \, Z_\mu \,
\sum_{i,j=1}^{3 + n_S} \bar \nu_i \gamma ^\mu \left(
P_L {C}_{ij} - P_R {C}_{ij}^* \right) \nu_j\,, \nonumber \\
& \mathcal{L}_{Z^0}^{\ell}\, = \,-\frac{g_w}{2 \cos \theta_w} \, Z_\mu \,
\sum_{\alpha=1}^{3}  \bar \ell_\alpha \gamma ^\mu \left(
{\bf C}_{V} - {\bf C}_{A} \gamma_5 \right) \ell_\alpha\,, \nonumber \\
& \mathcal{L}_{H^0}\, = \, -\frac{g_w}{4 M_W} \, H  \,
\sum_{i\ne j= 1}^{3 + n_S}    \bar \nu_i\,\left[{C}_{ij}\,\left(
P_L m_i + P_R m_j \right) +{C}_{ij}^\ast\left(
P_R m_i + P_L m_j \right) \right] \nu_j\ , \nonumber \\
& \mathcal{L}_{G^0}\, =\,\frac{i g_w}{4 M_W} \, G^0 \,
\sum_{i,j=1}^{3 + n_S} {C}_{ij}  \bar \nu_i \left[ {C}_{ij}
\left(P_R m_j  - P_L m_i  \right) + {C}_{ij}^\ast
\left(P_R m_i  - P_L m_j  \right)\right] \nu_j\,, \nonumber  \\
& \mathcal{L}_{G^\pm}\, =\, -\frac{g_w}{\sqrt{2} M_W} \, G^- \,
\sum_{\alpha=1}^{3}\sum_{j=1}^{3 + n_S} \mathcal{U}_{\alpha j}
\bar \ell_\alpha\left(
m_\alpha P_L - m_j P_R \right) \nu_j\, + \, \text{H.c.}\,, 
\end{align}
in which $\alpha, \rho = 1, \dots, 3$ denote the flavour of the charged leptons, while $i, j = 1, \dots, 3+n_S$ correspond to the physical (massive) 
neutrino states; $P_{L,R} = (1 \mp \gamma_5)/2$, $g_w$ is the weak coupling constant, and $\cos^2 \theta_w =  M_W^2 /M_Z^2$. 
Regarding the interaction of $Z$ bosons with charged leptons,
the SM vector and axial-vector currents can be cast in terms of the ${\bf C}_{V}$ and ${\bf C}_{A}$ coefficients,  respectively given by 
${\bf C}_{V} = -\frac{1}{2} + 2 \sin^2\theta_w$ and 
${\bf C}_{A} = -\frac{1}{2}$.
Moreover, the coefficients ${C}_{ij} $ are defined as: 
\begin{equation}
    {C}_{ij} = \sum_{\rho = 1}^3
  \mathcal{U}_{i\rho}^\dagger \,\mathcal{U}_{\rho j}^{\phantom{\dagger}}\:. 
\end{equation}
For completeness, the new interaction vertices (Feynman rules) are presented in Appendix~\ref{app:feynrules}.

\section{Revisiting cLFV neutral boson decays: analytic expressions}\label{sec:cLFV_HZ:exp}
In this section we present expressions for the relevant cLFV decays of $Z$ and Higgs bosons in the presence of $n_S$ singlet fermions. In general, the cLFV boson decays receive contributions at the one-loop level from boson vertex corrections, as well as from charged lepton self-energy diagrams. In the following, 
no approximation will be made leading to
the full amplitudes.
In this section, the results are obtained working in unitary gauge; for completeness, we also present the full expressions derived in Feynman-'t Hooft gauge in Appendix~\ref{app:ZHdecays:formulae}. 
\mathversion{bold}
\subsection{$Z$-boson decays: $Z\to \ell_\alpha^\mp\ell_\beta^\pm$}
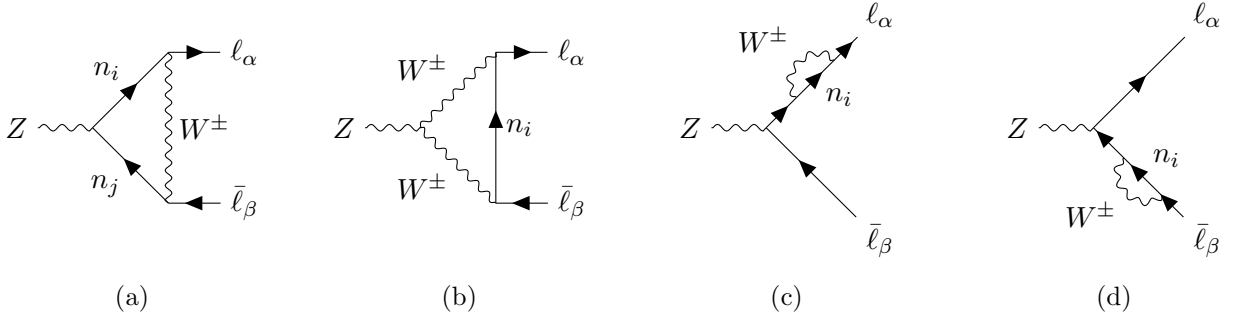
\begin{figure}[h!]
    \centering
    \begin{subfigure}[b]{0.24\textwidth}
    \centering
 \raisebox{5mm}{    \begin{tikzpicture}
    \begin{feynman}
    \vertex (a) at (0,0) {\(Z\)};
    \vertex (b) at (1,0);
    \vertex (c) at (2,1.);
    \vertex (d) at (2,-1);
    \vertex (e) at (3,1) {\( \ell_\alpha\)};
    \vertex (f) at (3,-1) {\( \bar\ell_\beta\)};
    \diagram* {
    (a) -- [boson] (b),
    (b) -- [fermion, edge label=\(n_i\)] (c),
    (c) -- [boson, edge label=\( W^\pm\)] (d),
    (d) -- [fermion, edge label=\(n_j\)] (b),
    (c) -- [fermion] (e),
    (f) -- [fermion] (d)
    };
    \end{feynman}
    \end{tikzpicture}
    }
            \caption*{(a)}
            \label{}
    \end{subfigure}
    \hfill
    \begin{subfigure}[b]{0.24\textwidth}
    \centering
\raisebox{5mm}{        \begin{tikzpicture}
    \begin{feynman}
    \vertex (a) at (0,0) {\(Z\)};
    \vertex (b) at (1,0);
    \vertex (c) at (2,1.);
    \vertex (d) at (2,-1.);
    \vertex (e) at (3,1.) {\( \ell_\alpha\)};
    \vertex (f) at (3,-1.) {\( \bar\ell_\beta\)};
    \diagram* {
    (a) -- [boson] (b),
    (b) -- [boson, edge label=\( W^\pm\)] (c),
    (c) -- [anti fermion, edge label=\(n_i\)] (d),
    (d) -- [boson, edge label=\( W^\pm\)] (b),
    (c) -- [fermion] (e),
    (f) -- [fermion] (d)
    };
    \end{feynman}
    \end{tikzpicture}
    }
            \caption*{(b)}
            \label{}
    \end{subfigure}
    \hfill
    \begin{subfigure}[b]{0.24\textwidth}
    \centering
    \begin{tikzpicture}
    \begin{feynman}
    \vertex (a) at (0,0) {\(Z\)};
    \vertex (b) at (1,0);
    \vertex (c) at (1.4, 0.4);
    \vertex (d) at (1.9,0.9);
    \vertex (e) at (2.5, 1.5) {\( \ell_\alpha\)};
    \vertex (f) at (2.5,-1.5) {\( \bar\ell_\beta\)};
    
    \diagram* {
    (a) -- [boson] (b),
    (b) -- [fermion] (c) -- [fermion, edge label'=\(n_i\)] (d) -- [fermion] (e),
    (b) -- [anti fermion] (f),
    (c) -- [boson, half left, edge label=\( W^\pm\)] (d)
    };   
    \end{feynman}
    \end{tikzpicture}
            \caption*{(c)}
            \label{}
    \end{subfigure}
    \hfill
    \begin{subfigure}[b]{0.24\textwidth}
    \centering
    \begin{tikzpicture}
    \begin{feynman}
    \vertex (a) at (0,0) {\(Z\)};
    \vertex (b) at (1,0);
    \vertex (c) at (1.4, -0.4);
    \vertex (d) at (1.9, -0.9);
    \vertex (e) at (2.5, 1.5) {\( \ell_\alpha\)};
    \vertex (f) at (2.5,-1.5) {\( \bar\ell_\beta\)};
    
    \diagram* {
    (a) -- [boson] (b),
    (b) -- [anti fermion] (c) -- [anti fermion, edge label=\(n_i\)] (d) -- [anti fermion] (f),
    (b) -- [fermion] (e),
    (c) -- [boson, half right, edge label'=\( W^\pm\)] (d)
    };   
    \end{feynman}
    \end{tikzpicture}
            \caption*{(d)}
            \label{}
    \end{subfigure}
    \caption{Feynman diagrams contributing to cLFV $Z$ decays (in unitary gauge).}
    \label{fig:cLFVZdecays:UG}
\end{figure}

\mathversion{normal}
\label{sec:cLFVZUG}
For on-shell charged leptons, the cLFV $Z$ vertex can in general be decomposed as
\begin{equation}
    \bar u_\alpha(p_\alpha) \Gamma_{\ell_\alpha\ell_\beta}^\mu (q) v_\beta(p_\beta) = \sum_{X = L, R} \bar u_\alpha(p_\alpha)\left[F_S^X q^\mu P_X + F_V^X \gamma^\mu P_X  + F_T^X \sigma^{\mu\nu} q_\nu P_X  \right]v_\beta(p_\beta)\,,
    \label{eqn:Zamplitude}
\end{equation}
in which $q^\mu = p_\alpha^\mu + p_\beta^\mu$ is the momentum of the $Z$ boson and $\sigma^{\mu\nu} = \frac{i}{2}[\gamma^\mu, \gamma^\nu]$; $F_{S,V,T}^{L,R}$ are scalar, vector and tensor form factors. For on-shell $Z$-boson decays, the scalar amplitudes do not contribute due to the Ward identity $q^\mu \epsilon_\mu^\ast(q) = 0$, but they can nevertheless appear if the $Z$ boson is an off-shell intermediate state. Furthermore, all amplitudes but $F_V^L$ vanish in the limit of massless charged leptons. We emphasise that in the calculation of the contributing diagrams, which are shown in Fig.~\ref{fig:cLFVZdecays:UG}, we do not neglect the final state charged lepton masses. Although the assumption of vanishing charged lepton masses is indeed justified for the cLFV decay rates, as we will discuss later on, the behaviour of the CP asymmetries significantly depends 
on the charged lepton masses, 
thus demanding a full computation\footnote{For previous works on cLFV $Z$ decays in the limit of vanishing lepton masses, see for instance~\cite{Ilakovac:1994kj,Illana:2000ic,DeRomeri:2016gum,Hernandez-Tome:2019lkb,Herrero:2018luu,Abada:2015zea,Abada:2014cca}.}.

\noindent
We separate the contributions to the invariant amplitudes according to the different topologies (the superscript refers to the diagrams of Fig.~\ref{fig:cLFVZdecays:UG})  as:
\begin{equation}
    F_{V, T, S}^L \,=\, F_{V, T, S}^{L \:(a)} + F_{V, T, S}^{L \:(b)} + F_{V, T, S}^{L \:(c+d)}\,.
\end{equation}
The different contributions are given by\footnote{Since the ``scalar'' amplitudes do not contribute to the on-shell decay, we list them display in Appendix~\ref{app:ZHdecays:formulae}.}
\begin{eqnarray}\label{eqn:FVLadU}
    F_V^{L\:(a)} &=& -\frac{g_w^3}{64 \pi^2 \,c_w \,M_W^2}\sum_{i,j}\, \mathcal U_{\alpha i}\,\mathcal U_{\beta j}^\ast\Bigg\{C_{ij}\left[(m_\alpha^2 - m_i^2 + 2 M_W^2)B_0^\alpha + (m_\beta^2 - m_j^2 + 2M_W^2)\,B_0^\beta \right. \nonumber\\
    &\phantom{=}& \left. - A_0 - M_W^2\, (6 - D)\, B_0^q  + m_\alpha^2 \,B_1^\alpha + m_\beta^2\, B_1^\beta\right. \nonumber\\
    &\phantom{=}& 
    + \Bigg( m_\alpha^2\, m_j^2 + m_\beta^2 \,m_i^2 - m_\alpha^2 \,m_\beta^2 - m_i^2\, m_j^2 - 
\nonumber \\
&& \quad \quad 
\left. 
-    2 M_W^2 \left(m_\alpha^2 + m_\beta^2 - m_i^2 - m_j^2 + \left(3 -\frac{D}{2}\right)M_W^2 + q^2\right)\right) \,C_0 \nonumber \\ 
    &\phantom{=}& \left. + \left(m_\alpha^2 \,m_j^2 + m_\beta^2\, m_i^2 - 2 m_\alpha^2 \,m_\beta^2 - 2 M_W^2 \,(m_\beta^2 + 3 m_\alpha^2)\right)\, C_1\right. \nonumber\\
    &\phantom{=}& \left. + \left(m_\alpha^2 \,m_j^2 + m_\beta^2 \,m_i^2 - 2 m_\alpha^2\, m_\beta^2 - 2 M_W^2\, (m_\alpha^2 + 3 m_\beta^2)\right) \,C_2\right.\nonumber\\
    &\phantom{=}& \left. - (m_\alpha^2 \,m_\beta^2 + (D-2)\,m_\alpha^2 \,M_W^2)\,C_{11} - (m_\alpha^2 \,m_\beta^2 + (D-2)\,m_\beta^2\, M_W^2)\,C_{22}\right.\nonumber\\
    &\phantom{=}&\left. - (2m_\alpha^2 \,m_\beta^2 + (D-2)\,(m_\alpha^2 + m_\beta^2)\,M_W^2)\,C_{12} - 2 M_W^2\,(D-2)\,C_{00}\right]\nonumber\\
    &\phantom{=}& + C_{ij}^\ast \,m_i \,m_j\left[ - B_0^q + (D-3)\,M_W^2 \,C_0 + m_\alpha^2\,C_{11} + m_\beta^2 \,C_{22}  \right.\nonumber\\
    &&  \quad 
    \left.
    + (m_\alpha^2 + m_\beta^2)\,C_{12} + 2 C_{00}\right]\Bigg\},\label{eqn:FVLaU}\\
    F_{V}^{R\:(a)} &=& -\frac{g_w^3}{64 \pi^2\, c_w \,M_W^2}m_\alpha\, m_\beta \sum_{i,j}\, \mathcal U_{\alpha i}\,\mathcal U_{\beta j}^\ast\Bigg\{-C_{ij}\,\left[- B_0^q + m_\alpha^2 \,C_{11} + m_\beta^2 \,C_{22} + (m_\alpha^2 + m_\beta^2)\, C_{12}\right.\nonumber\\
    &\phantom{=}&\left. + M_W^2\left((D-3)\,C_0 + (D-2)\,(2 C_2 + C_{22} + 2 C_1 + C_{11} + 2 C_{12})\right) + 2C_{00}\right]\nonumber\\
    &\phantom{=}& + m_i \,m_j \,C_{ij}^\ast\left[C_{11} + C_{22} + 2C_{12} \right]\Bigg\}\,,\label{eqn:FVRaU}\\
    F_{T}^{L\:(a)} &=& -\frac{i g_w^3}{64 \pi^2 \,c_w \,M_W^2} m_\alpha \sum_{i,j}\, \mathcal U_{\alpha i}\,\mathcal U_{\beta j}^\ast\Bigg\{C_{ij}\,\left[2 M_W^2\, C_0 + D\, M_W^2 \,C_1 +  (m_\beta^2 - m_j^2 + 2 M_W^2)\,C_2\right.\nonumber\\
    &\phantom{=}& \left. + (D-2)\,M_W^2 \,C_{11} + m_\beta^2\, C_{22} + (m_\beta^2 + (D-2)\, M_W^2)\,C_{12}\right]\nonumber\\
    &\phantom{=}& - m_i\, m_j \,C_{ij}^\ast[C_{12} + C_{11}]\Bigg\}\,,\label{eqn:FTLaU}\\
    F_T^{R\:(a)} &=& -\frac{i g_w^3}{64 \pi^2 \,c_w \,M_W^2} m_\beta \sum_{i,j}\, \mathcal U_{\alpha i}\,\mathcal U_{\beta j}^\ast\Bigg\{C_{ij}\,[2 M_W^2 \,C_0 + (m_\alpha^2 - m_i^2 + 2 M_W^2)\,C_1 + D \,M_W^2\, C_2\nonumber\\
    &\phantom{=}& + m_\alpha^2 \,C_{11} + (D-2)\,M_W^2 \,C_{22} + (m_\alpha^2 + (D-2)\,M_W^2)\,C_{12}] \nonumber\\
    &\phantom{=}& - m_i \,m_j \,C_{ij}^\ast\,[C_{12} + C_{22}]\Bigg\}\,,\label{eqn:FTRaU}
\end{eqnarray}
with the Passarino-Veltman functions~\cite{Passarino:1978jh} following the {\tt LoopTools} convention~\cite{Hahn:1998yk}:  $A_0 = A_0(M_W^2)$, $B_{0,1}^\alpha = B_{0,1}(m_\alpha^2, M_W^2, m_i^2)$, $B_{0,1}^\beta = B_{0,1}(m_\beta^2, M_W^2, m_j^2)$, $B_0^q = B_0(q^2, m_i^2, m_j^2)$,  and $C_{rs} = C_{rs}(m_\alpha^2, q^2, m_\beta^2,$ $M_W^2, m_i^2, m_j^2)$ (with $rs = 0, 1, 2, 11, 22, 12, 00$); we integrate in $D=4-2\varepsilon$ dimensions.
\begin{eqnarray}
    F_V^{L\:(b)} &=& \frac{g_w^3 c_w}{64 \pi^2 \,M_W^4}\sum_i \,\mathcal U_{\alpha i}\,\mathcal U_{\beta i}^\ast\,\Bigg\{(4 M_W^2 \,q^2 - 8 M_W^4)\,B_0^q + (4 M_W^2 - 2 q^2)\,B_{00}^q - 4 M_W^2\, A_0\nonumber\\
    &\phantom{=}& + (4 M_W^4 - 2 m_i^2 \,M_W^2)\,(B_0^\alpha + B_0^\beta) - 2 M_W^2\,(m_\alpha^2 \,B_1^\alpha + m_\beta^2\, B_1^\beta)\nonumber\\
    &\phantom{=}& + \Big[(m_\alpha^2 + m_\beta^2)\,\left(2 m_i^2\, M_W^2 - 4M_W^4 - m_i^2 \,q^2\right) + 4 m_i^2\, M_W^2\,(q^2 - 2 M_W^2) + 8 M_W^6\Big]C_0\nonumber\\
    &\phantom{=}& + \Big[m_\alpha^2\left(2(D-8) \,M_W^4 + 2 m_i^2\,(2 M_W^2 - q^2) + m_\beta^2\,(2 M_W^2 - q^2) + 2 M_W^2 \,q^2\right) \nonumber\\
    &\phantom{=}&  + m_\beta^2\left(m_i^2\,(2 M_W^2 - q^2) - 2 M_W^2\,(2 M_W^2 - q^2)\right)\Big]C_1\nonumber\\
    &\phantom{=}& +\Big[m_\beta^2\left(2(D-8) \,M_W^4 + 2 m_i^2\,(2 M_W^2 - q^2) + m_\alpha^2\,(2 M_W^2 - q^2) + 2 M_W^2 \,q^2\right) \nonumber\\
    &\phantom{=}&  + m_\alpha^2\left(m_i^2\,(2 M_W^2 - q^2) - 2 M_W^2\,(2 M_W^2 - q^2)\right)\Big]C_2\nonumber\\
    &\phantom{=}& +\Big[m_\alpha^2\, m_\beta^2\,(2 M_W^2 - q^2) + m_\alpha^2( 2 m_i^2 \,M_W^2 - m_i^2\, q^2 + 2(D-2)\,M_W^4)\Big]C_{11}\nonumber\\
    &\phantom{=}& +\Big[m_\alpha^2 \,m_\beta^2\,(2 M_W^2 - q^2) + m_\beta^2\,( 2 m_i^2 \,M_W^2 - m_i^2\, q^2 + 2(D-2)\,M_W^4)\Big]\,C_{22}\nonumber\\
    &\phantom{=}& +\Big[2 m_\alpha^2 \,m_\beta^2\,(2 M_W^2 - q^2) +(m_\alpha^2 + m_\beta^2)\,\left(2 m_i^2 \,M_W^2 - m_i^2\, q^2 + 2(D-2)\,M_W^4\right)\Big]\,C_{12}\nonumber\\
    &\phantom{=}& +\Big[4(D-2)\,M_W^4 + 2 m_i^2 \,(2 M_W^2 - q^2)\Big]\,C_{00}\Bigg\}\,,\label{eqn:FVLbU}\\
    F_V^{R\:(b)} &=& \frac{g_w^3 c_w}{64 \pi^2 \,M_W^4}m_\alpha m_\beta\sum_i \,\mathcal U_{\alpha i}\,\mathcal U_{\beta i}^\ast\,\Bigg\{[4 M_W^2 - 2 q^2]\,(m_i^2\,C_0 + C_{00}) \nonumber\\
    &\phantom{=}& + [(m_\alpha^2 + 3 m_i^2)\,(2 M_W^2 - q^2) + 2(D-2)\,M_W^4]\,C_1\nonumber\\
    &\phantom{=}& + [(m_\beta^2 + 3 m_i^2)\,(2 M_W^2 - q^2) + 2(D-2)\,M_W^4]\,C_2\nonumber\\
    &\phantom{=}& + [(m_\alpha^2 + m_i^2)\,(2M_W^2 - q^2) + 2(D-2)\,M_W^4]\,C_{11}\nonumber\\
    &\phantom{=}& + [(m_\beta^2 + m_i^2)\,(2M_W^2 - q^2) + 2(D-2)\,M_W^4]\,C_{22}\nonumber\\
    &\phantom{=}& + [(m_\alpha^2 + m_\beta^2 + 2 m_i^2)\,(2M_W^2 - q^2) + 4(D-2)\,M_W^4]\,C_{12}\Bigg\}\,,\label{eqn:FVRbU}\\
    F_T^{L\:(b)} \!\!\! \!&=& -\frac{ig_w^3 c_w}{64 \pi^2 \,M_W^4}m_\alpha \sum_i \,\mathcal U_{\alpha i}\,\mathcal U_{\beta i}^\ast\,\Bigg\{m_i^2\,[2 M_W^2 - q^2] \,C_0 + 2[m_i^2\,(2M_W^2 - q^2) + (D-4)\,M_W^4]\,C_1 \nonumber\\
    \!\!\! \!&\phantom{=}& + [(m_i^2 + m_\beta^2 - 2 M_W^2)\,(2M_W^2 - q^2)]\,C_2 + [m_i^2\,(2M_W^2 - q^2) + 2(D-2)\,M_W^4]\,C_{11}\nonumber\\
    \!\!\! \!&\phantom{=}& + [m_\beta^2\,(2M_W^2 - q^2)]\,C_{22} + [(m_i^2 + m_\beta^2)\,(2M_W^2 - q^2) + 2(D-2)\,M_W^4]\,C_{12}\Bigg\}\,,\label{eqn:FTLbU}\\
    F_T^{R\:(b)} &=& -\frac{ig_w^3 c_w}{64 \pi^2\, M_W^4}m_\beta \sum_i \,\mathcal U_{\alpha i}\,\mathcal U_{\beta i}^\ast\,\Bigg\{m_i^2\,[2 M_W^2 - q^2] \,C_0 + [(m_i^2 + m_\alpha^2 - 2 M_W^2)\,(2M_W^2 - q^2)]\,C_1\nonumber\\
   &\phantom{=}& + 2[m_i^2\,(2M_W^2 - q^2) + (D-4)\,M_W^4]\,C_2 + [m_\alpha^2\,(2M_W^2 - q^2)]\,C_{11}\nonumber\\
   &\phantom{=}&+ [m_i^2\,(2M_W^2 - q^2) + 2(D-2)\,M_W^4]\,C_{22} + [(m_i^2 + m_\alpha^2)\,(2M_W^2 - q^2) \nonumber \\
  &\phantom{=}& 
  +2(D-2)\,M_W^4]\,C_{12}\Bigg\}\,,\label{eqn:FTRbU}
\end{eqnarray}
with the Passarino-Veltman functions $A_0 = A_0(M_W^2)$, $B_{0,00}^q = B_{0,00}(q^2, M_W^2, M_W^2)$,  $B_{0,1}^{\alpha, \beta} = B_{0,1}(m_{\alpha, \beta}^2,$ $ m_i^2, M_W^2)$ 
and $C_{rs} = C_{rs}(m_\alpha^2, q^2, m_\beta^2, m_i^2, M_W^2, M_W^2)$ (with $rs = 0, 1, 2, 11, 22, 12, 00$). 

\noindent 
Finally, the self-energy type contributions are given by
\begin{eqnarray}
    F_V^{L\:(c+d)} &=& \frac{g_w^3 (1 - 2 s_w^2)}{64 \pi^2 \,c_w \,M_W^2 (m_\alpha^2 - m_\beta^2)}\sum_i \, \mathcal U_{\alpha i} \, \mathcal U_{\beta i}^\ast \Bigg\{(m_\alpha^2 - m_\beta^2)\,A_0 \nonumber\\
    &\phantom{=}& + \Big[m_\alpha^2\,(m_i^2 - (D-2)\,M_W^2) - (m_i^2 - M_W^2)\,(m_i^2 + (D-2) \,M_W^2)\Big]\,B_0^\alpha\nonumber\\
    &\phantom{=}& - \Big[m_\beta^2\,(m_i^2 - (D-2)\,M_W^2) - (m_i^2 - M_W^2)\,(m_i^2 + (D-2)\, M_W^2)\Big]\,B_0^\beta\nonumber\\
    &\phantom{=}& + m_\alpha^2\Big[ m_\alpha^2 - m_i^2 - (D-2)\,M_W^2\Big]\,B_1^\alpha\nonumber\\
    &\phantom{=}& - m_\beta^2\Big[ m_\beta^2 - m_i^2 - (D-2)\,M_W^2\Big]\,B_1^\beta\Bigg\} \,,\label{eqn:FVLcdU}\\
    F_V^{R\:(c+d)} &=& \frac{g_w^3 (2 s_w^2)}{64\pi^2 \,c_w \,M_W^2 }\frac{m_\alpha m_\beta}{(m_\alpha^2 - m_\beta^2)}\sum_i \, \mathcal U_{\alpha i} \, \mathcal U_{\beta i}^\ast \Bigg\{[m_\alpha^2 - m_i^2 - M_W^2]\,B_0^\alpha - [m_\beta^2 - m_i^2 - M_W^2]\,B_0^\beta\nonumber\\
    &\phantom{=}& + [m_\alpha^2 - m_i^2 - (D-2)\,M_W^2]\,B_1^\alpha -  [m_\beta^2 - m_i^2 - (D-2)\,M_W^2]\,B_1^\beta\Bigg\}\,,\label{eqn:FVRcdU}
\end{eqnarray}
with $A_0 = A_0(M_W^2)$, and $B_{0,1}^{\alpha, \beta} = B_{0,1}(m_{\alpha, \beta}^2, m_i^2, M_W^2)$, and $F_T^{L\:(c+d)} = F_T^{R\:(c+d)} = 0$.

In Eqs.~(\ref{eqn:FVLadU}-\ref{eqn:FVRcdU}), $m_{\alpha(\beta)}$ are the charged lepton masses while $m_{i(j)}$ are the physical masses of the neutrinos; we further introduced the abbreviations $c_w(s_w)\equiv\cos{\theta_w}(\sin{\theta_w})$. 

Note that in all above contributions, the terms proportional to $A_0(M_W^2)$ and to $B_{0,00}(q^2, M_W^2, M_W^2)$ do not contribute to the amplitude since they do not depend on the internal neutrino masses (and therefore cancel due to the (semi-)unitarity\footnote{We recall that only the upper $3\times(3+n_S)$ block of $\mathcal{U}$ enters in the interaction vertices.} of $\mathcal U$).
Among the distinct contributions listed above, terms proportional to the $B_{0,1}$ and $C_{00}$ functions contain ultraviolet (UV) divergences which are regulated dimensionally (we integrate in $(4-2\varepsilon)$ dimensions). In the case of on-shell $Z$-boson decays ($q^2 = M_Z^2$), the sum over all contributions is, however, manifestly finite as shown in Appendix~\ref{app:ZHdecays:formulae}.
We thus set $q^2 = M_Z^2$ and safely take the limit $D \to 4$.

A few additional comments are in order:
firstly, we note here that all contributions, except $F_V^L$, are generically suppressed by at least one power of the charged lepton masses, and can be therefore safely neglected in what regards their contribution to the decay rate.
Secondly, note that functions of the form $B_{0,1}(m_{\alpha,\beta}^2, M_W^2, m_i^2)$ (or similar) contain potentially problematic large logarithms $\sim \log(m_i^2/M_W^2)$ be it in the limit of very small or very large neutrino masses.
However, in the sum over all contributions (and internal states) these logarithms cancel and, as expected, the light neutrino contribution to the amplitude is negligible.
Finally, we have numerically checked that our results agree with those of~\cite{DeRomeri:2016gum, Abada:2014cca,  Illana:2000ic}, which were obtained in the limit of vanishing charged lepton masses.

Neglecting suppressed contributions, the decay rate can be written as\footnote{The full decay rate (including the suppressed contributions) is given in Appendix~\ref{app:ZHdecays:formulae}.}
\begin{eqnarray}\label{eq:Zwidth:compact}
 \Gamma(Z\to \ell_\alpha^-\ell_\beta^+) &\simeq & \frac{1}{48\pi^2 M_Z}\sqrt{1 - \frac{(m_\alpha + m_\beta)^2}{M_Z^2}}\sqrt{1 - \frac{(m_\alpha - m_\beta)^2}{M_Z^2}}\times\nonumber \\
 &&     \times \left[2 M_Z^2 - (m_\alpha^2 + m_\beta^2) - \frac{(m_\alpha^2 - m_\beta^2)^2}{M_Z^2}\right]\,|F_V^L|^2.
\end{eqnarray}
We notice that concerning the evaluation of the Passarino-Veltman functions we use the public Fortran code 
{\tt LoopTools}~\cite{Hahn:1998yk} (wrapped into a dedicated python code).
From a numerical point of view, the observable evaluated is 
the CP averaged decay rate
\begin{equation}
    \Gamma (Z\to \ell_\alpha^\pm \ell_\beta^\mp) = \frac{1}{2}\left[\Gamma(Z\to \ell_\alpha^+\ell_\beta^-) + \Gamma(Z\to \ell_\alpha^-\ell_\beta^+)\right]\,,
\end{equation}
as current search results do not distinguish the charges of the final state leptons. 

\mathversion{bold}
\subsection{Higgs decays: $H \to \ell_\alpha^\mp \ell_\beta^\pm$}\label{sec:Hdecays:th}
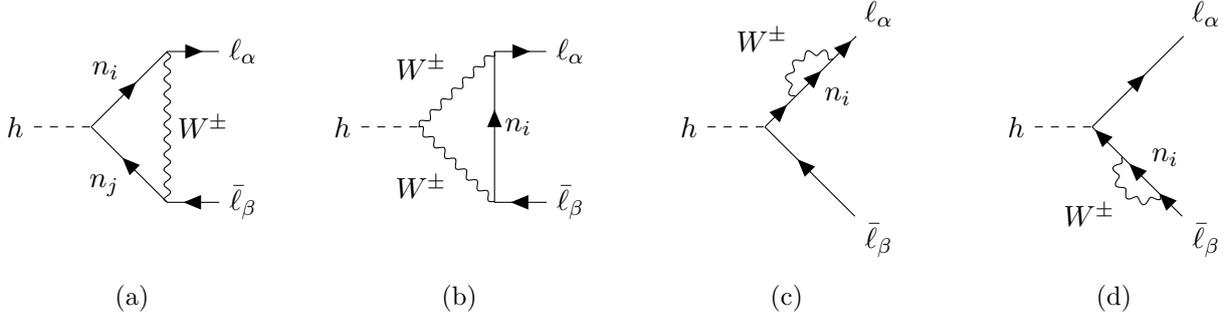
\begin{figure}
    \centering
    \begin{subfigure}[b]{0.24\textwidth}
    \centering
 \raisebox{5mm}{    \begin{tikzpicture}
    \begin{feynman}
    \vertex (a) at (0,0) {\(h\)};
    \vertex (b) at (1,0);
    \vertex (c) at (2,1.);
    \vertex (d) at (2,-1);
    \vertex (e) at (3,1) {\( \ell_\alpha\)};
    \vertex (f) at (3,-1) {\( \bar\ell_\beta\)};
    \diagram* {
    (a) -- [scalar] (b),
    (b) -- [fermion, edge label=\(n_i\)] (c),
    (c) -- [boson, edge label=\( W^\pm\)] (d),
    (d) -- [fermion, edge label=\(n_j\)] (b),
    (c) -- [fermion] (e),
    (f) -- [fermion] (d)
    };
    \end{feynman}
    \end{tikzpicture}
    }
            \caption*{(a)}
            \label{}
    \end{subfigure}
    \hfill
\begin{subfigure}[b]{0.24\textwidth}
    \centering
\raisebox{5mm}{        \begin{tikzpicture}
    \begin{feynman}
    \vertex (a) at (0,0) {\(h\)};
    \vertex (b) at (1,0);
    \vertex (c) at (2,1.);
    \vertex (d) at (2,-1.);
    \vertex (e) at (3,1.) {\( \ell_\alpha\)};
    \vertex (f) at (3,-1.) {\( \bar\ell_\beta\)};
    \diagram* {
    (a) -- [scalar] (b),
    (b) -- [boson, edge label=\( W^\pm\)] (c),
    (c) -- [anti fermion, edge label=\(n_i\)] (d),
    (d) -- [boson, edge label=\( W^\pm\)] (b),
    (c) -- [fermion] (e),
    (f) -- [fermion] (d)
    };
    \end{feynman}
    \end{tikzpicture}
    }
            \caption*{(b)}
            \label{}
    \end{subfigure}
    \hfill
    \begin{subfigure}[b]{0.24\textwidth}
    \centering
    \begin{tikzpicture}
    \begin{feynman}
    \vertex (a) at (0,0) {\(h\)};
    \vertex (b) at (1,0);
    \vertex (c) at (1.4, 0.4);
    \vertex (d) at (1.9,0.9);
    \vertex (e) at (2.5, 1.5) {\( \ell_\alpha\)};
    \vertex (f) at (2.5,-1.5) {\( \bar\ell_\beta\)};
    
    \diagram* {
    (a) -- [scalar] (b),
    (b) -- [fermion] (c) -- [fermion, edge label'=\(n_i\)] (d) -- [fermion] (e),
    (b) -- [anti fermion] (f),
    (c) -- [boson, half left, edge label=\( W^\pm\)] (d)
    };   
    \end{feynman}
    \end{tikzpicture}
            \caption*{(c)}
            \label{}
    \end{subfigure}
    \hfill
    \begin{subfigure}[b]{0.24\textwidth}
    \centering
    \begin{tikzpicture}
    \begin{feynman}
    \vertex (a) at (0,0) {\(h\)};
    \vertex (b) at (1,0);
    \vertex (c) at (1.4, -0.4);
    \vertex (d) at (1.9, -0.9);
    \vertex (e) at (2.5, 1.5) {\( \ell_\alpha\)};
    \vertex (f) at (2.5,-1.5) {\( \bar\ell_\beta\)};
    
    \diagram* {
    (a) -- [scalar] (b),
    (b) -- [anti fermion] (c) -- [anti fermion, edge label=\(n_i\)] (d) -- [anti fermion] (f),
    (b) -- [fermion] (e),
    (c) -- [boson, half right, edge label'=\( W^\pm\)] (d)
    };   
    \end{feynman}
    \end{tikzpicture}
            \caption*{(d)}
            \label{}
    \end{subfigure}
    \caption{Feynman diagrams contributing to cLFV Higgs decays (in unitary gauge).}
    \label{fig:cLFVHiggsdecays:UG}
\end{figure}

\mathversion{normal}
In the presence of $n_S$ sterile fermions, one finds new contributions leading to cLFV $H \to \ell_\alpha^\mp \ell_\beta^\pm$ ($\alpha \neq \beta$). Working in unitary gauge, one can identify three types of diagrams as depicted in Fig.~\ref{fig:cLFVHiggsdecays:UG}: corrections to the $H\ell \ell$ vertex due to two neutrinos (cf. diagram (a)), two $W^\pm$ bosons and one neutrino (diagram (b)), as well as corrections to the fermion legs (diagrams (c) and (d)). 

We have carried out the evaluation of the new contributions, leading to the following expression for the branching ratio:
\begin{equation}
    \text{BR}(H \to \ell_\alpha^- \ell_\beta^+) \,=\,
    \frac{\Gamma(H \to \ell_\alpha^- \ell_\beta^+)}{\Gamma_H}\,,
\end{equation}
where $\Gamma_H$ denotes the total Higgs width.
The cLFV width can be cast as 
\begin{eqnarray}\label{eq:cLFVHdecay}
    \Gamma(H \to \ell_\alpha \ell_\beta) & = &
    \frac{1}{16\pi \, M_H}\, 
    \sqrt{\left( 1 - \left( \frac{m_{\ell_\alpha} + m_{\ell_\beta}}{M_H}\right)^2\right)\, 
    \left( 1 - \left( \frac{m_{\ell_\alpha} - m_{\ell_\beta}}{M_H}\right)^2\right)
    } \, \nonumber \\
    && 
    \times \left[
    \left( M_H^2 - m_{\ell_\alpha}^2 - m_{\ell_\beta}^2\right)\,
    \left( |\Delta F_L|^2 +  |\Delta F_R|^2 \right)
   \right.  \nonumber \\
    && \left. \phantom{m_{\ell_\beta}^2}\quad 
    -4\, m_{\ell_\alpha}\, m_{\ell_\beta}\, 
    \text{Re}\left( \Delta F_L \, \Delta F_R^*\right)\right]\,, 
\end{eqnarray}
with $M_H$ the SM Higgs mass, and 
\begin{equation}
\Delta F_{L(R)} \, =\, \sum_{i} F_{L(R)}^{i} \,,
\quad i= \text{(a),\,(b),\,(c) }\text{and } \text{(d)}\,,
\end{equation}
as illustrated in Fig.~\ref{fig:cLFVHiggsdecays:UG}. Below, we collect the expressions for each of the form factors, $F_{L(R)}^{i}$, cast in terms of the relevant entries of the leptonic mixing matrix $\mathcal{U}_{\alpha i}$, the masses of the external leptons, $m_{\alpha(\beta)}$, and finally of the physical masses of the neutral fermions running in the loop, $m_{i(j)}$ $(i,j=1,..., 3+n_S)$. As before, our results are given in terms of Passarino-Veltman functions following {\tt LoopTools} conventions; for previous works - relying on different Passarino-Veltman basis -, see~\cite{Arganda:2014dta,Arganda:2004bz,Hernandez-Tome:2020lmh,Thao:2017qtn}.
\begin{eqnarray}
F_L^{(a)} &=& \frac{-g_w^3}{64\pi^2}\frac{m_\alpha}{M_W^3} \sum_{i,j=1}^{3+n_S} \mathcal{U}_{\alpha i} \,\mathcal{U}_{\beta j}^*\left\{C_{ij} \left[m_j^2 \,B_{0}^q - m_i^2 \,B_{1}^\alpha - (D-3) \,m_j^2 \,M_W^2 \, C_{0} \right.\right.\nonumber\\
&\phantom{=}& + \left. \left[m_\alpha^2 \,m_j^2+m_i^2\, \left(m_\beta^2-2 m_j^2\right)- (D-2)\, M_W^2 \left(m_i^2+m_j^2\right)\right] \,C_{1} \right] \nonumber\\
&\phantom{=}& + \, C_{ij}^* m_i \,m_j \left[B_{0}^q - B_{1}^\alpha - (D-3)\, M_W^2 \,C_{0}\right.\nonumber\\
&\phantom{=}&\left.\left. + \left(m_\alpha^2+m_\beta^2-m_i^2-m_j^2- 2(D-2) \,M_W^2\right)\, C_{1}\right] 
 \right\} \, ,\label{eqn:FLa}\\
F_R^{(a)} &=& \frac{-g_w^3}{64\pi^2}\frac{m_\beta}{M_W^3}\sum_{i,j=1}^{3+n_S} \mathcal{U}_{\alpha i}\, \mathcal{U}_{\beta j}^* \left\{
C_{ij}\left[ m_i^2 \,B_{0}^q  - m_j^2 \,B_{1}^\beta  - (D-3)\, m_i^2\, M_W^2 \,C_{0}   \right.\right. \nonumber\\
&\phantom{=}& + \left.\left[m_\alpha^2\, m_j^2+m_i^2 \,\left(m_\beta^2-2 m_j^2\right)- (D-2)\, M_W^2\, \left(m_i^2+m_j^2\right)\right]\,C_{2} \right]\nonumber\\
&\phantom{=}& +  \, C_{ij}^* m_j  \,m_i \left[ B_{0}^q - B_{1}^\beta - (D-3) \,M_W^2 \,C_{0} \right.\nonumber\\
&\phantom{=}& + \left.\left. \left(m_\alpha^2+m_\beta^2-m_i^2-m_j^2- 2(D-2) \,M_W^2\right)\,C_{2} \right] \right\} \, \label{eqn:FRa},
\end{eqnarray}
with the Passarino-Veltman functions $B_0^q = B_0\left(q^2,m_i^2,m_j^2\right)$, $B_{1}^{\beta} = B_{1}\left(m_\beta^2,M_W^2,m_j^2\right)$, $B_1^\alpha = B_1\big(m_\alpha^2,M_W^2,$ $m_i^2\big)$ and $C_{rs} = C_{rs}\left(m_\alpha^2,q^2,m_\beta^2,M_W^2,m_i^2,m_j^2\right)$ (with $rs = 0, 1, 2$).
\begin{eqnarray}
F_L^{(b)} &=& 
\frac{-g_w^3}{64\pi^2}\frac{m_\alpha}{M_W^3} \sum_{i=1}^{3+n_S} \mathcal{U}_{\alpha i}\, \mathcal{U}_{\beta i}^* \left\{ - m_i^2\, B_{0}^\alpha - \left(m_i^2-2 M_W^2\right)\, B_{0}^\beta - 2 M_W^2\, B_{0}^q 
\right. \nonumber\\
&\phantom{=}& - m_i^2 \,B_{1}^\alpha - m_\beta^2 \,B_{1}^\beta - \left(2 M_W^2+q^2\right) \,B_{1}^q - A_{0}(M_W)  \nonumber\\
&\phantom{=}& + \left[-2 m_\alpha^2 \,M_W^2+m_i^2 \,q^2+2 M_W^4\right] \,C_{0} \nonumber\\
&\phantom{=}& + \left[-2 M_W^2 \left(m_\alpha^2+m_\beta^2-m_i^2\right)+m_i^2 \,q^2+ 2(D-2)\, M_W^4\right]\,C_{1} \nonumber\\
&\phantom{=}& + \left.  m_\beta^2 \left(q^2-2 M_W^2\right)\, C_{2}\right\} \, ,\\
F_R^{(b)} &=&
\frac{-g_w^3}{64\pi^2}\frac{m_\beta}{M_W^3} \sum_{i=1}^{3+n_S} \mathcal{U}_{\alpha i} \,\mathcal{U}_{\beta i}^* \left\{ - \left(m_i^2-2 M_W^2\right) \,B_{0}^\alpha - m_i^2 \,B_{0}^\beta + q^2 \,B_{0}^q \right. \nonumber\\
&\phantom{=}& - m_\alpha^2 \,B_{1}^\alpha -m_i^2 \,B_{1}^\beta + \left(2 M_W^2 +q^2 \right)\,B_{1}^q - A_{0}(M_W) \nonumber\\
&\phantom{=}& + \left[m_i^2 \,q^2 -2 M_W^2\, m_\beta^2 + 2 M_W^4 \right] \,C_{0} \nonumber\\
&\phantom{=}& + m_\alpha^2 \left(q^2-2 M_W^2\right) \,C_{1} \nonumber\\
&\phantom{=}& + \left. \left[2 m_i^2 \,M_W^2 +m_i^2 \,q^2 + 2(D-2)\, M_W^4 - 2 M_W^2\left(m_\alpha^2+m_\beta^2\right) \right]
\,C_{2}  \right\} \, ,
\end{eqnarray}
with $B_{0,1}^{\alpha,\beta} = B_{0,1}\left(m_{\alpha, \beta}^2,m_i^2,M_W^2\right) $, $B_{0,1}^{q} = B_{0,1}\left(q^2,M_W^2,M_W^2\right)$ and $C_{rs} \equiv C_{rs}\left(m_\alpha^2,q^2,m_\beta^2,m_i^2,M_W^2,M_W^2\right)$ (with $rs = 0, 1, 2$).
\begin{eqnarray}
F_L^{(c+d)} &=& \frac{-g_w^3}{64 \pi^2}\frac{m_\alpha}{M_W^3 (m_\alpha^2-m_\beta^2)} \sum_{i=1}^{3+n_S}\mathcal{U}_{\alpha i}\, \mathcal{U}_{\beta i}^* 
\left\{m_\beta^2 \left(-m_\alpha^2+m_i^2+M_W^2\right)\, B_{0}^\alpha
\right.\nonumber \\
&\vphantom{=}&+ m_\beta^2\left(-m_\alpha^2+m_i^2+ (D-2) \,M_W^2\right) \,B_{1}^\alpha \nonumber \\
&\vphantom{=}&+ \left[m_\beta^2 - m_i^2 - (D-2) \,M_W^2 \right]\,A_{0}(m_i^2) + \left[-m_\beta^2+m_i^2+ (D-3) \,M_W^2\right] A_{0}(M_W^2) + D \,A_{00}(M_W^2) \nonumber \\
&\vphantom{=}&+ \left[(D-3)\,m_i^2 \,M_W^2 + m_i^4 -m_\beta^2 \,m_i^2 + (D-2)\,m_\beta^2\, M_W^2 -(D-2) \,M_W^4  \right]\,B_{0}^\beta\nonumber \\
&\vphantom{=}&+\left.m_\beta^2 \left(-m_\beta^2+m_i^2+(D-2)\, M_W^2\right) B_{1}^\beta
\right\} \, ,\\
F_R^{(c+d)} &=& \frac{-g_w^3}{64\pi^2}\frac{m_\beta}{M_W^3 (m_\alpha^2-m_\beta^2)} \sum_{i=1}^{3+n_S} \mathcal{U}_{\alpha i}\, \mathcal{U}_{\beta i}^* \left\{
- m_\alpha^2 \left(-m_\beta^2+m_i^2+M_W^2\right) \,B_{0}^\beta \right.
\nonumber\\
&\vphantom{=}& -m_\alpha^2 \left(-m_\beta^2+m_i^2+(D-2) \,M_W^2\right) \,B_{1}^\beta \nonumber\\
&\vphantom{=}&- \left[m_\alpha^2 - m_i^2 - (D-2) \,M_W^2  \right] A_{0}(m_i^2) -
\left[-m_\alpha^2 + m_i^2 + (D-3)\, M_W^2\right]\,A_{0}(M_W^2) - D\, A_{00}(M_W^2) \nonumber\\
&\vphantom{=}&- \left[(D-3)\,m_i^2\, M_W^2 + m_i^4 
 - m_\alpha^2 \,m_i^2 + (D-2)\, m_\alpha^2 \, M_W^2 - (D-2) \,M_W^4  \right] \,B_{0}^\alpha \nonumber\\
&\vphantom{=}&- \left. m_\alpha^2 \left(-m_\alpha^2 + m_i^2 +  (D-2)\, M_W^2\right) \,B_{1}^\alpha \right\} \, ,
\end{eqnarray}
where $B_{0,1}^{\alpha,\beta}= B_{0,1}\left(m_{\alpha, \beta}^2,m_i^2,M_W^2\right)$.

\medskip
The following section will be devoted to numerically studying the 
rates for the cLFV Higgs and $Z$ decays, in particular emphasising the role of the leptonic CPV Dirac and Majorana phases.

\section{cLFV neutral boson decays: the impact of leptonic CP phases}\label{sec:results}
We now proceed to address how the presence of the Dirac and Majorana CP violating phases can modify the predicted rates for cLFV $Z$ and Higgs decays. We first discuss ``simplified scenarios", without aiming at a complete phenomenological discussion; a thorough phenomenological study will be carried out in subsequent sections.

\subsection{Simplified scenarios}\label{sec:simple_plots}
We begin by considering some illustrative scenarios, whose aim is to offer a first insight into the role of phases in neutral boson decays, and a direct comparison of the CP conserving limit with the CPV case (non-vanishing Dirac and/or Majorana phases). Notice that at this stage we do not apply any constraints, so that one might encounter scenarios which are already experimentally excluded.
This first exploratory study relies on simple assumptions: 
we set $\theta_{\alpha 4} = \theta_{\alpha 5}$ and, concerning the new sterile states, we assume that they are heavy and mass-degenerate, above the electroweak (EW) scale, typically of the order of a few~TeV, $m_4 = m_5 \gg \Lambda_\text{EW}$. 

\begin{figure}[t!]
    \centering
\mbox{   \hspace*{-5mm}  \includegraphics[width=0.51\textwidth]{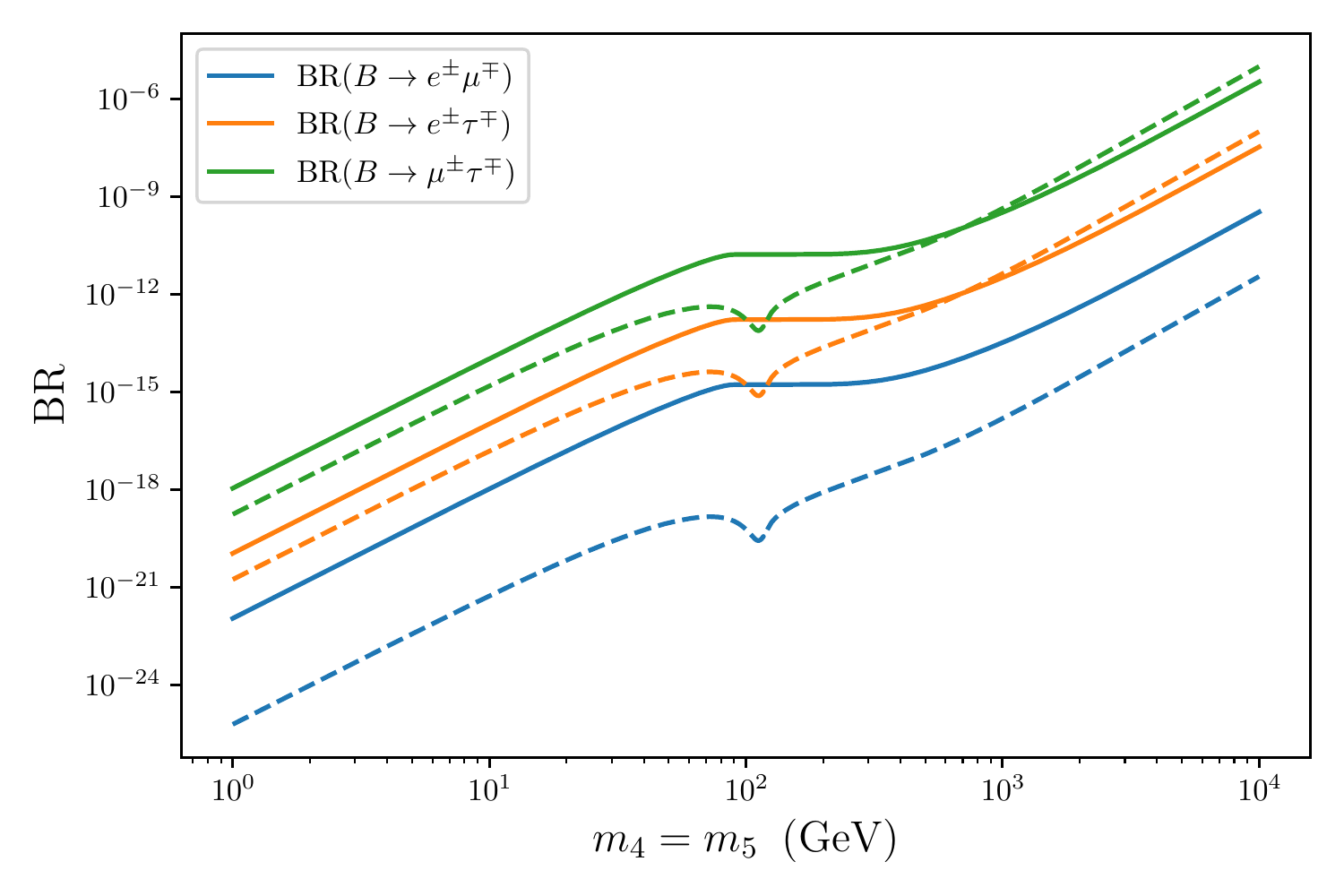}\hspace*{2mm}     }
    \caption{ Rates of cLFV Higgs and $Z$ decays as a function of the degenerate heavy sterile mass, $m_4=m_5$ (in~GeV), for vanishing CPV phases. 
    We fix $\theta_{1j}=10^{-3}$, $\theta_{2j}=0.01$ and $\theta_{3j}=0.1$ ($j=4,5$). We display
BR($B \to e \mu$) (blue), BR($B \to e \tau $) (orange) and BR($B \to  \mu \tau$) (green), for $B=Z$ (full lines) and 
$B=H$ (dashed lines).
    }
    \label{fig:Higgs_Z_cLFV.m45}
\end{figure}

In particular, we consider a simple benchmark choice for the values of the mixing angles,
$\sin\theta_{14}=\sin\theta_{15}= 10^{-3}$, 
$\sin\theta_{24}=\sin\theta_{25}= 0.01$ and 
$\sin\theta_{34}=\sin\theta_{35}= 0.1$, 
as well as three representative values of the (degenerate)  heavy fermion masses $m_4 = m_5 =1, 5, 10$~TeV. 
In Fig.~\ref{fig:Higgs_Z_cLFV.m45} we display the cLFV rates for the $B \to \ell_\alpha \ell_\beta$ decay as a function of the degenerate masses of the heavy sterile states ($m_{4,5}$), in the CP conserving limit with all CPV phases set to zero.
Notice that in this figure (and generically in what follows), the observable associated with the cLFV decays of a $B$-boson refers to the average of the two combinations of final state charges,  
\begin{equation}
    \mathrm{BR}(B\to \ell_\alpha^\pm \ell_\beta^\mp) =  
    \dfrac{1}{2}\, \frac{\Gamma(B\to \ell_\alpha^+\ell_\beta^-) + \Gamma(B\to \ell_\alpha^-\ell_\beta^+)}{\Gamma_B}\,,
\end{equation}
with $B = H, Z$. In our analysis, we further take the following values for the neutral boson decay widths,
\begin{equation} \label{eq:HZwidths}
    \Gamma_Z\,=\, 2.4952 \text{ GeV} \quad \text{and} \quad 
    \Gamma_H\,=\, 4.115 \text{ MeV}\,,
\end{equation}
with $\Gamma_Z$ as given in~\cite{ParticleDataGroup:2020ssz}, and the SM prediction for $\Gamma_H$ from the~\href{https://twiki.cern.ch/twiki/bin/view/LHCPhysics/CERNYellowReportPageBR}{CERN Yellow Report}~\cite{LHCHiggsCrossSectionWorkingGroup:2016ypw} (for $M_H = 125.1\:\mathrm{GeV}$).

As seen from Fig.~\ref{fig:Higgs_Z_cLFV.m45}, the sterile states lead to significant contributions to the cLFV decay rates, both for the $Z$ and the Higgs; the dependence of the rates on the heavy states' mass confirms the findings of previous studies (see, e.g.~\cite{Illana:2000ic,Arganda:2014dta,Thao:2017qtn})\footnote{We have numerically verified that our predictions for the Higgs cLFV rates are in agreement with the results of~\cite{Arganda:2014dta,Thao:2017qtn}.}. In particular, and although in the present study no approximation is made regarding the mass of the final state leptons, the predictions for the cLFV $Z$ decay rates are in excellent agreement with the results of~\cite{DeRomeri:2016gum, Abada:2014cca,  Illana:2000ic}, obtained in the limit in which $m_{\alpha, \beta} \to 0$.

Although the decays are not formally identical (scalar vs.~vector boson decays), the common topology of the one-loop contributions leads to similar behaviours for the observables. 
In both cases, and in the limit of large $m_{4,5}$, the diagrams with two virtual neutrinos (diagrams (a) of 
Figs.~\ref{fig:cLFVZdecays:UG} and~\ref{fig:cLFVHiggsdecays:UG}), provide the dominant contribution.

\begin{figure}[t!]
    \centering
\mbox{   \hspace*{-5mm}  \includegraphics[ width=0.51\textwidth]{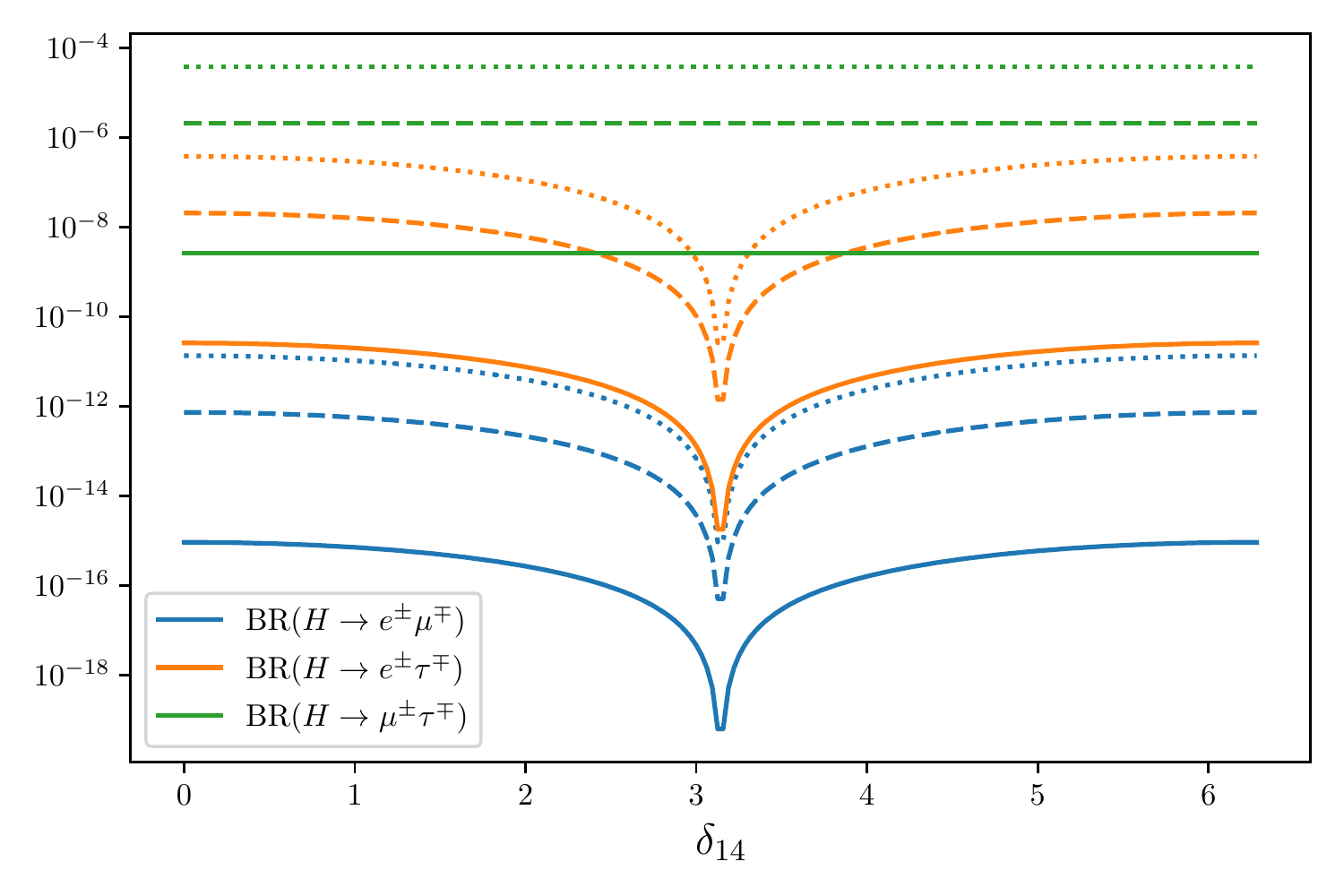}\hspace*{2mm} 
    \includegraphics[width=0.51\textwidth]{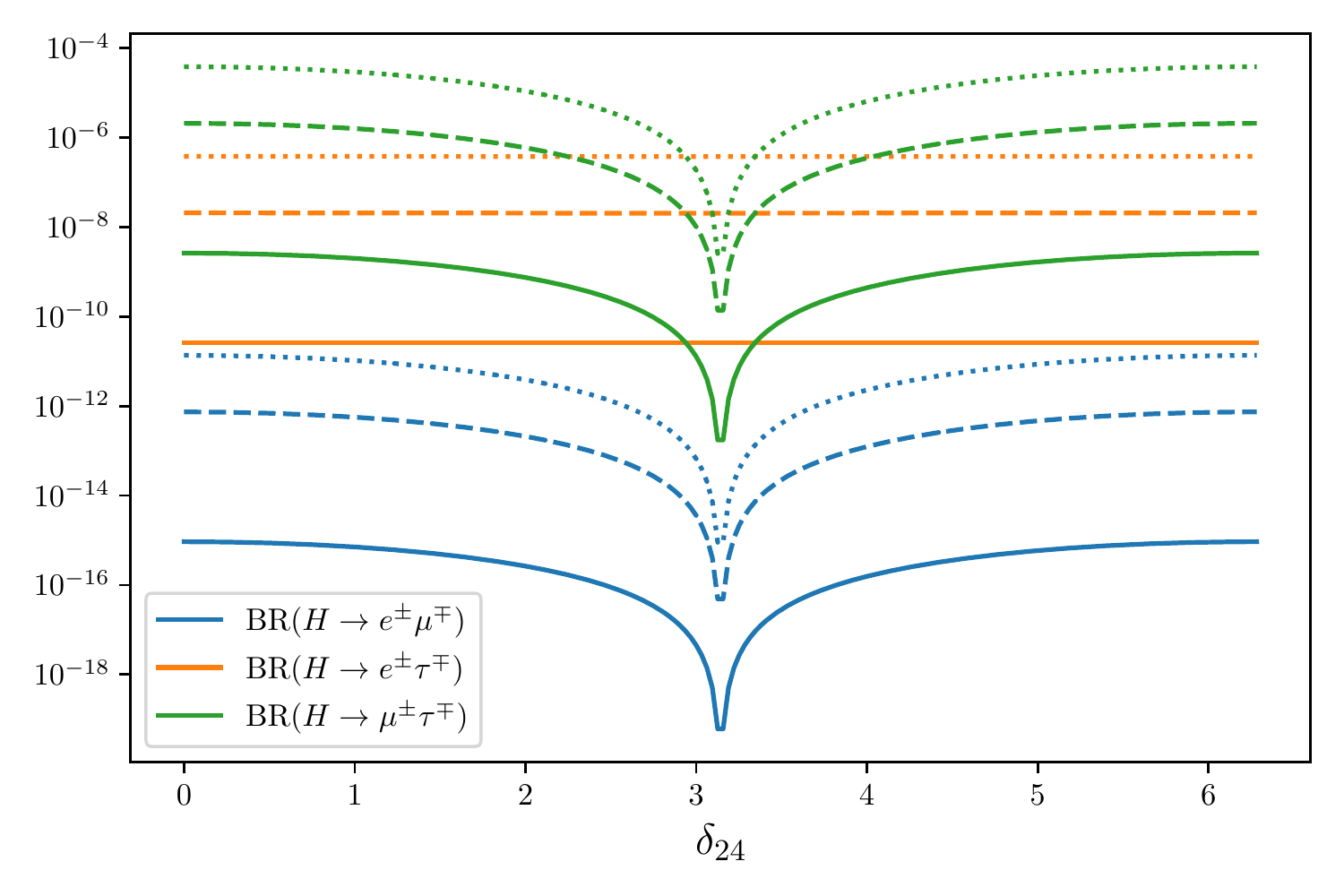}}
\mbox{   \hspace*{-5mm}  \includegraphics[ width=0.51\textwidth]{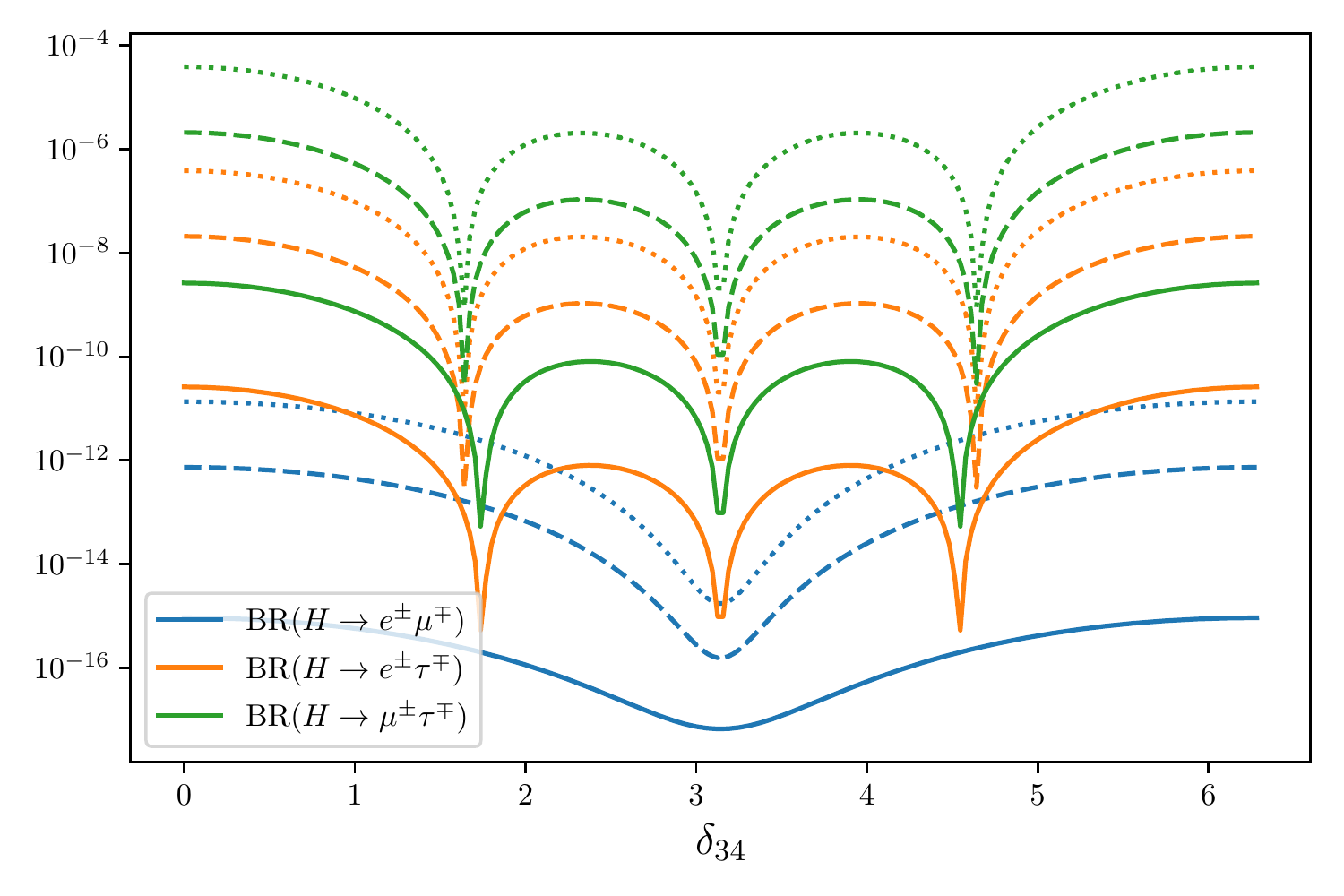}\hspace*{2mm} 
    \includegraphics[ width=0.51\textwidth]{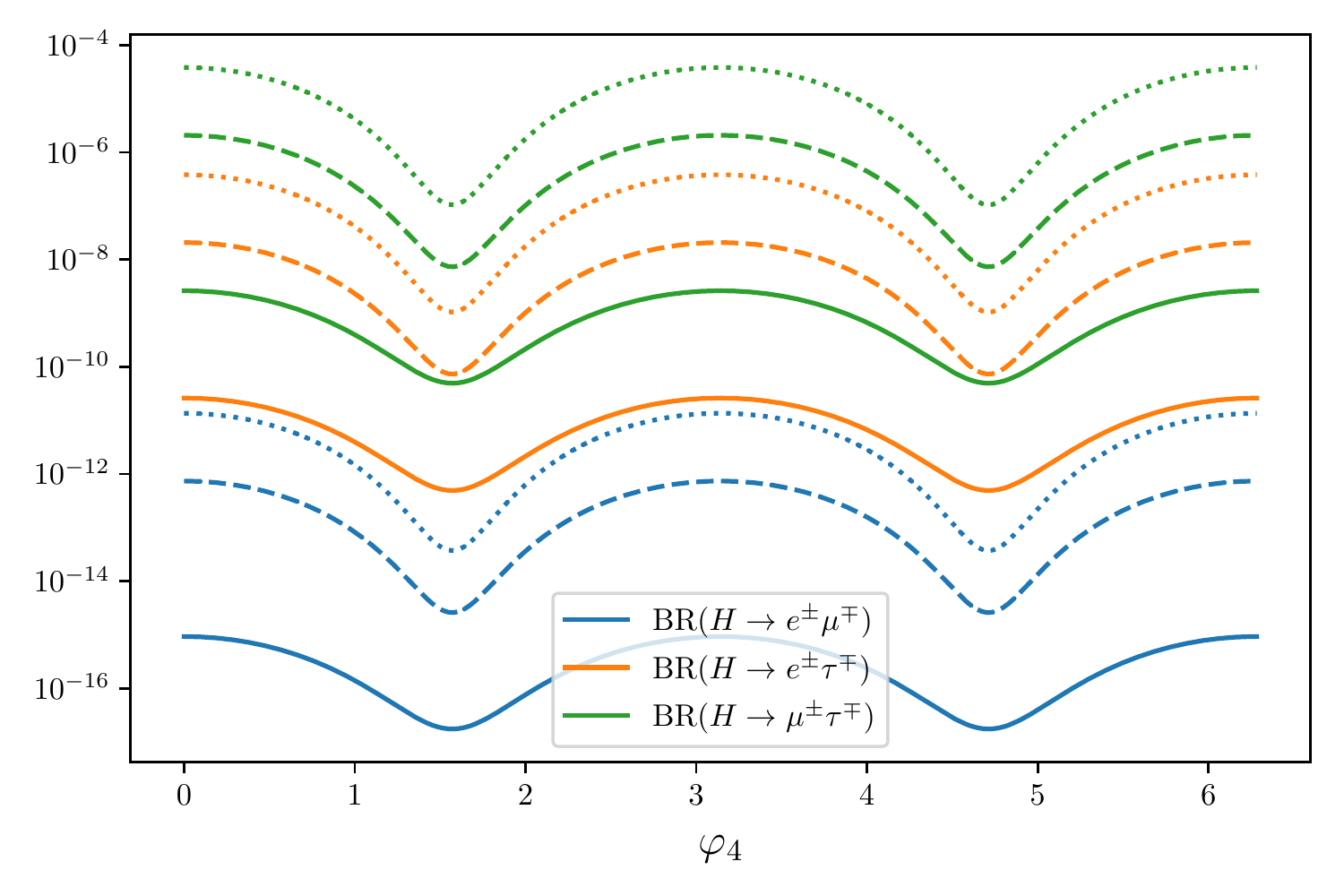}}
    \caption{Rates of cLFV Higgs decays as a function of the CP violating Dirac and Majorana phases. From left to right and top to bottom, dependence on $\delta_{14}$, $\delta_{24}$, $\delta_{34}$ and $\varphi_4$ (all phases set to zero in each case, except for the one displayed). 
    The colour code denotes the flavour composition of the final state lepton pair: $e\mu$ (blue), $e\tau$ (orange) and $\mu \tau$ (green).
    In both panels, solid, dashed and dotted lines respectively correspond to the following heavy neutral fermion masses: $m_4=m_5=1, 5, 10~\text{TeV}$.
    }
    \label{fig:Higgs_cLFV_delta14.24.34.phi4}
\end{figure}

In Fig.~\ref{fig:Higgs_cLFV_delta14.24.34.phi4} we now display the (individual) effects of Dirac and Majorana CPV phases for the distinct cLFV rates $H \to \ell_\alpha \ell_\beta$. As mentioned before, we also consider three regimes for the degenerate heavy fermion masses, $m_4=m_5=1, 5, 10~\text{TeV}$.
While the effects of the  Dirac phases $\delta_{14}$ ($\delta_{24}$) are absent for 
$\mu \tau$ ($e \tau$) final states - a consequence of the very simplifying hypotheses for the mixing angles - all cLFV Higgs decays exhibit a clear dependence on $\delta_{34}$. This is a consequence of the contributions proportional to $C_{ij} m^2_i$, which introduce 
$s_{34,35}$-enhanced terms, see Eqs.~(\ref{eqn:FLa},\ref{eqn:FRa}). In all cases, Dirac phases open the possibility of a strong suppression in the decay rates for values $\sim \pi$ (albeit for different cLFV decays, this behaviour had already been identified in~\cite{Abada:2021zcm}). 
The cLFV Higgs decay rates are also sensitive to Majorana phases, which contribute via the terms proportional to $C_{ij}^*$, see Eqs.~(\ref{eqn:FLa},\ref{eqn:FRa}).
Although not displayed here, the behaviour of the BR($H \to \ell_\alpha \ell_\beta$) with respect to other phases ($\delta_{i5}$ and $\varphi_5$) is identical to what is illustrated 
in Fig.~\ref{fig:Higgs_cLFV_delta14.24.34.phi4}.

\begin{figure}[t!]
    \centering
\mbox{   \hspace*{-5mm}  \includegraphics[width=0.51\textwidth]{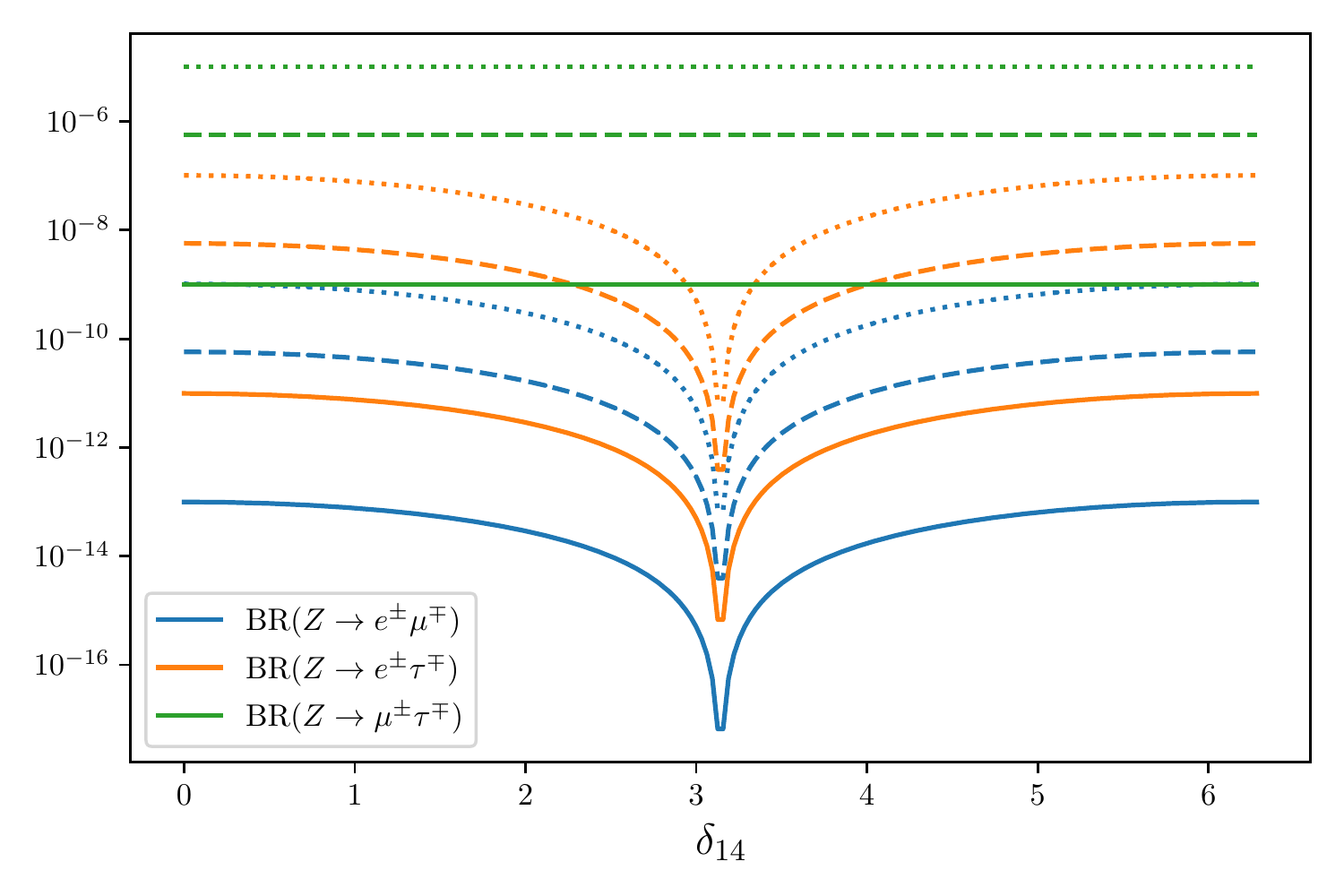}\hspace*{2mm} 
    \includegraphics[ width=0.51\textwidth]{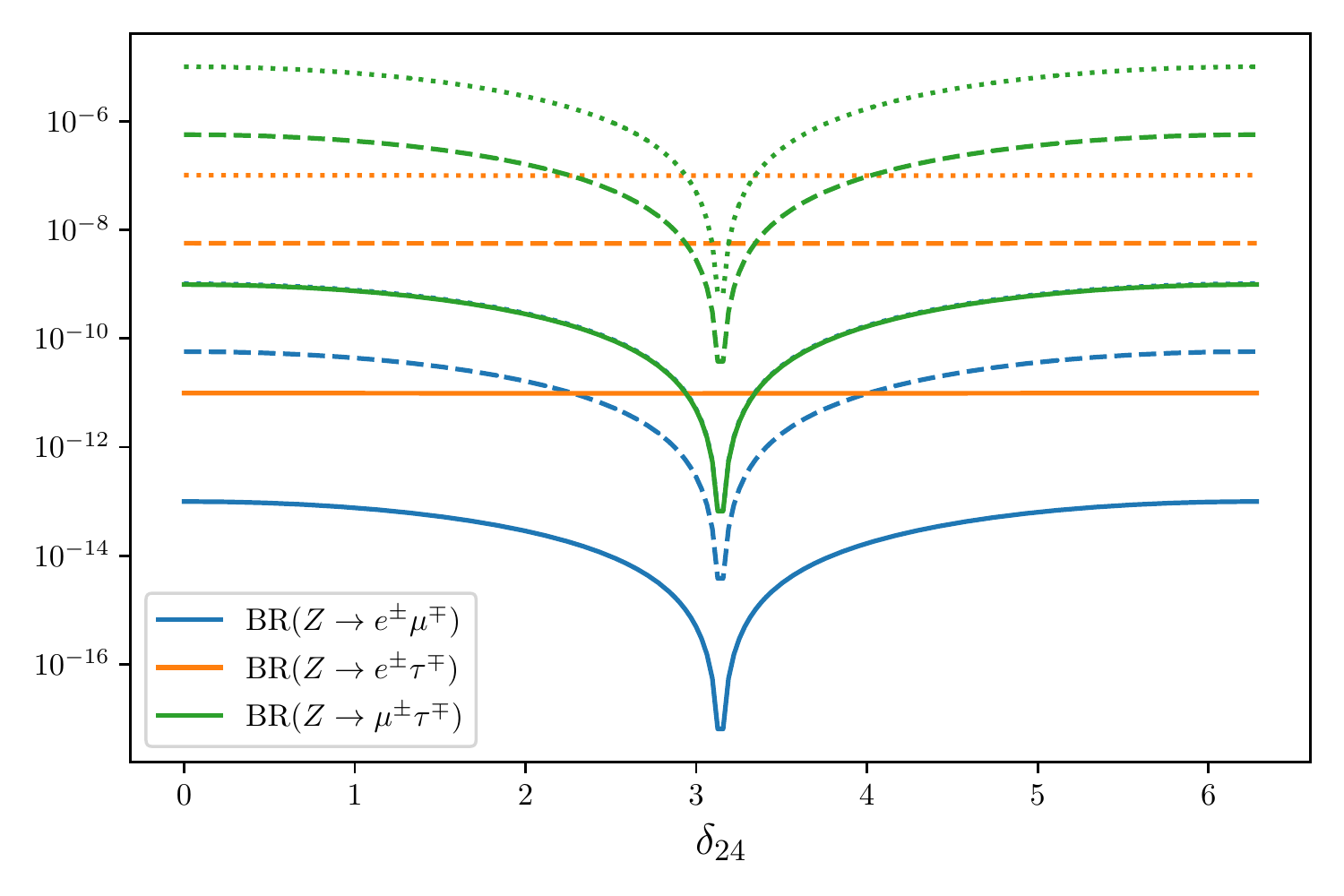}}
\mbox{   \hspace*{-5mm}  \includegraphics[ width=0.51\textwidth]{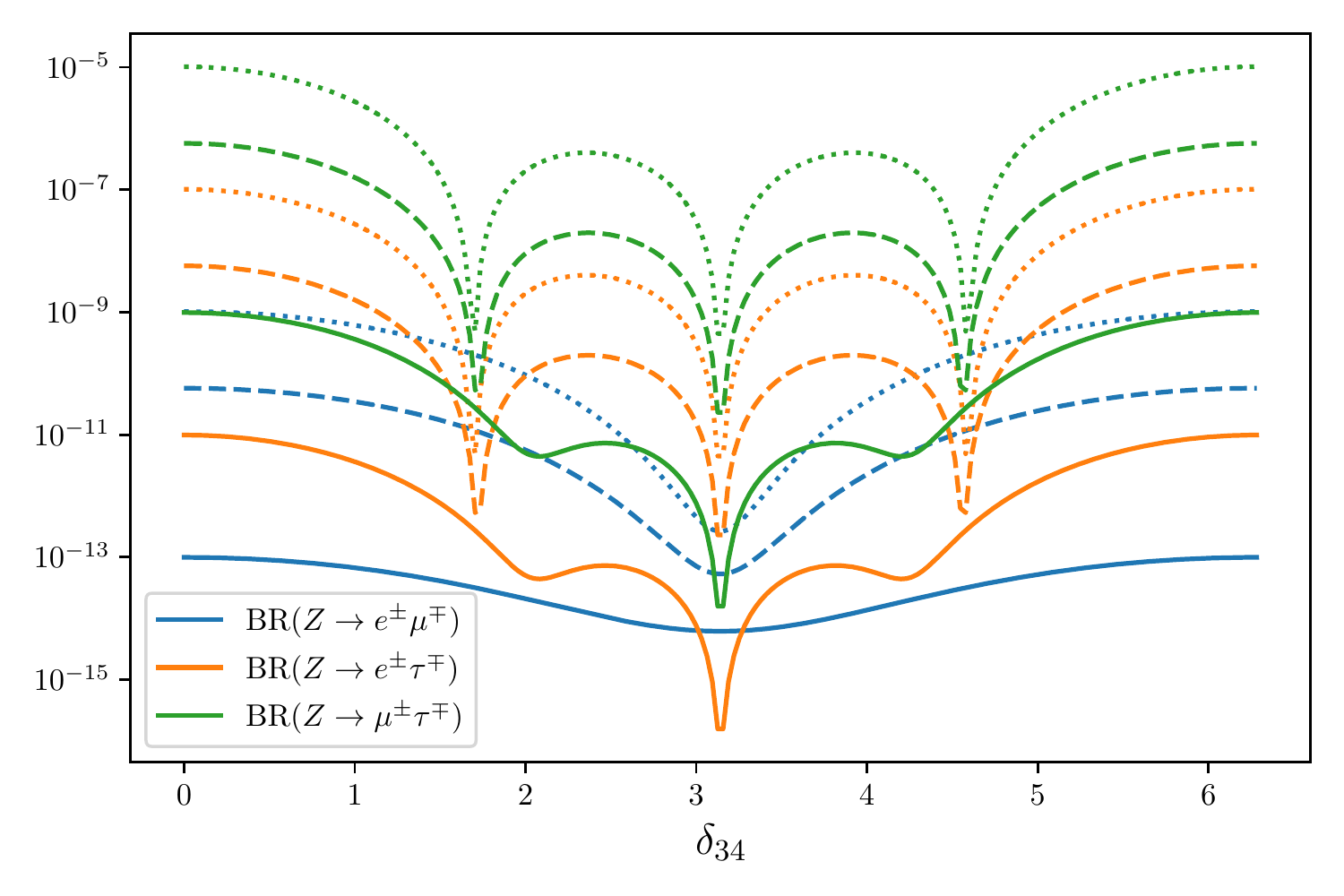}\hspace*{2mm} 
    \includegraphics[ width=0.51\textwidth]{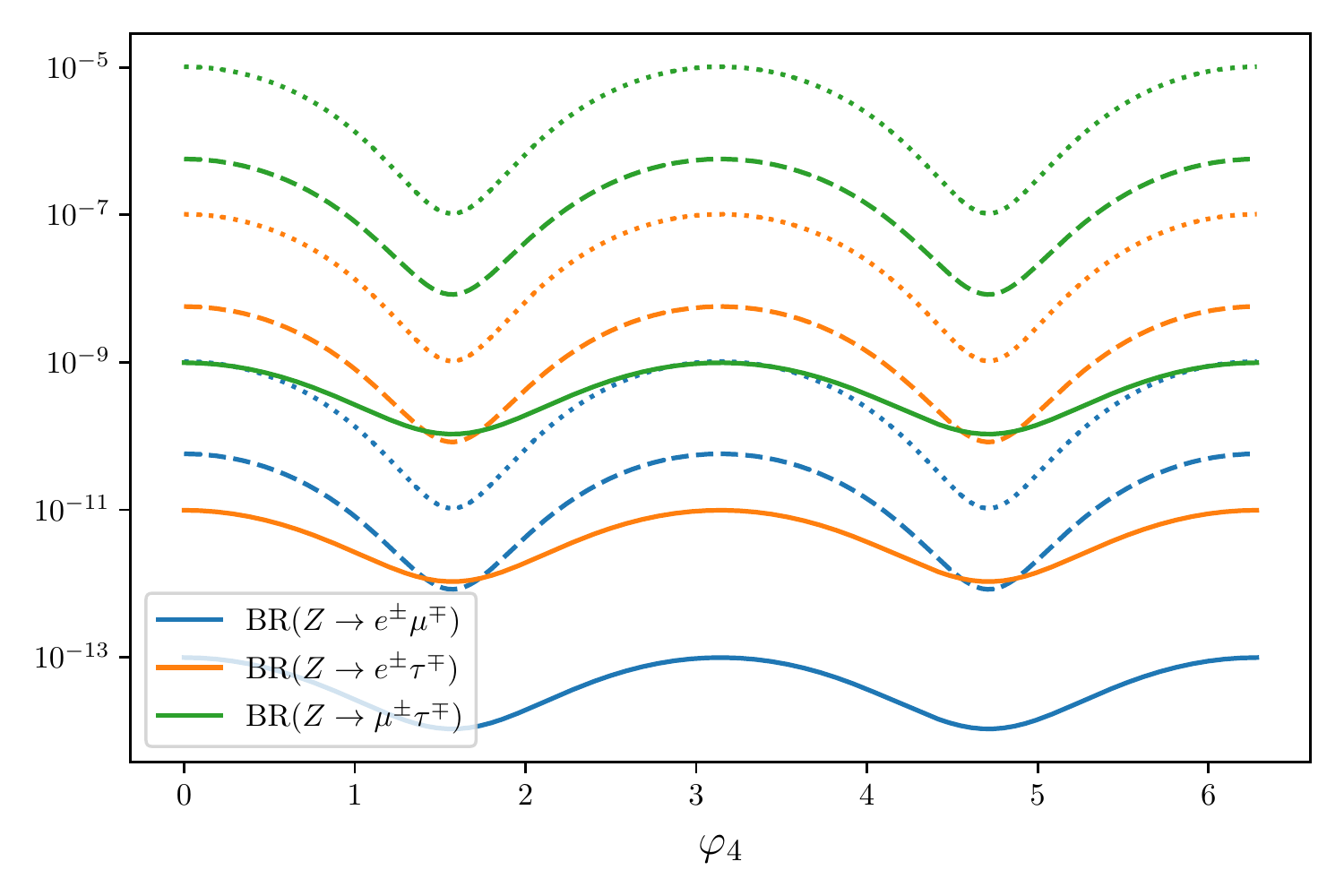}}
    \caption{Rates of cLFV $Z$-boson decays as a function of the CP violating Dirac and Majorana phases. From left to right and top to bottom, dependence on $\delta_{14}$, $\delta_{24}$, $\delta_{34}$ and $\varphi_4$ (all other phases set to zero in each case). 
    The colour code denotes the flavour composition of the final state lepton pair: $e\mu$ (blue), $e\tau$ (orange) and $\mu \tau$ (green).
    In all panels, solid, dashed and dotted lines respectively correspond to the following heavy fermion masses: $m_4=m_5=1, 5, 10~\text{TeV}$.
    }
    \label{fig:Z_cLFV_delta14.24.34.phi4}
\end{figure}

\bigskip
As already noticed in~\cite{Abada:2021zcm}, cLFV $Z$ decays exhibit a clear dependence on both Dirac and Majorana CP violating phases.
For completeness we display in Fig.~\ref{fig:Z_cLFV_delta14.24.34.phi4} a study analogous to that of Fig.~\ref{fig:Higgs_cLFV_delta14.24.34.phi4}, considering the impact of Dirac and Majorana phases on BR($Z \to \ell_\alpha \ell_\beta$), for distinct masses of the heavy mediators.
Moreover, these decays are closely related to several other cLFV leptonic decays, in particular to processes receiving contributions from $Z$-penguin topologies. 
While in the present study - and as already mentioned - no approximations were made leading to the computation of the cLFV $Z$ decay rates, the findings are in very good agreement with the previous study of~\cite{Abada:2021zcm}.
Finally, notice that for both cases of $Z$ and Higgs cLFV decays the effects of the phases are amplified for increasing masses of the sterile states.

\begin{figure}[t!]
    \centering
\mbox{   \hspace*{-5mm}  \includegraphics[width=0.51\textwidth]{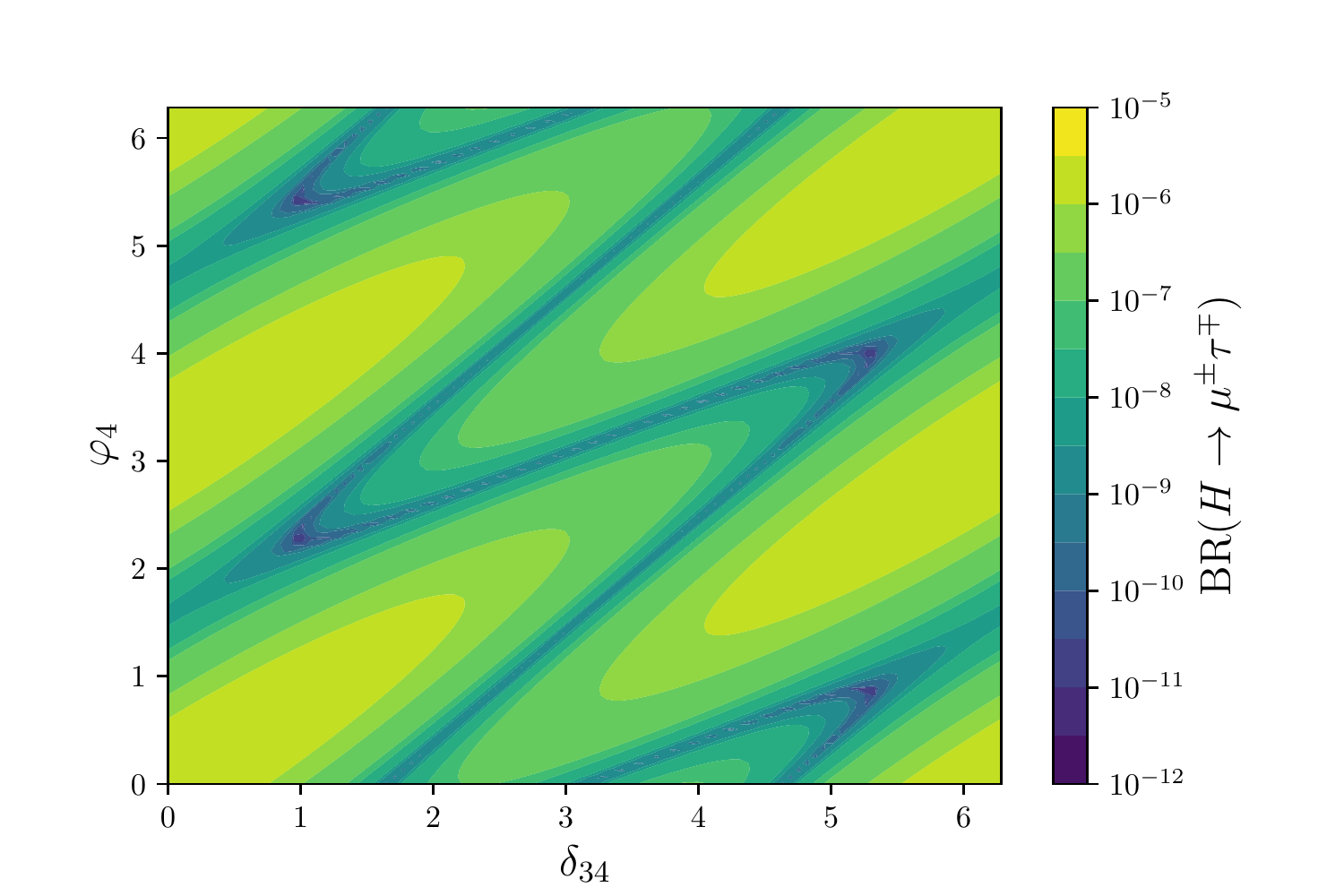}\hspace*{2mm} 
    \includegraphics[width=0.51\textwidth]{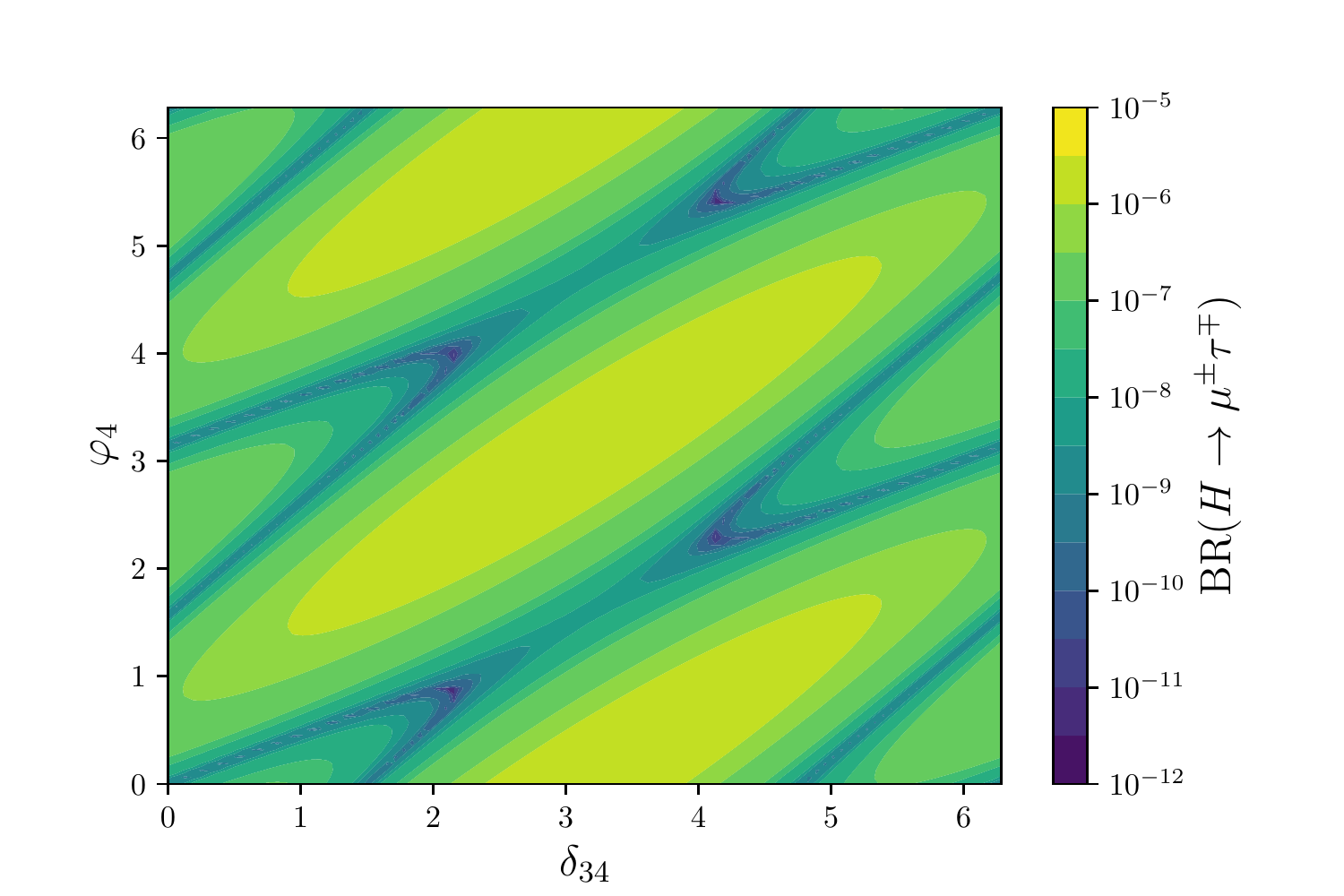}}
    \caption{Contour plots for cLFV Higgs decays, $H \to \tau \mu$. We fix $|\theta_{1j}|=10^{-3}$, $|\theta_{2j}|=0.01$ and $|\theta_{3j}|=0.1$ ($j=4,5$), and take $m_4=m_5=5$~TeV, varying the CPV phases.
    The colour scheme denotes the associated value of BR($H \to \tau \mu$) in the plane spanned by 
    $(\delta_{34}-\varphi_4)$. On the left panel, we take $s_{i4} = s_{i5}$ for $i = 1-3$, while on the right $s_{24} = -s_{25}$ (with $s_{14}=s_{15}$ and $s_{34}=s_{35}$). 
    }
    \label{fig:contour_Higgs}
\end{figure}

\bigskip
In order to illustrate the joint effect of Dirac and Majorana phases, we display in Fig.~\ref{fig:contour_Higgs} the iso-surfaces for BR($H \to \tau \mu$) as spanned by the CPV phases $\delta_{34}$ and $\varphi_4$, for fixed values of the mixing angles and sterile states' masses. 
As visible, different regimes for the CPV phases can account for variations on the predicted values for the rates by as much as 7 orders of magnitude, due to the impact of the (destructive) interference effects. 
The comparison of left and right panels also reveals the possibility of constructive interferences: in this case, effects already emerge for opposite-sign mixings (illustrated by $s_{24} = -s_{25}$), leading to a significant increase in the branching ratio.

\subsection{Phenomenological impact}\label{sec:pheno_num}
In this section we complete and generalise the discussion of the previous one, carrying out a comprehensive phenomenological analysis to fully assess the effect of the CPV Dirac and Majorana phases on $Z$- and Higgs boson cLFV decays, in the framework of a minimal low-energy extension of the SM via 2 heavy neutral leptons. 

We thus now take into account all available experimental constraints which include, among others, limits on the active-sterile mixings, negative results of direct and indirect searches for the sterile states, as well as EW precision tests; likewise, bounds on searches for other cLFV transitions - such as radiative and three-body leptonic decays as well as rare transitions in muonic atoms - are also taken into account (see Appendix~\ref{app:constraints} for the constraints on HNL extensions of the SM, and Appendix~\ref{app:cLFVobservables} for the relevant expressions for the cLFV observables, as well as the associated current bounds and future sensitivities.). 
The parameter space of the model is also now thoroughly explored: no assumptions are made on relations between mixing angles, nor on the heavy mediator masses. 
The former are now randomly and independently varied and drawing samples from log-uniform distributions, further randomly varying their signs, while the latter are no longer taken to be degenerate (only sufficiently close in mass to allow for interference effects\footnote{For fixed
$m_4$, random values of $m_5$ are obtained from 
half-normal distributions with the scale set to a value representative of the width of the sterile states.}).

In summary, the ranges of the parameters to be here explored are
\begin{eqnarray}\label{eqn:6d_scan_ranges}
     && \: m_5 - m_4 \:\, \in\,[10~\text{MeV}, 1~\text{TeV}]\,,\nonumber \\
    && |\sin\theta_{14,15}|\in\, [6.0\times 10^{-5}, 6.0\times10^{-3}]\,,\nonumber \\
    && |\sin\theta_{24,25}|\in\, [1.9\times 10^{-4}, 0.036]\,,\nonumber \\
    && |\sin\theta_{34,35}|\in\, [8.3\times 10^{-4}, 0.13]\,.
\end{eqnarray}
These ranges\footnote{The upper bounds are chosen to comply (mostly) with current experimental data; regimes with smaller mixing angles were not considered as for the lower bounds these already correspond to predictions well beyond future experimental sensitivities.} correspond to regimes for which the CP conserving cases comply with experimental bounds (cf. Appendix~\ref{app:constraints}). From the $10^4$ points thus selected, and for each tuple of mixing angles, we then vary all CPV phases associated with the sterile states, i.e. $\delta_{\alpha 4, 5}, \varphi_{4, 5}\,\in\,[0, 2\pi]$ (with $\alpha = e\,,\mu\,,\tau$), drawing 100 values for each of the 8 from a uniform distribution.

In the following discussion, the obtained theoretical predictions are then compared with the existing current bounds (and prospects for future sensitivities) of the $Z$- and Higgs boson cLFV decays; these are summarised in Table~\ref{tab:cLFV_ZH}. 
\renewcommand{\arraystretch}{1.3}
\begin{table}[h!]
    \centering
    \hspace*{-7mm}{\small\begin{tabular}{|c|c|c|}
    \hline
    Observable & Current bound & Future sensitivity  \\
    \hline\hline
    $\mathrm{BR}(Z\to e^\pm\mu^\mp)$ & \quad$< 4.2\times 10^{-7}$\quad (ATLAS~\cite{Aad:2014bca}) & \quad$\mathcal O (10^{-10})$\quad (FCC-ee~\cite{Abada:2019lih})\\
    $\mathrm{BR}(Z\to e^\pm\tau^\mp)$ & \quad$< 4.1\times 10^{-6}$\quad (ATLAS~\cite{ATLAS:2021bdj}) & \quad$\mathcal O (10^{-10})$\quad (FCC-ee~\cite{Abada:2019lih})\\
    $\mathrm{BR}(Z\to \mu^\pm\tau^\mp)$ & \quad$< 5.3\times 10^{-6}$\quad (ATLAS~\cite{ATLAS:2021bdj}) & \quad $\mathcal O (10^{-10})$\quad (FCC-ee~\cite{Abada:2019lih})\\
    \hline
    \hline
    $\mathrm{BR}(H\to e^\pm\mu^\mp)$ & \quad$< 6.1\times 10^{-5}$\quad\cite{ParticleDataGroup:2020ssz}  & $<1.2\times10^{-5}$ (FCC-ee~\cite{Qin:2017aju})\\
    $\mathrm{BR}(H\to e^\pm\tau^\mp)$ & \quad$< 4.7\times 10^{-3}$\quad\cite{ParticleDataGroup:2020ssz}  &$< 1.6\times 10^{-4}$ (FCC-ee~\cite{Qin:2017aju})\\
    $\mathrm{BR}(H\to \mu^\pm\tau^\mp)$ & \quad$< 2.5\times 10^{-3}$\quad\cite{ParticleDataGroup:2020ssz}  & $<1.4\times 10^{-4}$ (FCC-ee~\cite{Qin:2017aju})\\
    \hline
    \end{tabular}}
    \caption{Current experimental bounds and future sensitivities on cLFV $Z$ and Higgs decays (all limits are given at $90\%\:\mathrm{C.L.}$).}
    \label{tab:cLFV_ZH}
\end{table}
\renewcommand{\arraystretch}{1.}

\begin{figure}[t!]
    \centering
    \mbox{   \hspace*{-5mm} 
    \includegraphics[width=0.51\textwidth]{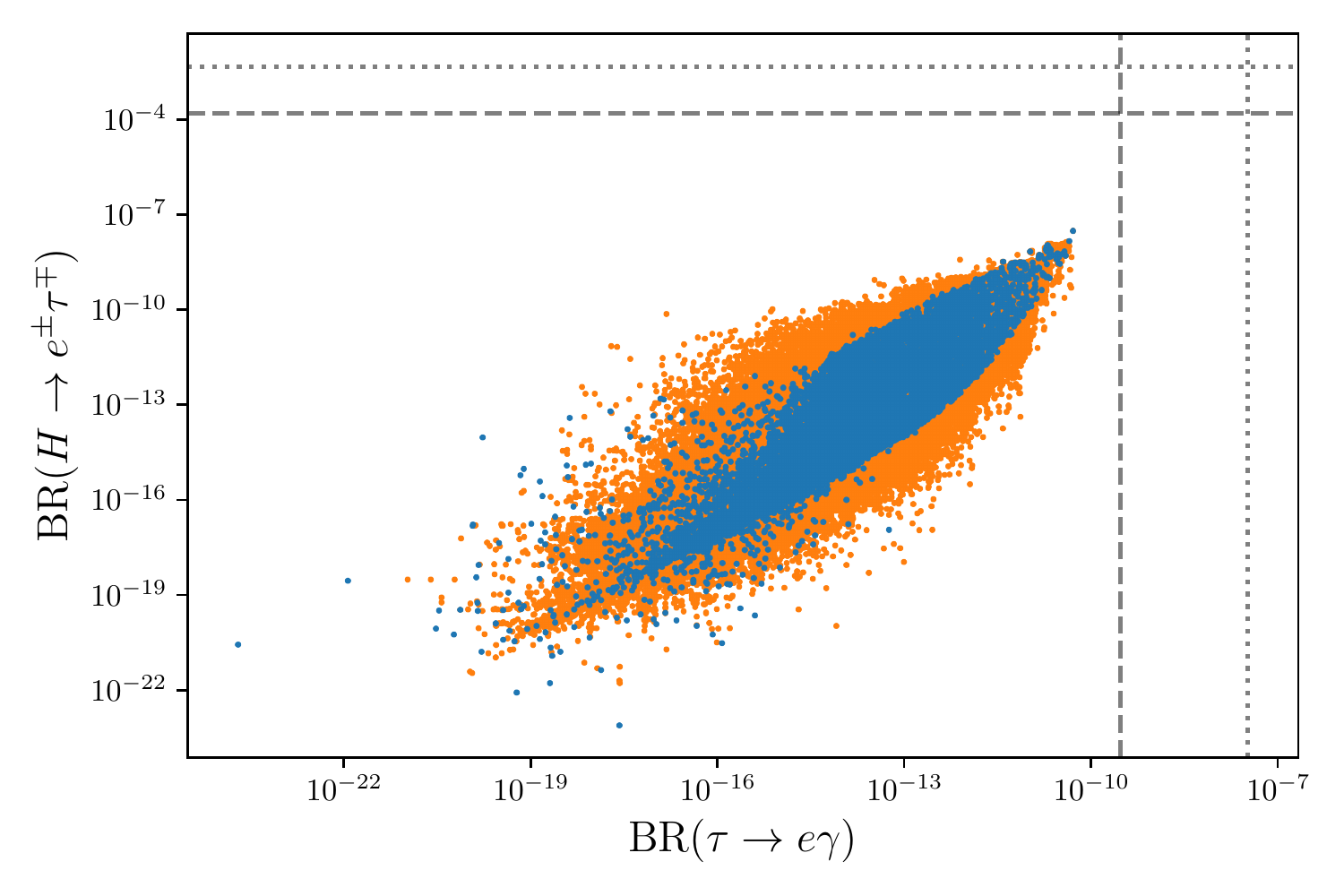}\hspace*{2mm} 
    \includegraphics[width=0.51\textwidth]{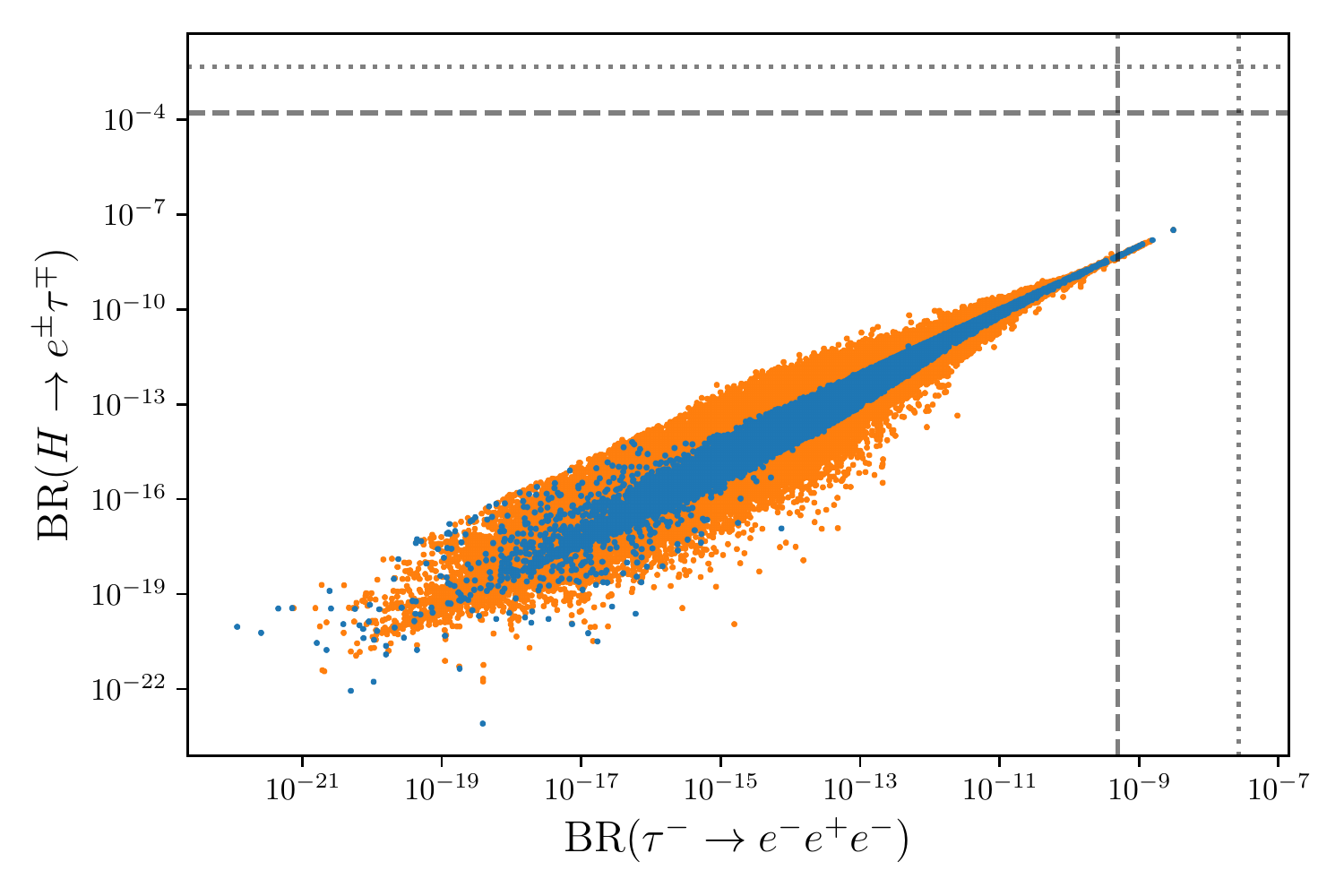}}
\mbox{   \hspace*{-5mm}  \includegraphics[width=0.51\textwidth]{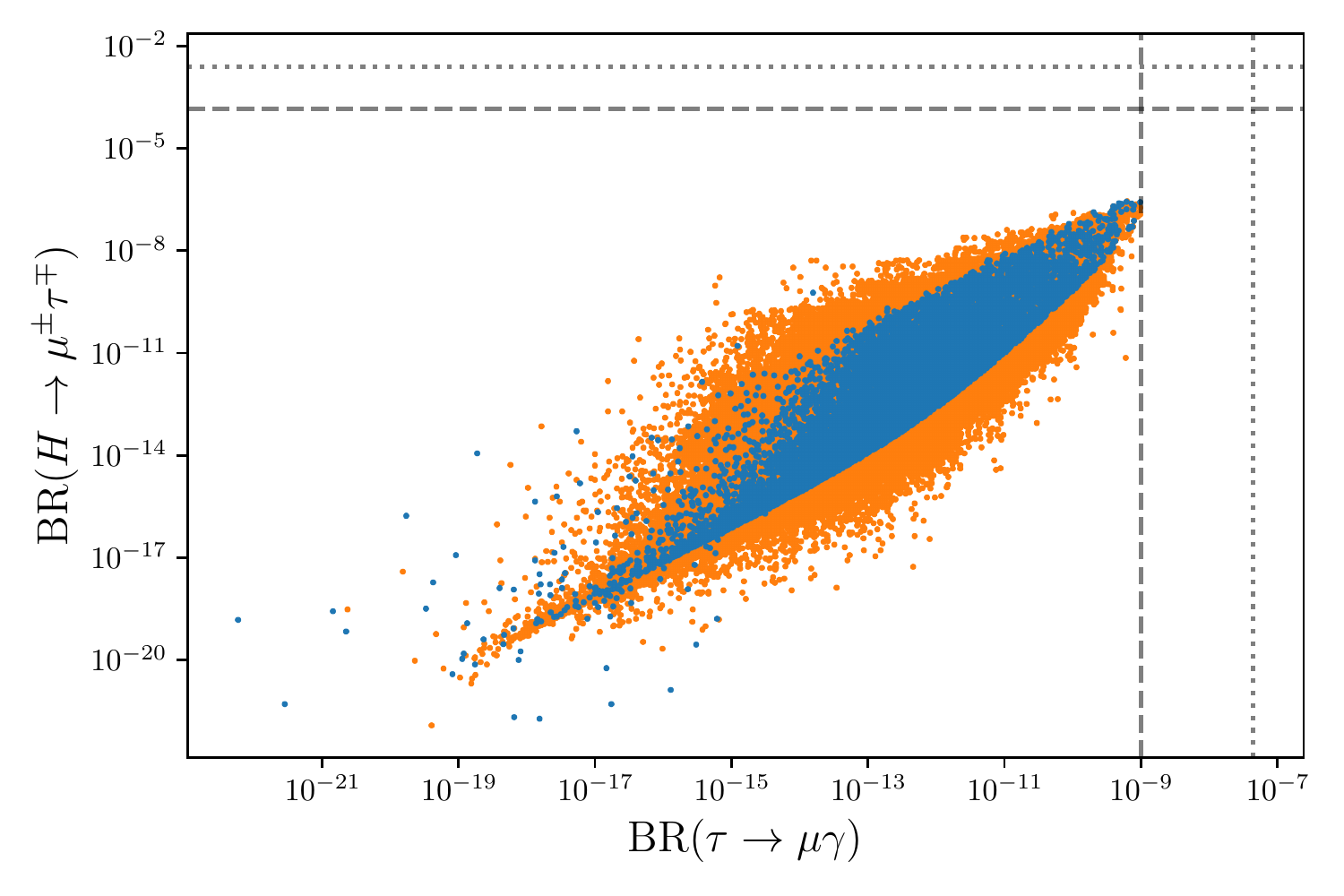}\hspace*{2mm} 
    \includegraphics[width=0.51\textwidth]{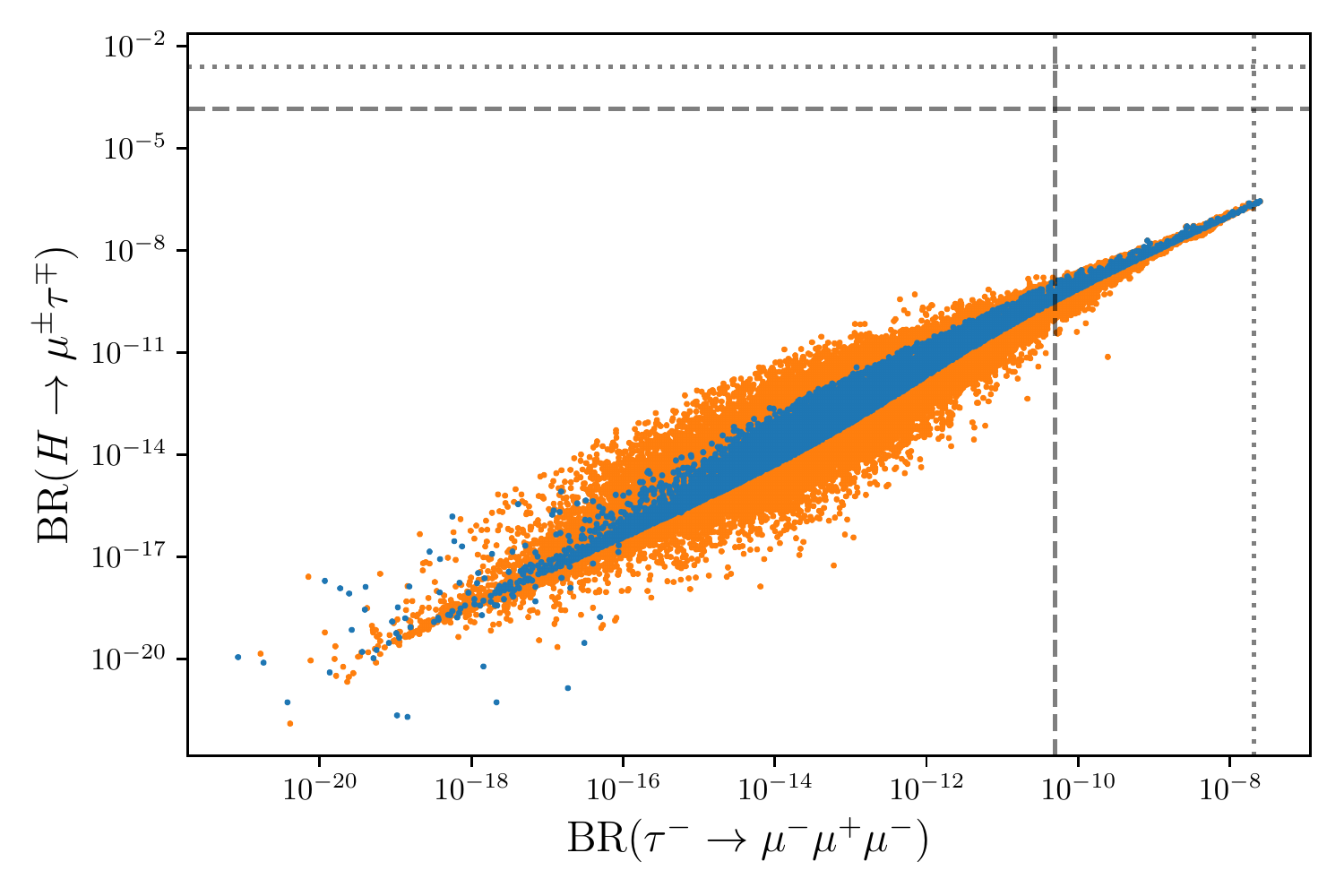}}
    \caption{General overview of cLFV $H \to \ell_\alpha \tau$ vs.~cLFV tau-lepton decays in the ``3+2 model'' parameter space. All active-sterile mixing angles, as well as Dirac and Majorana CP phases, are randomly varied. In all panels, $m_4=5$~TeV, with $m_5-m_4 \in [10~\text{MeV}, 1~\text{TeV}]$. Blue points correspond to vanishing phases, while orange denotes random values of all phases ($\delta_{\alpha i}$ and $\varphi_i$, with $\alpha = e,\mu,\tau$ and $i=4,5$). Dotted (dashed) lines denote current bounds (future sensitivity) as given in Table~\ref{tab:cLFV_ZH}. (For additional information, see detailed description in the text.)    }
    \label{fig:higgs_tau_scatter}
\end{figure}

\bigskip
A first overview of the prospects for cLFV Higgs decays is presented in Fig.~\ref{fig:higgs_tau_scatter}, in which 
we display the rates of $H \to \ell_\alpha \tau$ ($\alpha=e,\mu$) vs.~radiative and 3-body cLFV tau-lepton decays. We emphasise once again that now all active-sterile mixing angles, as well as Dirac and Majorana CP phases, are randomly varied, without any underlying simplifying assumptions.
Notice that in what follows we will mostly focus on final state lepton pairs including one tau-lepton, as the decay for $e\mu$ final states is associated with comparatively smaller rates. 

As expected from the underlying topologies of the different decays, 
$H \to \ell_\alpha \tau$ processes exhibit a significant correlation with $\tau \to 3 \ell_\alpha$ (notice that for heavy sterile states, the latter decays are in general dominated by $Z$-penguin contributions, which are also topologically similar to Higgs decays); this is in contrast to the radiative tau-lepton decays, 
$\tau \to \mu \gamma$.
Although CPV phases (both Majorana and Dirac) lead to extensive interference effects, their impact is less striking  than what had been found for purely leptonic processes (e.g. $\mu-e$ conversion vs. $\mu \to 3 e$, see~\cite{Abada:2021zcm}). 

Another conclusion to be drawn is that in the framework of this SM extension via 2 heavy neutral leptons (with masses around a few TeVs), cLFV Higgs decays are in all cases beyond future sensitivity; although larger rates could in principle be possible, the associated active-sterile mixing regimes are disfavoured from bounds on several electroweak precision observables (including invisible $Z$ decays, among others) and cLFV $\tau \to 3 \mu$ decays.
In turn, this suggests that a near future observation of 
$H \to \ell_\alpha \tau$ decays would strongly disfavour this minimal class of HNL extensions of the SM.

\begin{figure}[t!]
    \centering
\mbox{   \hspace*{-5mm}  \includegraphics[width=0.51\textwidth]{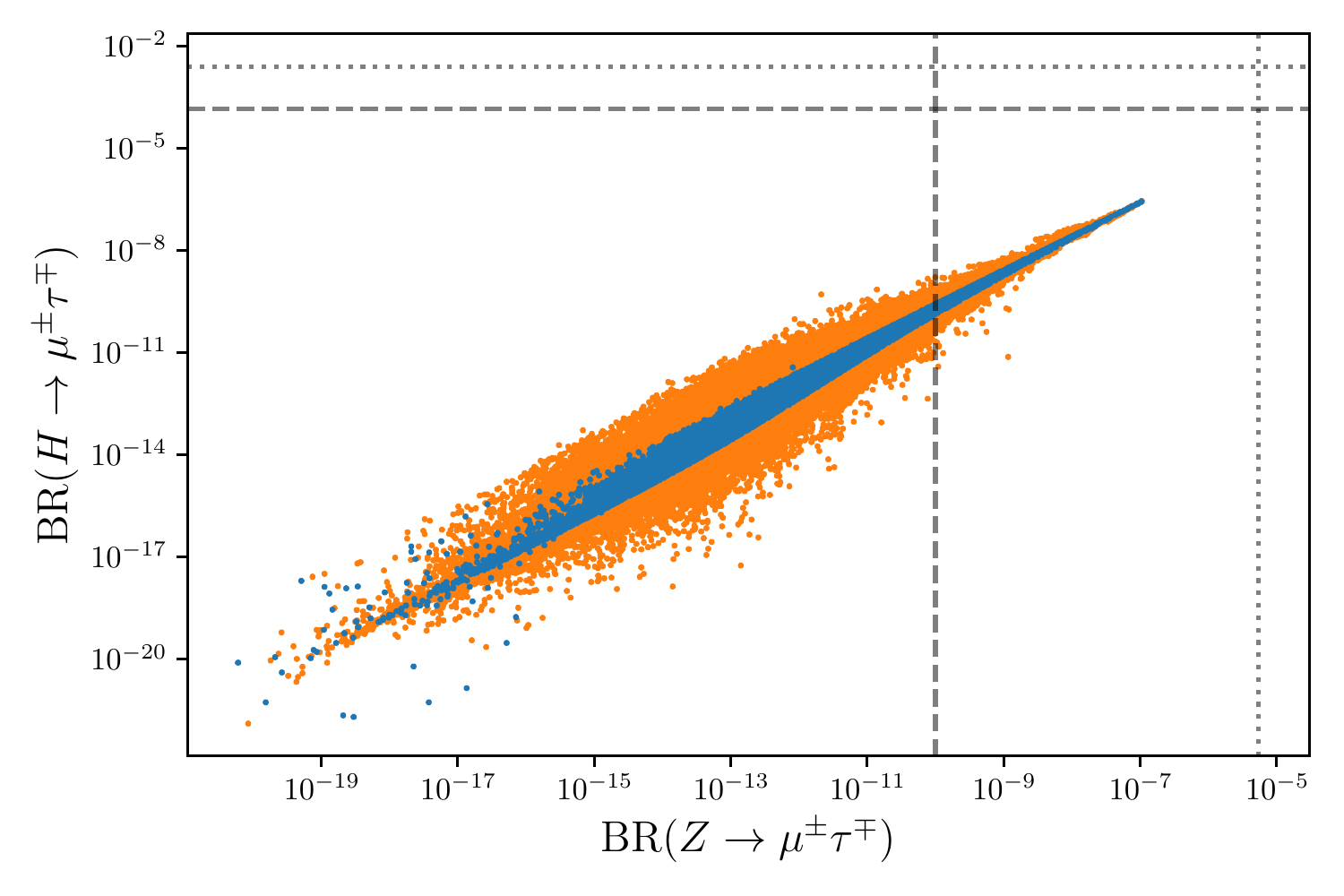}\hspace*{2mm} 
    \includegraphics[width=0.51\textwidth]{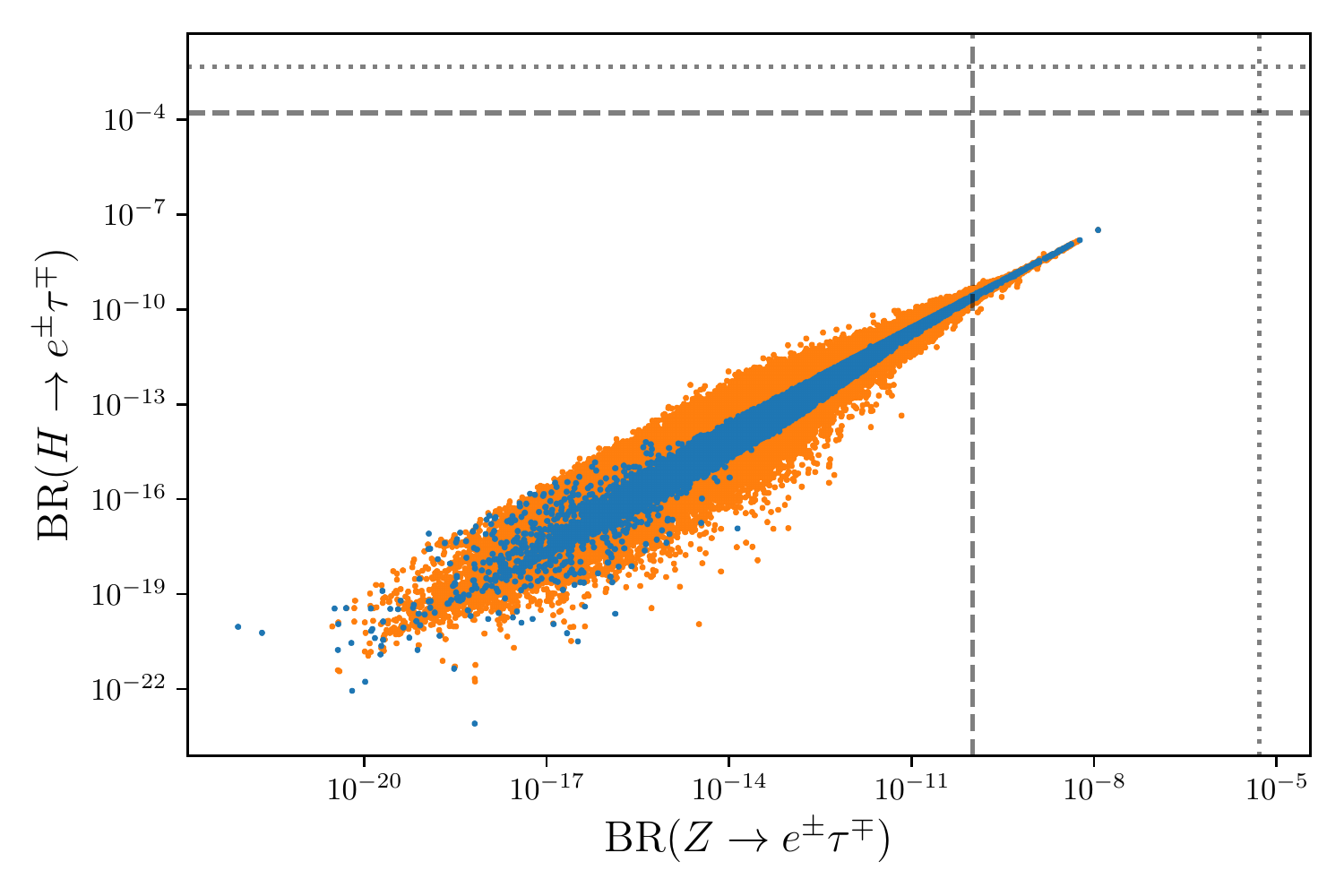}}
    \caption{ General overview of cLFV $H \to \ell_\alpha \tau$ and $Z \to \ell_\alpha \tau$ (with $\ell_\alpha = e, \mu$) in the ``3+2 model'' parameter space. 
    All active-sterile mixing angles, as well as Dirac and Majorana CP phases, are randomly varied. In both panels, $m_4=5$~TeV, with $m_5-m_4 \in [10~\text{MeV}, 1~\text{TeV}]$. Blue points correspond to vanishing phases, while orange denotes random values of all phases ($\delta_{\alpha i}$ and $\varphi_i$, with $\alpha = e,\mu,\tau$ and $i=4,5$). Dotted (dashed) lines denote current bounds (future sensitivity) as given in Table~\ref{tab:cLFV_ZH}. (For additional information, see detailed description in the text.)
    }
    \label{fig:higgs_z_scatter}
\end{figure}

A comparison between the prospects of cLFV Higgs and $Z$ decays is presented in Fig.~\ref{fig:higgs_z_scatter}, in which we display 
$H \to \ell_\alpha \tau$ vs. $Z \to \ell_\alpha \tau$, for  $\ell_\alpha = e, \mu$. Both $Z \to e \tau$ and $Z \to \mu \tau$
decays are within future sensitivity of FCC-ee (running at the $Z$-pole), the latter offering the most promising prospects. 
As already anticipated, both observables exhibit a strong correlation in the CP conserving case (especially for $\mu \tau$ final states); such correlated behaviour is indeed affected by the presence of CP violating phases, which can induce both constructive and destructive interference. Nonetheless, and as mentioned before, the observed loss of correlation is milder than for purely leptonic processes~\cite{Abada:2021zcm}. 

\section{CP asymmetries in cLFV boson decays}\label{sec:asymmetries}

As mentioned in the Introduction, and extensively discussed in the present work, in models in which the SM is extended via heavy Majorana fermions, the ``effective" $Z$ and Higgs vertices to leptons are modified, allowing for both flavour and CP violation.
In view of the capabilities and clean environment of a future FCC-ee (tera-$Z$ factory), one can then consider to which extent such a minimal BSM construction could be at the source of non-vanishing contributions to CP-asymmetries, in particular concerning $Z$-boson decays,
$\mathcal{A}_{CP}(Z\to\ell_\alpha\ell_\beta)$. The latter are defined as
\begin{equation}\label{eq:def:ACP:Z}
    \mathcal{A}_{CP}(Z\to\ell_\alpha\ell_\beta)\, =\, 
    \frac{
    \Gamma(Z \to \ell_\alpha^-\ell_\beta^+) -
    \Gamma(Z \to \ell_\alpha^+\ell_\beta^-)
    }{\Gamma(Z \to \ell_\alpha^-\ell_\beta^+) +
    \Gamma(Z \to \ell_\alpha^+\ell_\beta^-)}\,.
\end{equation}

\begin{figure}[t!]
    \centering
\mbox{   \hspace*{-5mm}  \includegraphics[width=0.51\textwidth]{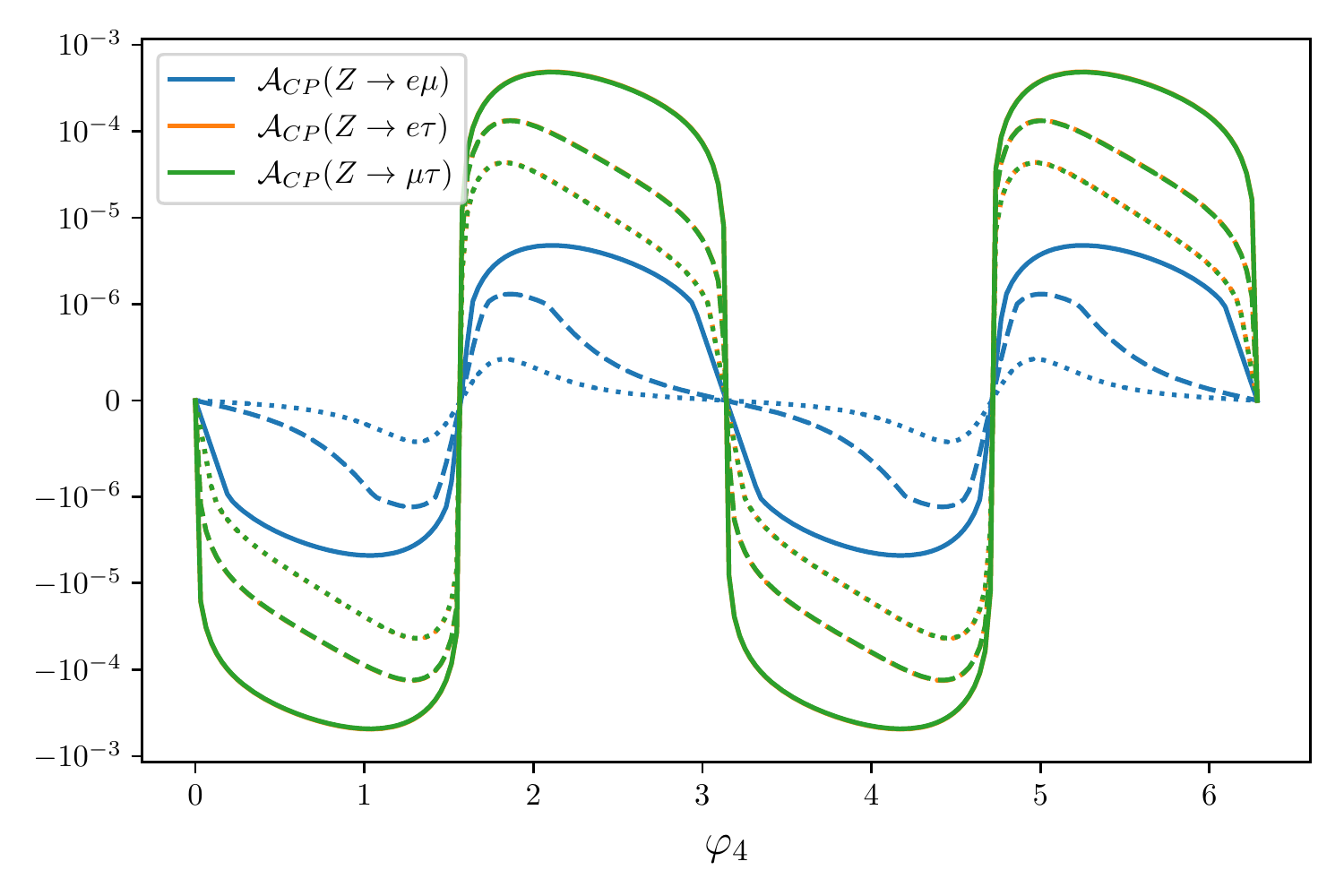}\hspace*{2mm} 
    \includegraphics[width=0.51\textwidth]{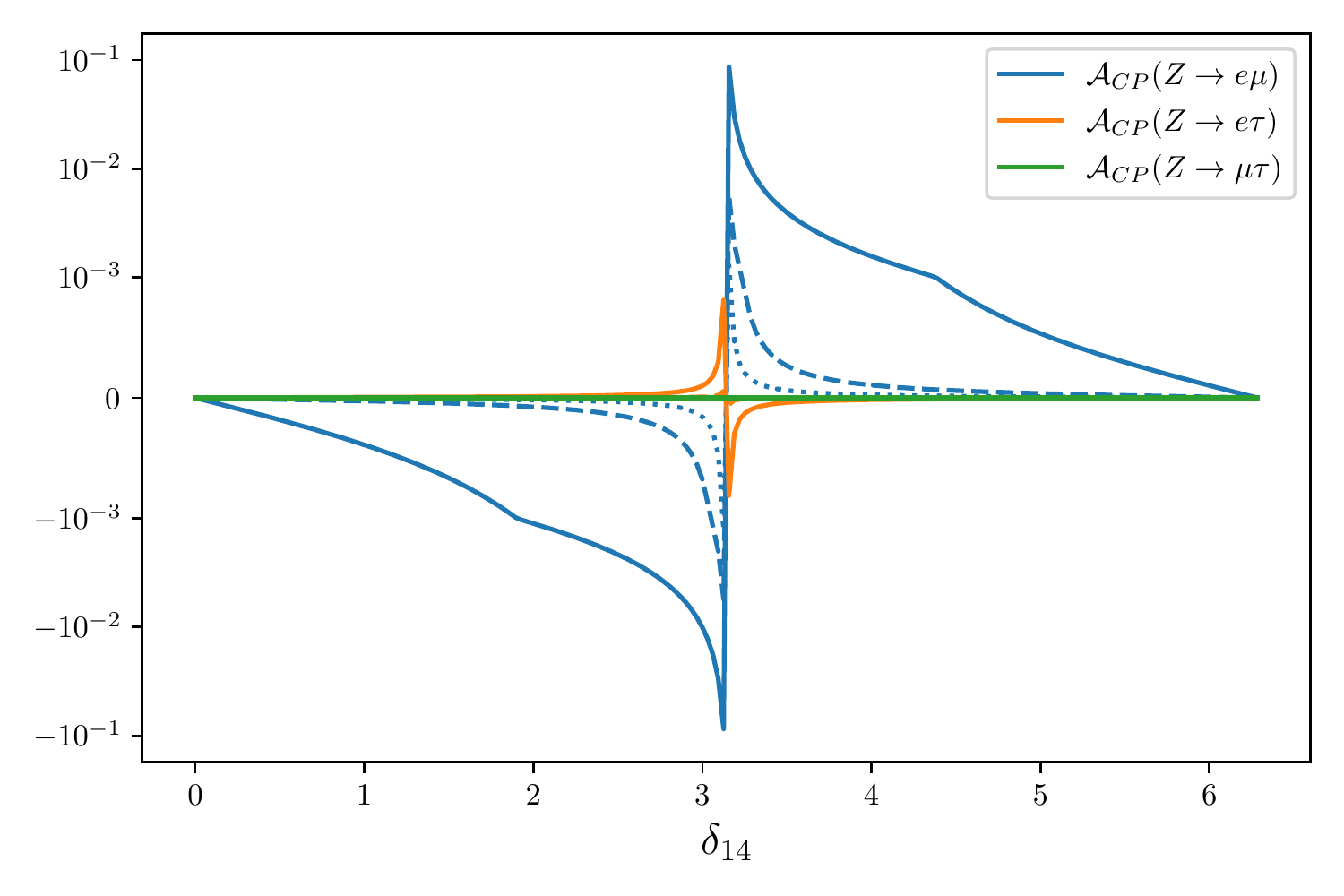}}\\
    \mbox{   \hspace*{-5mm}  \includegraphics[width=0.51\textwidth]{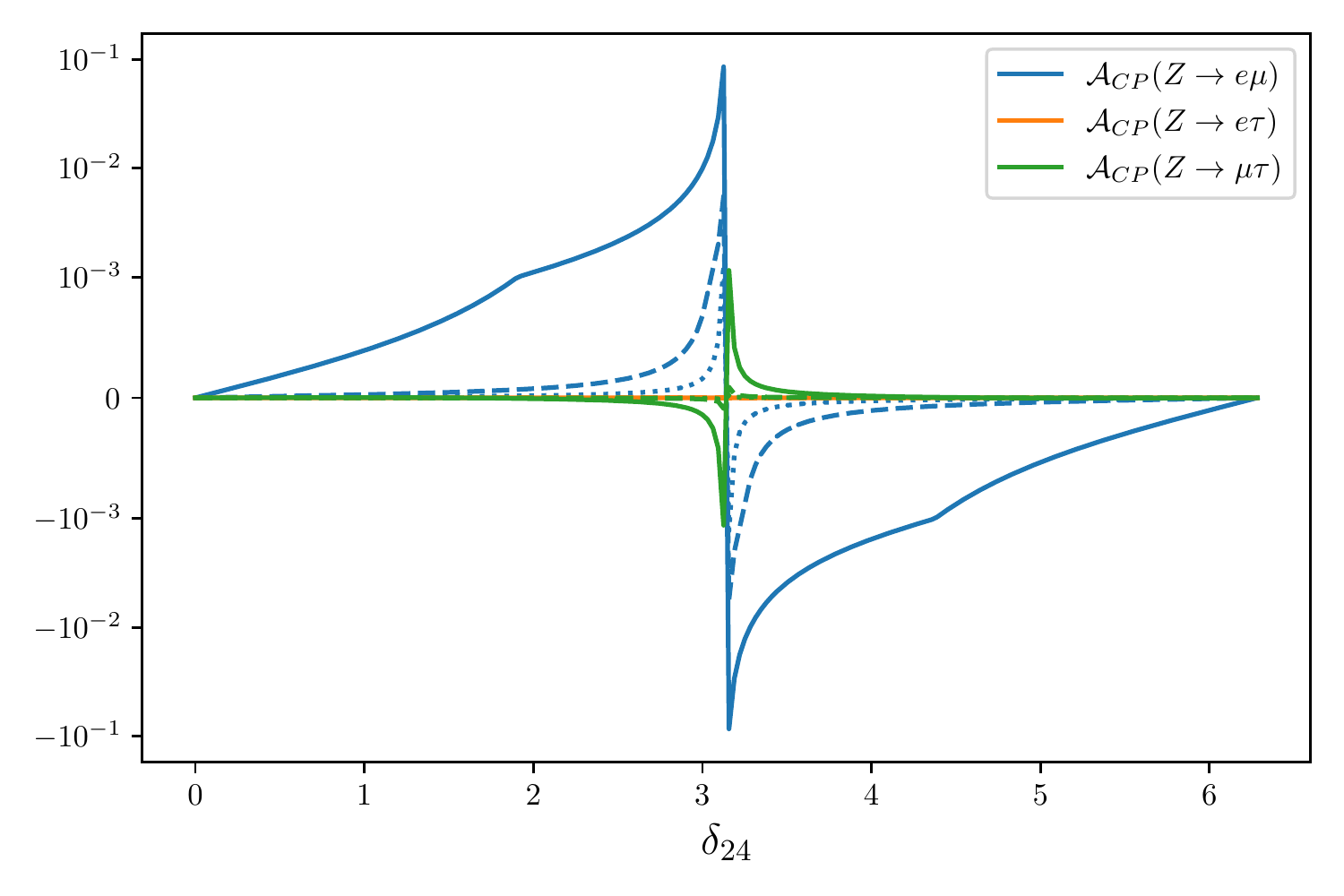}\hspace*{2mm} 
    \includegraphics[width=0.51\textwidth]{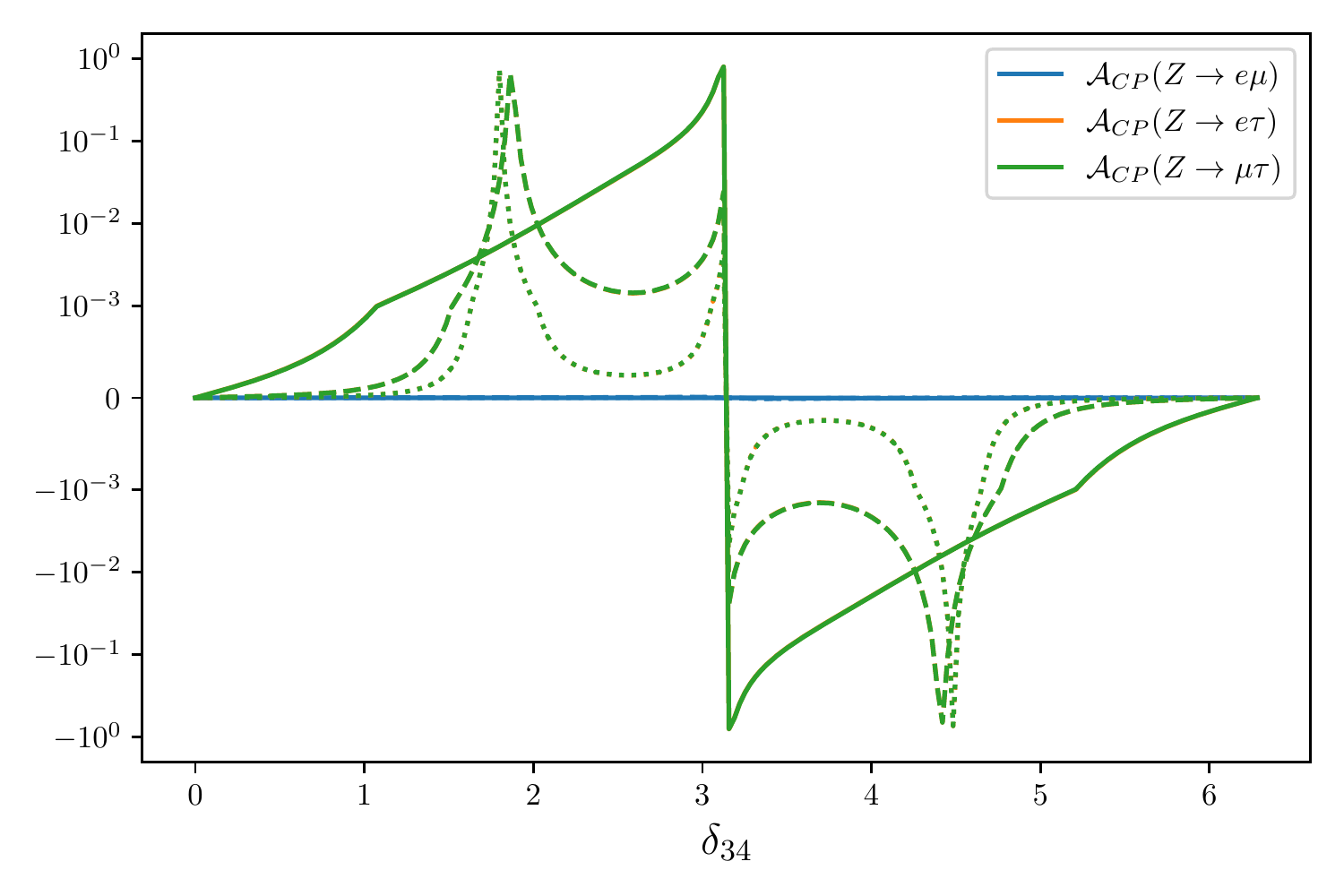}}
    \caption{CP asymmetry in cLFV $Z \to \ell_\alpha \ell_\beta$ decays as a function of the Majorana and Dirac CP violating phases. From left to right, top to bottom, $\mathcal{A}_{CP}(Z\to\ell_\alpha\ell_\beta)$ dependence on $\varphi_4$, 
    $\delta_{14}$, $\delta_{24}$ and $\delta_{34}$
    (with all other phases set to zero in each case). The colour code denotes the flavour composition of the final state lepton pair: $e\mu$ (blue), $e\tau$ (orange) and $\mu \tau$ (green).
    In all panels, solid, dashed and dotted lines respectively correspond $m_4=m_5=1, 5, 10~\text{TeV}$. We fix $\theta_{1j}=10^{-3}$, $\theta_{2j}=0.01$ and $\theta_{3j}=0.1$ ($j=4,5$).
    }
    \label{fig:Z_cLFV_ACP}
\end{figure}

As can be seen by the illustrative results of Fig.~\ref{fig:Z_cLFV_ACP}, obtained under the same simple assumptions of the preliminary discussion of Section~\ref{sec:simple_plots}, one can indeed have non-negligible contributions to $\mathcal{A}_{CP}(Z\to\ell_\alpha\ell_\beta)$, induced by both Majorana and Dirac CPV phases. Individually, the former have a reduced impact, while the latter (especially $\delta_{34}$, through the $C_{ij}$ terms) can lead to very large asymmetries. As expected, and in general terms, the most promising channels appear to be those leading to final states containing one tau-lepton.
It is also worth mentioning that the asymmetries are significantly dominated by the contributions to the cLFV $Z$ decays stemming from the diagrams with 2 neutrinos in the loop (see Fig.~\ref{fig:cLFVZdecays:UG}~(a)).
Also notice that the CP asymmetries computed in the limit of vanishing charged lepton masses are typically predicted to be larger; moreover their behaviour with respect to the CPV phases significantly differs. 

\begin{figure}[t!]
    \centering
    \mbox{   \hspace*{-12mm}          \includegraphics[width=0.56\textwidth]{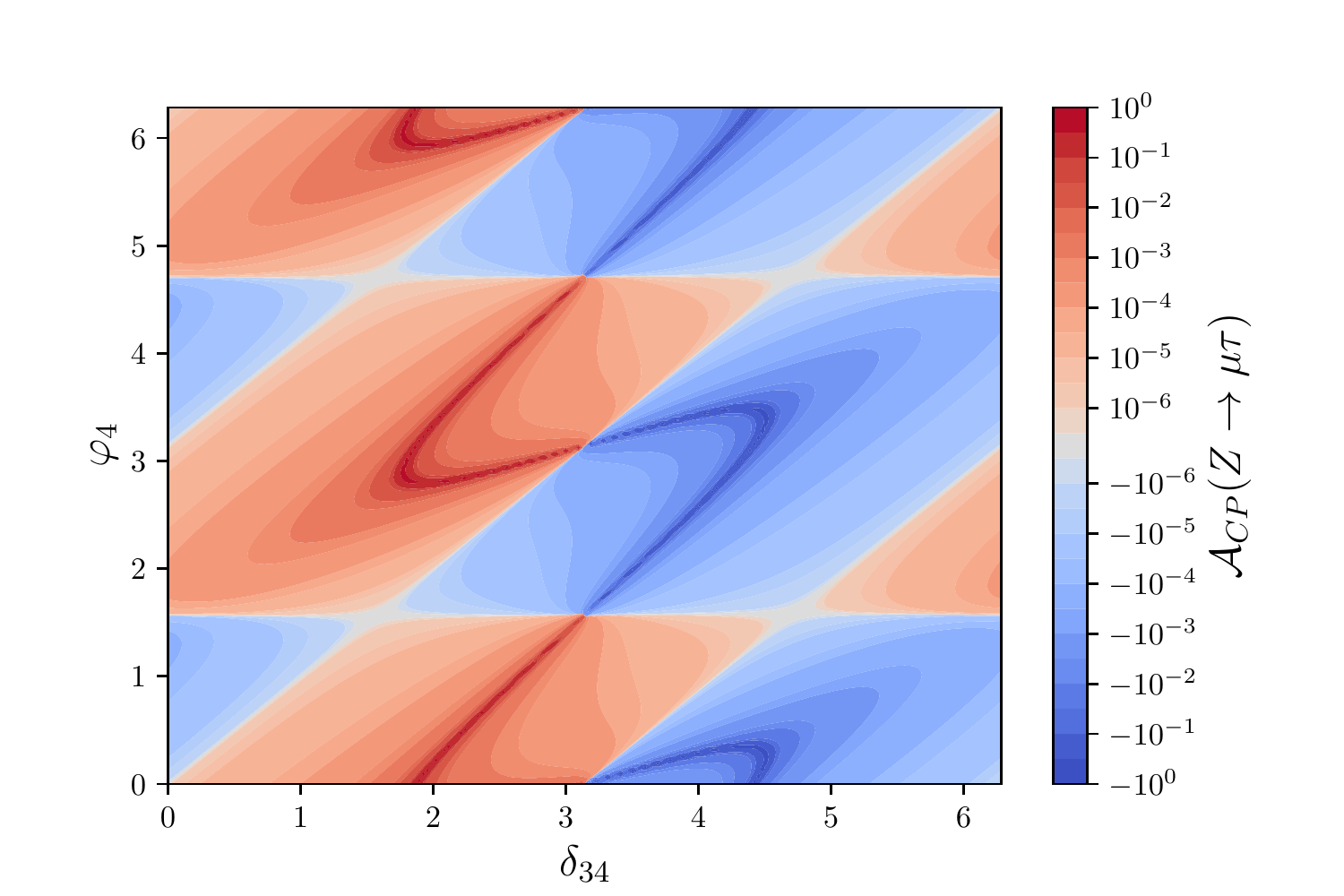}\hspace*{-5mm} 
    \includegraphics[width=0.56\textwidth]{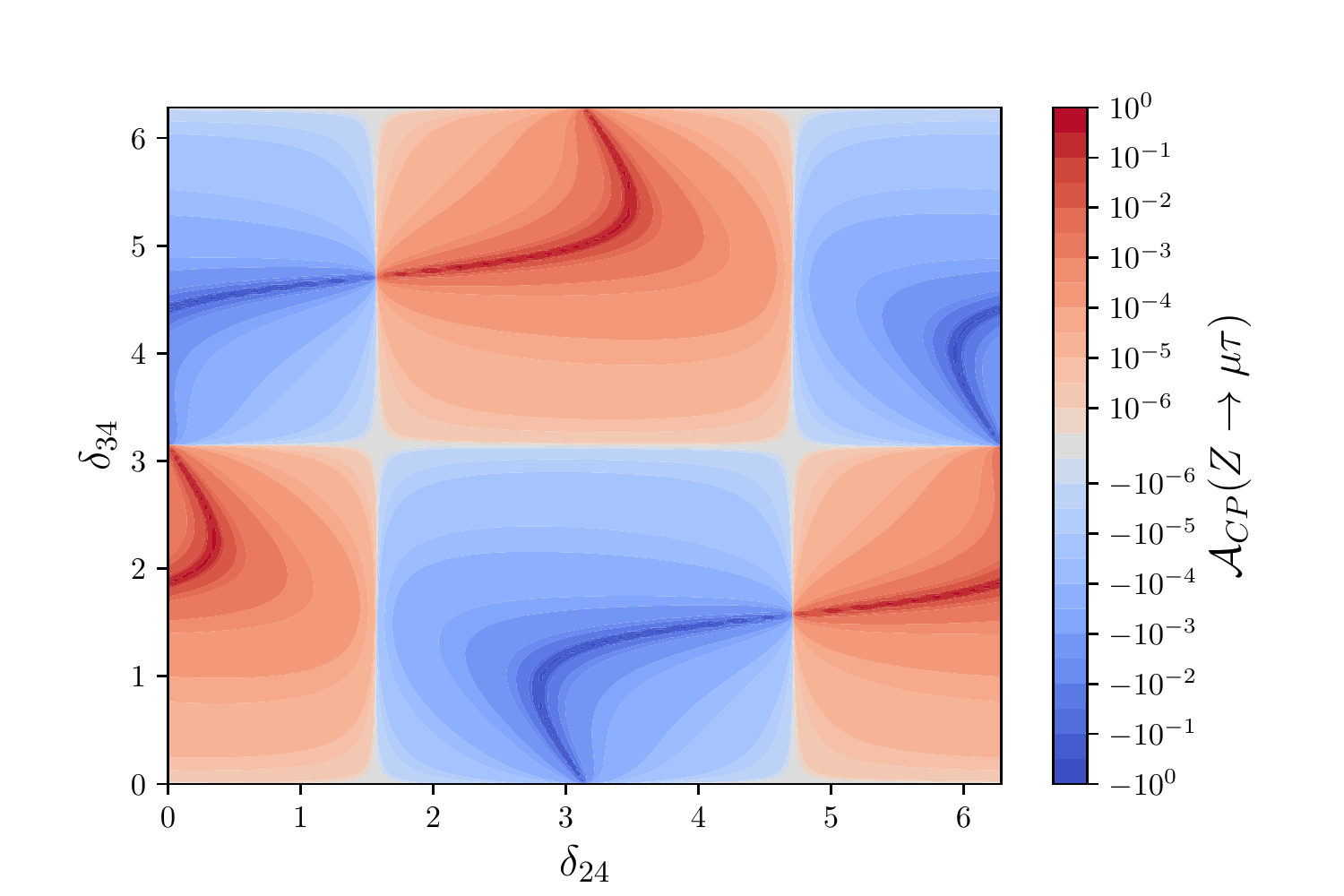}}
    \caption{Contour plots for the 
    CP asymmetry $\mathcal{A}_{CP}(Z\to\mu \tau)$, in the plane spanned by 
    $(\delta_{34}-\varphi_4)$ (left) and the plane spanned by 
    $(\delta_{24}-\delta_{34})$ (right).
    We take $m_4=m_5=5$~TeV, fix $\theta_{1j}=10^{-3}$, $\theta_{2j}=0.01$ and $\theta_{3j}=0.1$ ($j=4,5$).
    The colour scheme denotes the associated value of 
     $\mathcal{A}_{CP}(Z\to\mu \tau)$. 
    }
    \label{fig:contour_ACP}
\end{figure}

In Fig.~\ref{fig:contour_ACP}, we explore the joint effects of Dirac and Majorana phases on $\mathcal{A}_{CP}(Z\to\mu \tau)$, still under simplifying assumptions for the mixing angles and masses of the heavy states. As can be seen, provided that Dirac phases are present, the Majorana phases have a significant impact (which was not the case should they be the only source of CPV, as seen from 
Fig.~\ref{fig:Z_cLFV_ACP}).

\begin{figure}[t!]
    \centering
\mbox{   \hspace*{-5mm}  \includegraphics[width=0.51\textwidth]{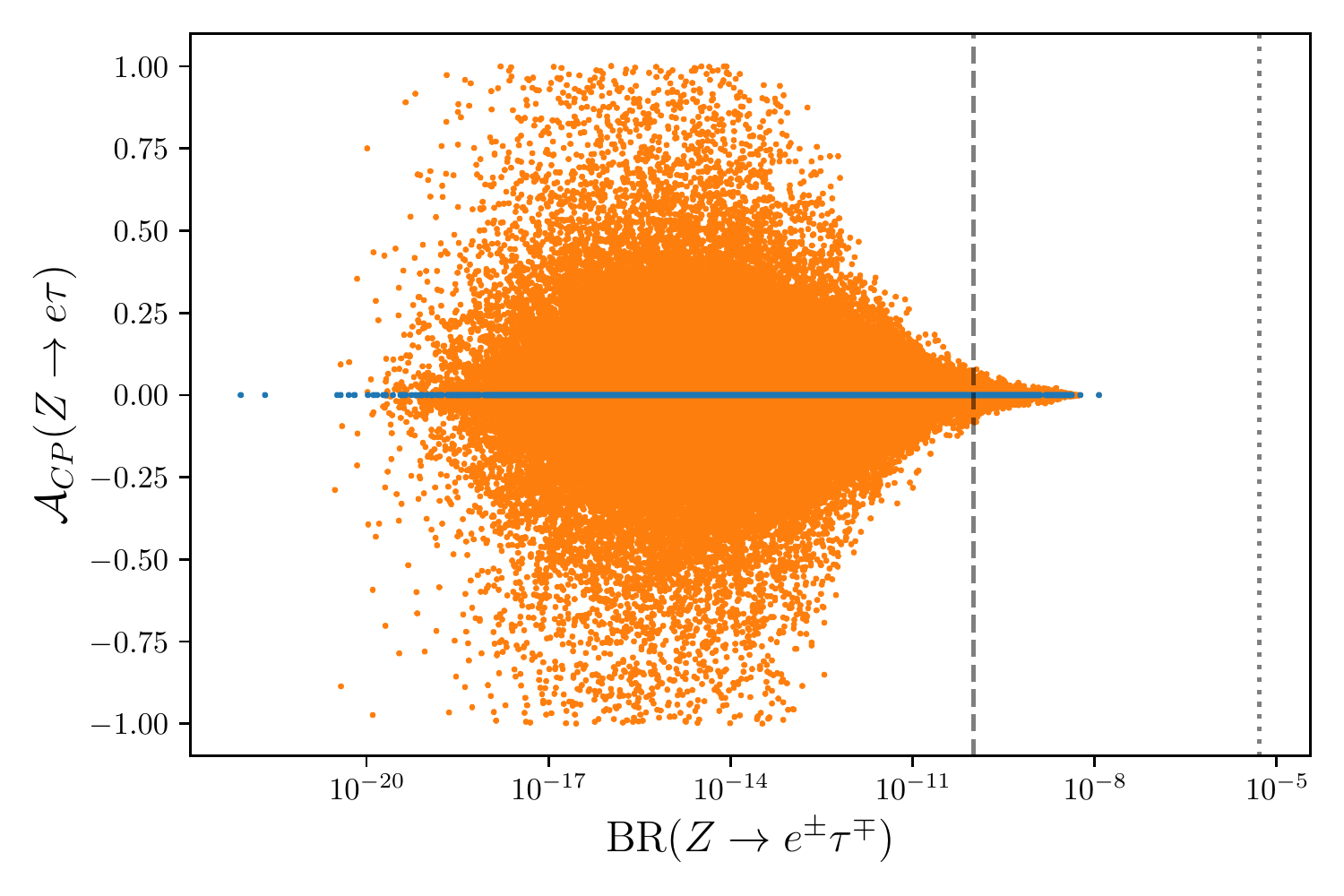}\hspace*{2mm} 
    \includegraphics[width=0.51\textwidth]{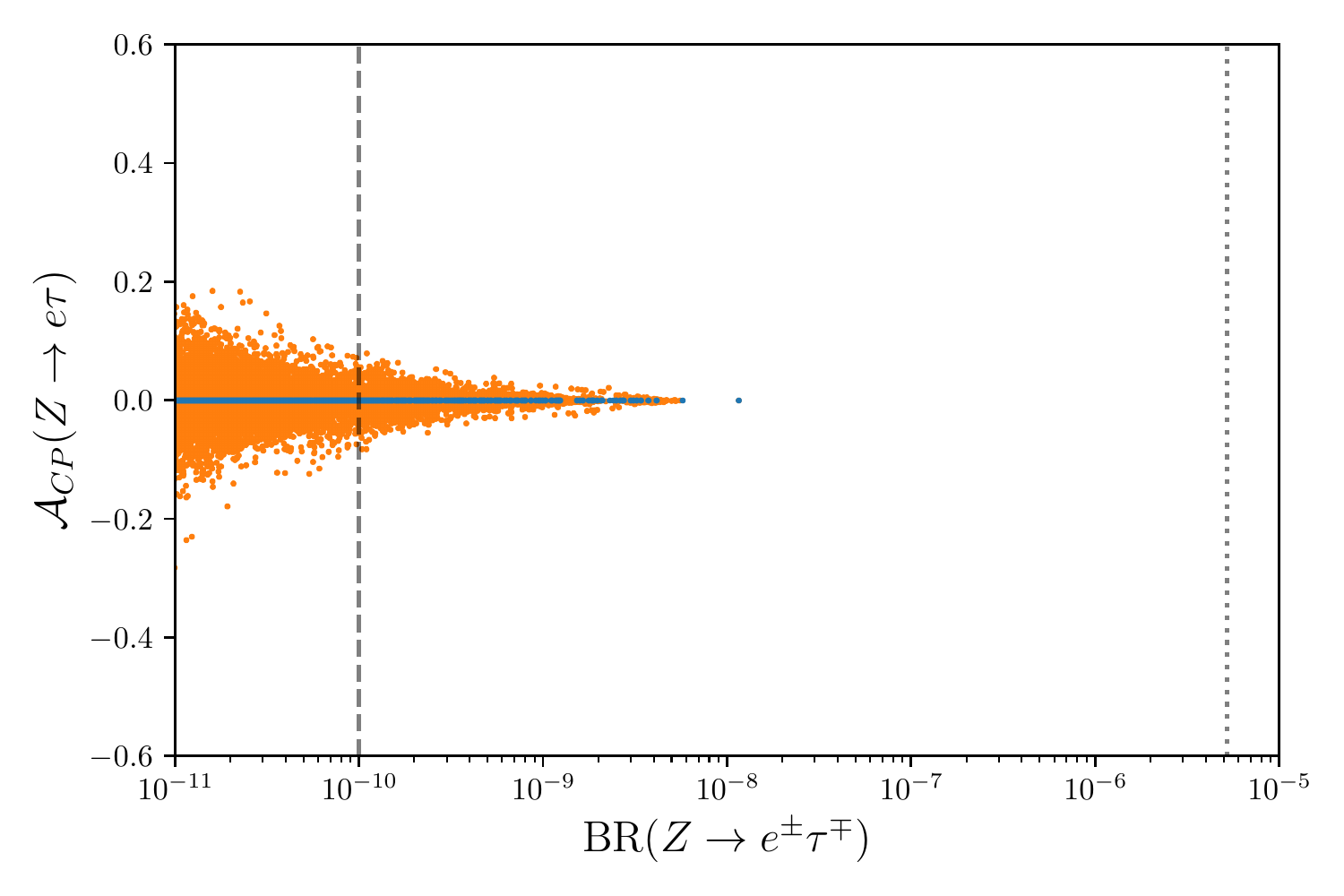}}
    \mbox{   \hspace*{-5mm}  \includegraphics[width=0.51\textwidth]{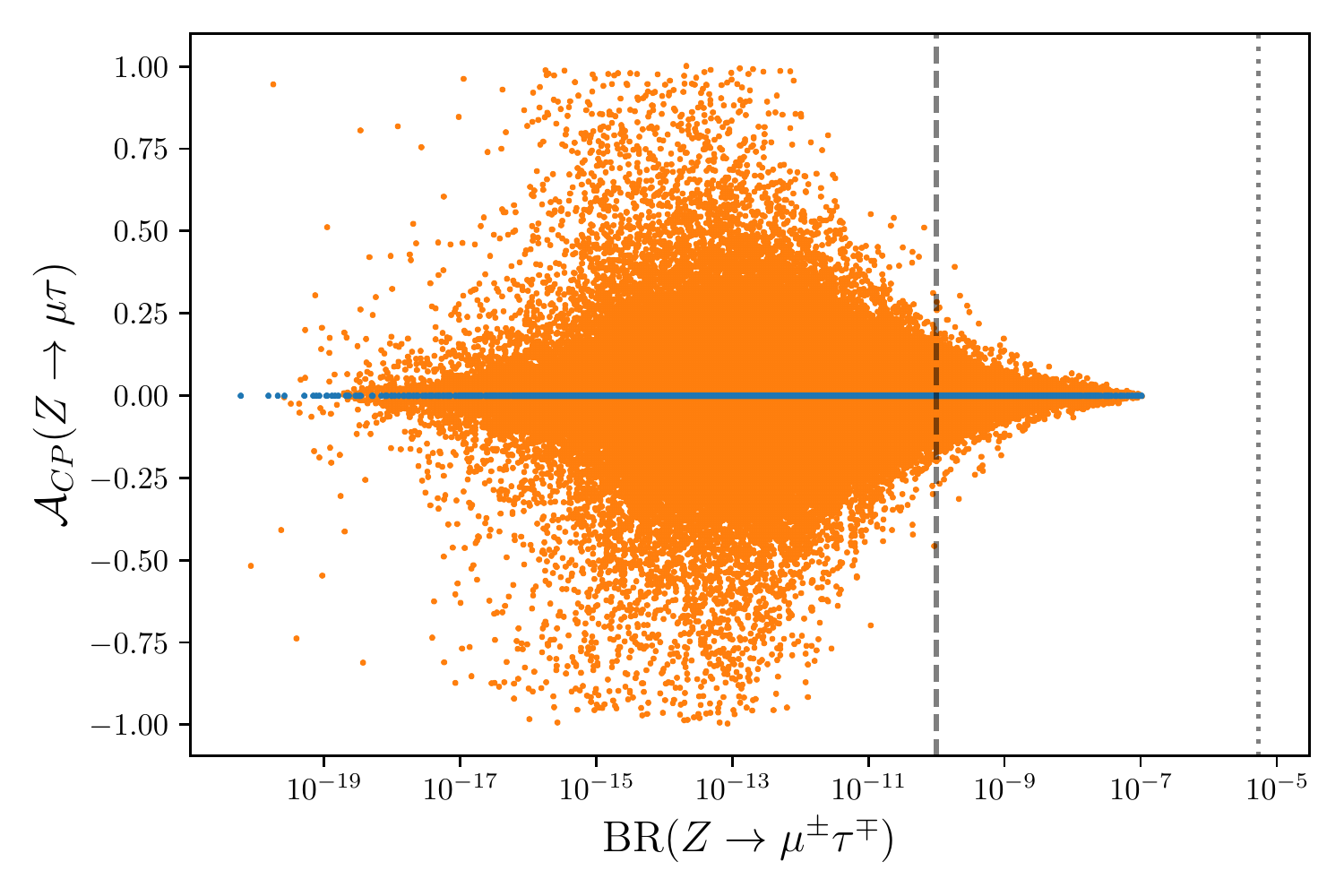}\hspace*{2mm} 
    \includegraphics[width=0.51\textwidth]{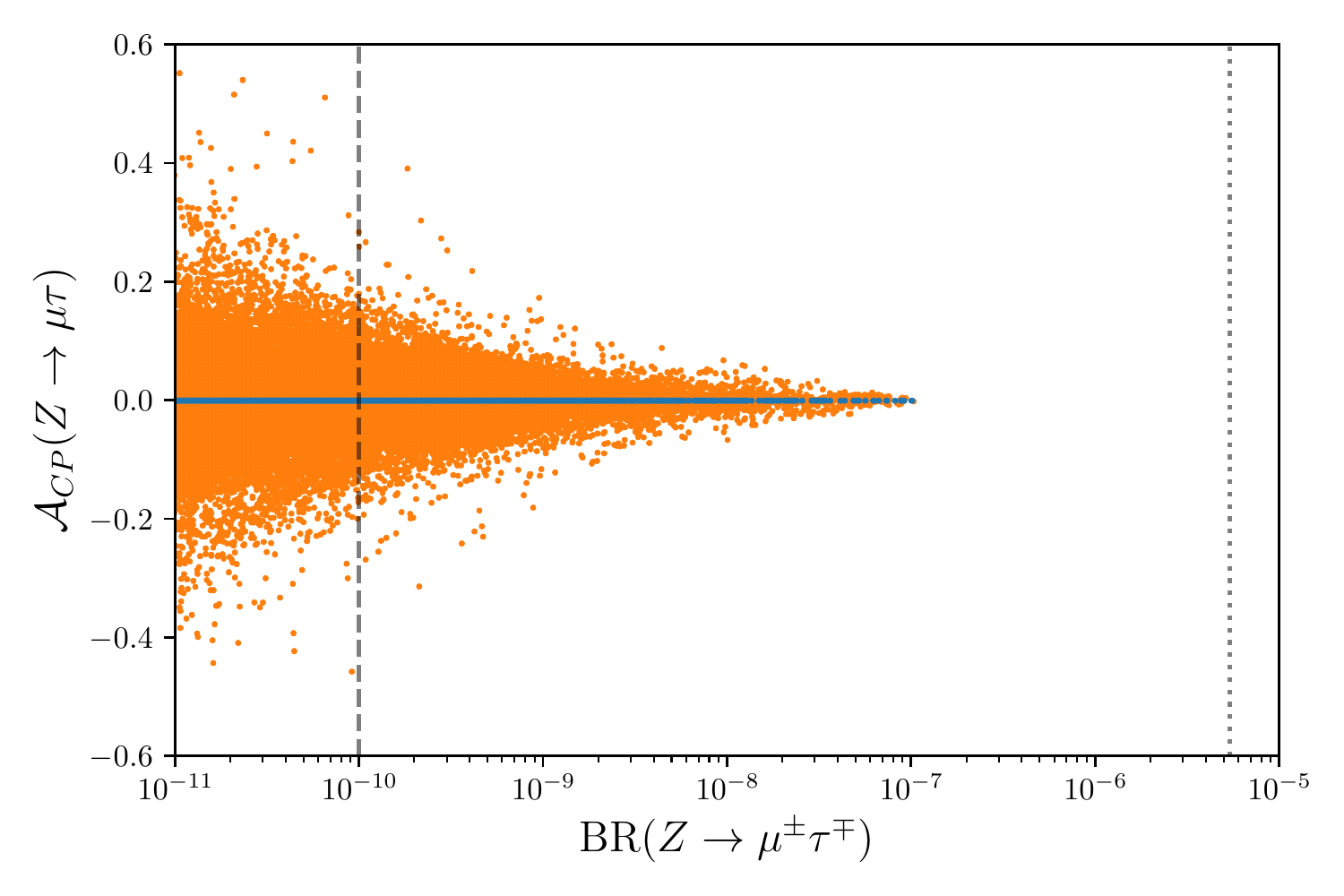}}
    \caption{Prospects for the CP asymmetry $\mathcal{A}_{CP} (Z \to \ell_\alpha \tau)$ as a functtion of the associated cLFV 
    decay rates, BR($Z \to \ell_\alpha \tau$). On the top row, $\ell_\alpha = e$, on the bottom one $\ell_\alpha =\mu$. The right panels depict a closer (zoomed) view of the points which have BR($Z \to \ell_\alpha \tau$) potentially within experimental sensitivity. All active-sterile mixing angles, as well as Dirac and Majorana CP phases, are randomly varied. In all panels, $m_4=5$~TeV, with $m_5-m_4 \in [10~\text{MeV}, 1~\text{TeV}]$. Blue points correspond to vanishing phases, while orange denote random values of all phases ($\delta_{\alpha i}$ and $\varphi_i$, with $\alpha = e,\mu,\tau$ and $i=4,5$). Dotted (dashed) lines denote current bounds (future sensitivity) as given in Table~\ref{tab:cLFV_ZH}.
    }
    \label{fig:z_ACP_scatter}
\end{figure}

\medskip
A thorough phenomenological study has been conducted for the CP asymmetries in the most promising channels of the cLFV $Z$ decays, in particular those leading to final states containing one tau lepton (in view of the better prospects for observation of the decay itself).
In Fig.~\ref{fig:z_ACP_scatter} we display the CP asymmetries $\mathcal{A}_{CP} (Z \to \ell_\alpha \tau)$ vs.~the associated cLFV 
decay rates, BR($Z \to \ell_\alpha \tau$), for $\ell_\alpha = e, \mu$. (See Section~\ref{sec:pheno_num} for details on the underlying numerical scan of the parameter space.)

Once all the relevant experimental and phenomenological constraints are taken into account, one can still have very large asymmetries (up to 100\%) in both cases; notice however that such regimes are associated with rates for the cLFV process that are significantly beyond future sensitivities. 
For regimes associated with BR($Z\to \ell_\alpha \tau$) within future sensitivity, the expected CP asymmetries are typically of order of a few \% to 30\% (lower in the case of $e\tau$ final states). 

\begin{figure}[t!]
    \centering
\mbox{   \hspace*{-5mm}  \includegraphics[width=0.51\textwidth]{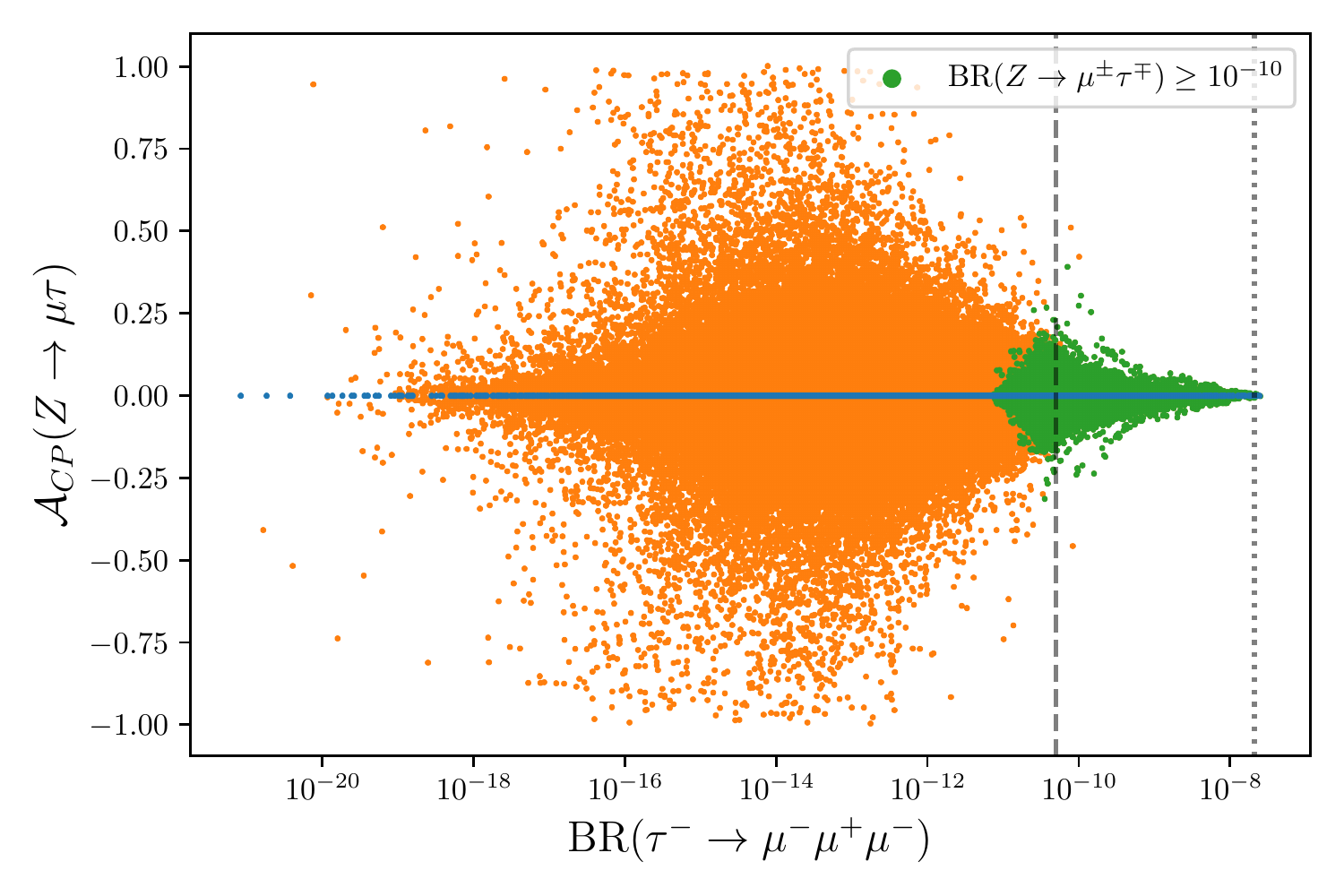}\hspace*{2mm} 
    \includegraphics[width=0.51\textwidth]{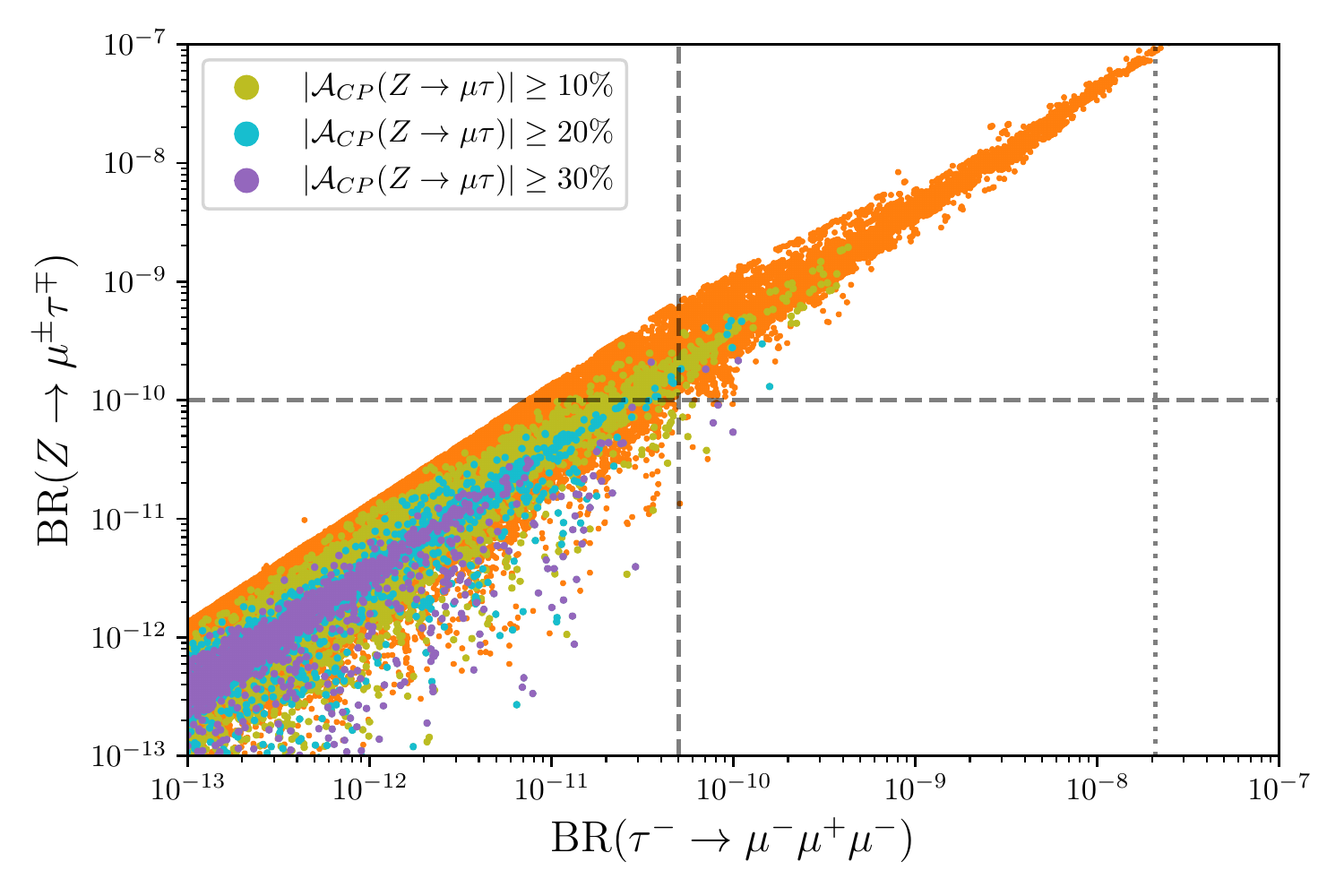}}
    \caption{On the left, prospects for the CP asymmetry $\mathcal{A}_{CP} (Z \to \mu \tau)$ vs.~the cLFV 
    decay rate BR($\tau\to \mu\mu\mu$), for $m_4=5$~TeV, with $m_5-m_4 \in [10~\text{MeV}, 1~\text{TeV}]$. Orange and blue points as indicated in Fig.~\ref{fig:z_ACP_scatter}. In green, points associated with $\mathrm{BR}(Z\to\mu^\pm\tau^\mp) \geq 10^{-10}$, thus within future sensitivity of FCC-ee.
    On the  right, $\mathrm{BR}(Z\to\mu^\pm\tau^\mp)$ vs.~$\mathrm{BR}(\tau\to\mu\mu\mu)$. Colour code as before, with olive green, cyan and purple points respectively denoting  an associated CP asymmetry $|\mathcal A_{CP}(Z\to\mu\tau)|\geq 10\%\,, 20\%\, \text{ and } 30\%$.
    }
    \label{fig:z_ACP_others}
\end{figure}

In the $\mu\tau$ sector, one can in fact simultaneously test the presence of additional heavy neutral leptons (and their CP violating phases) 
via several observables: this is displayed in Fig.~\ref{fig:z_ACP_others}, in which we emphasise the joint behaviour of three observables, 
$\mathcal{A}_{CP} (Z \to \mu \tau)$, $\mathrm{BR}(Z\to\mu^\pm\tau^\mp)$ and the pure leptonic decays BR($\tau\to \mu\mu\mu$). As can be seen from both panels (complementary views of the observables), for regimes leading to both $\mathrm{BR}(Z\to\mu^\pm\tau^\mp)$ and $\mathrm{BR}(\tau\to\mu\mu\mu)$ within future sensitivities\footnote{For lighter HNL masses (e.g. $\mathcal{O}(1\text{ TeV})$) one can still have sizeable asymmetries, as large as 20\%; however, the associated rates for the cLFV decays are smaller, respectively below $10^{-8}$ and $10^{-9}$, for $Z\to\mu\tau$ and $\tau \to 3\mu$ decays. }, one can still have $\mathcal{A}_{CP} (Z \to \mu \tau)$ as large as 20\% (cyan points). Although not a ``smoking gun", the joint observation of the three observables could be interpreted as highly suggestive of such a  extension of the SM via {\it at least} 2 heavy Majorana fermions (featuring new CPV phases).
(Notice that for minimal ``3+1" extensions via a single HNL, the cLFV rates are in general unaffected by the presence of CPV phases.)

\bigskip
Finally, and to further highlight the impact of (potential) measurements of the CP asymmetries, we have selected the following CP conserving ($P_1$) and CP violating ($P_2$) benchmark points, specified as follows:
\begin{eqnarray}
    P_1 &:& m_4 = 5 \:\mathrm{TeV}, 
    m_5 = 5.1 
    \:\mathrm{TeV},\nonumber \\
    &\phantom{:}&
    s_{14} = -0.0028\,,\, s_{15} = 0.0045\,,\, s_{24} = -0.0052\,,\, s_{25} = -0.0037\,,\, s_{34} = -0.052\,,\, s_{35} = -0.028\,,
    \nonumber \\
    &\phantom{:}& \delta_{ij} = \varphi_{i} =0\,,\\
    P_2 &:& m_4 = 5 \:\mathrm{TeV}, m_5 =5.1     
    \:\mathrm{TeV},     \nonumber \\
    &\phantom{:}&
    s_{14} = 0.00020\,,\, s_{15} = -7.1\times 10^{-5}\,,\, s_{24} = -0.0024\,,\, s_{25} = 0.029\,,\, s_{34} = -0.073\,,\, s_{35} = -0.037\,,\nonumber\\
    &\phantom{:}& \delta_{14} =0.71\,,\, \delta_{15} = 5.21\,,\, \delta_{24} = 2.06\,,\, \delta_{25} = 4.78\,,\, \delta_{34} = 3.80\,,\, \delta_{35} = 4.74\,, \nonumber \\
    &\phantom{:}&
        \varphi_{4} = 1.77\,,\, \varphi_5 = 4.33\,.
\end{eqnarray}
Both benchmark points (CP conserving and CP violating) lead to common cLFV predictions as displayed in Table~\ref{tab:pred}, thus rendering them indistinguishable if cLFV signals are observed in the future:
in particular, notice that only the predictions for $\mu\to eee$, $\mathrm{CR}(\mu-e, \:\mathrm{Al})$, $\tau\to \mu\mu\mu$ and $Z\to\mu\tau$ lie within future sensitivities. Notice that for $P_2$, which is associated with smaller mixing angles, CPV phases are at the source of constructive interferences, leading to predictions similar to $P_1$ (larger active-sterile mixings).

\renewcommand{\arraystretch}{1.5}
\begin{table}[h!]
    \centering
    \hspace*{-7mm}{\small\begin{tabular}{|l|c|c|c|c|}
    \hline
    Observable &
    $\mu\to eee$ & 
    $\mu-e \:(\mathrm{Al})$ & 
    $\tau\to \mu\mu\mu$ & 
    $Z\to\mu\tau$    \\
    \hline
    $P_{1,2}$ prediction &
    $2\times 10^{-15}$ & 
    $5\times10^{-14}$ & 
    $1\times 10^{-10}$ & 
    $2\times 10^{-10}$\\
    \hline
    \end{tabular}}
    \caption{cLFV predictions (cLFV ``profile") associated with benchmark points $P_1$ and $P_2$ (see text for details); observables not displayed are associated with rates lying beyond future sensitivity.}
    \label{tab:pred}
\end{table}
\renewcommand{\arraystretch}{1.}

From a phenomenological point of view, and since 
they share the same cLFV ``profile", these two scenarios are in essence indistinguishable, and little can be learnt regarding the presence of leptonic CP violating phases. However, the CP asymmetries in $Z$-boson decays offer a clear distinction between them: $P_2$ leads to the following predictions for the CP asymmetries
\begin{equation}
    \mathcal A_{CP}(Z\to e\mu) = -1\%\,,\quad\quad\mathcal A_{CP}(Z\to e\tau) = -5\%\,,\quad\quad\mathcal A_{CP}(Z\to \mu\tau) = 30\%\,.
\end{equation}
In particular, $\mathcal A_{CP}(Z\to \mu\tau)$ is potentially observable, thus allowing to disentangle between CP conserving and CPV scenarios.
Similarly to what is currently pursued for (semi-)leptonic LFV decays of mesons (e.g. $B\to K^{(*)}\ell_\alpha^\pm\ell_\beta^\mp$) - as motivated by leptoquark models -, we thus recommend a measurement of individual cLFV $Z$ decays (i.e. with charge ``tags'' of the final state charged leptons). 
An independent measurement of both $Z\to \mu^-\tau^+$ and $Z\to \mu^+\tau^-$ rates would significantly help in disentangling CP conserving from CP violating scenarios of HNL models of new physics.

\bigskip
CP asymmetries in cLFV Higgs boson decays 
have also been considered. However, and in comparison with $\mathcal{A}_{CP} (Z \to \ell_\alpha \tau)$, the scalar decays are far less promising. 
As discussed in the previous section, the associated branching ratios (for regimes fulfilling all imposed constraints) are typically beyond future experimental reach; the latter issue non-withstanding, one would nevertheless be led to maximal values of $\mathcal{A}_{CP} (H \to \ell_\alpha \tau)$ much smaller than those found for $Z$ decays, at most $\mathcal O(10^{-12})$. Such a difference can be understood by considering the structure of the ``effective Higgs vertex'' (see Eq.~(\ref{eq:cLFVHdecay})), where the only significant source of a CP asymmetry would stem\footnote{Notice that, as shown by the expressions detailed in Section~\ref{sec:Hdecays:th}, one has $F_L^{\alpha \beta} \simeq F_R^{\beta\alpha}$; therefore any variation in $F_L$ is compensated by one in $F_R$ in the decay rate. No such effect is found for $Z$ decays.} from the interference term ($\propto \text{Re} (\Delta F_L \Delta F_R^*)$); however this would be suppressed by the smallness of the light lepton masses. 

\section{Conclusions}
\label{sec:concs}
Barring symmetries to enforce CP conservation, leptonic CPV phases are a generic feature of SM extensions via heavy neutral fermions. Arising in UV-complete models from various sources (among them Yukawa couplings), in the low-energy (effective) realisations both Dirac and Majorana phases can be present
as intrinsic physical degrees of freedom of these SM extensions.

Following a first assessment of the role of CP violating phases on
cLFV observables~\cite{Abada:2021zcm}, here we have carried out a thorough investigation of the impact of Dirac and Majorana phases on cLFV neutral boson decays. 
We have thus revisited Higgs and $Z$-boson decay rates, computing the 
widths for $Z \to \ell_\alpha \ell_\beta$ and 
$H \to \ell_\alpha \ell_\beta$ without any simplifying approximations. The full computations were done in both Feynman-’t Hooft and unitary gauges, and the
complete expressions are given for future application in a user-friendly way.

Focusing on minimal SM extensions by heavy Majorana sterile fermions, the impact of the CPV phases on a number of cLFV decays was considered. 
In particular, we have illustrated the results for a minimal toy-model, in which 2 heavy Majorana sterile states are added to the SM content (without any assumption on the underlying 
mechanism of neutrino mass generation). Despite its minimality and simplicity, these constructions can be interpreted as a low-energy phenomenological limit of a complete high-energy model; the impact of the heavy states is a consequence of their masses, and of their mixings to the active states (including CP violating phases). 
Even if inferred for such a minimal SM extension, 
the conclusions of the present study are generic for heavy neutral lepton models (in which the new leptonic mixings are at the source of lepton flavour violation). In this sense, the results are comprehensive and generic, 
since the parameters are not related nor are certain regimes protected (or precluded) by any imposed flavour symmetry. 

\medskip
Confirming the findings of previous studies, cLFV $Z$ and Higgs decays are very sensitive to the presence of the additional heavy neutral leptons; moreover, the CP violating phases have a clear impact on the decay rates, leading to both destructive and constructive interferences (again with the Dirac phases allowing for striking reductions - cancellations - of the widths).
Following a first illustrative study, we have conducted a thorough phenomenological investigation, further emphasising the effect of CPV phases on (possibly) correlated behaviours between observables. 

While cLFV branching ratios for $Z \to \ell_\alpha \ell_\beta$ are potentially within reach of a future FCC-ee (especially for heavy states with masses $\mathcal{O} (1 \text{ TeV})$ or above), the prospects for the analogous Higgs decays are not very promising, as a consequence of the poorer future sensitivity. 
In the CP conserving limit, and as expected in view of the (leading) contributing topologies, one observes clearly correlated patterns between $B \to \ell_\alpha \ell_\beta$ and 
$\ell_\alpha \to 3 \ell_\beta$ (for $B=Z, H$). The presence of the CP phases (Dirac and Majorana) does affect the correlated patterns, although in a less striking manner than what occurs for the correlation of purely leptonic cLFV observables (cf.~findings of~\cite{Abada:2021zcm}).

\medskip
Motivated by the clean environment and future sensitivity of an FCC-ee (running at the $Z$-pole) for $Z \to \ell_\alpha \ell_\beta$ transitions, we have also considered a distinct class of observables: the CP asymmetries associated with the latter cLFV boson decays. 

Non-negligible - large - CP asymmetries in $Z$ decays turn out to be a generic feature of HNL extensions encompassing {\it at least} 2 heavy states. 
For final states composed of a tau and a light charged lepton,
with decay rates potentially within future sensitivity, one can have very large CP asymmetries; in particular, these can be as large as 20--30\% for the case of $\mathcal{A}_{CP}(Z\to\mu \tau)$, interestingly in association with sizeable rates for $\tau\to 3 \mu$ (also within future sensitivity).

As we have strongly emphasised, the presence of CP violating phases can open the possibility of having identical cLFV signatures in association with very distinct regimes of active-sterile mixings (a consequence of the interference effects due to the phases). 
Exploring the CP asymmetries in cLFV $Z$ decays then offers a complementary probe to clearly assess the presence of CP violation. As highlighted, and despite not definitive, the joint observation of cLFV leptonic decays, together with 
cLFV $Z$ decays and the associated CP asymmetry would strongly hint towards the presence of HNL-mediated flavour and CP violation. 
Moreover, and while a given cLFV observation can either stem from minimal extensions by only one sterile fermion (``3+1" model), the observation of a CP asymmetry is a clear signal of the presence of non-minimal HNL realisations (at least a ``3+2"-like SM extension).

If on the one hand it is clear that CP violating phases
should be in general taken into account upon comparison between prediction and observation in the context of cLFV HNL extensions of the SM, 
$\mathcal{A}_{CP}(Z\to\ell_\alpha \ell_\beta)$ might hold the key to clearly establishing the presence of leptonic CP violation; in turn, this might have strong implications regarding leptogenesis (relying on complete models including heavy sterile states).
Whenever possible, data from individual channels (i.e. $\ell^+_\alpha \ell^-_\beta$ and $\ell^-_\alpha \ell^+_\beta$) should thus be analysed and compared.

\section*{Acknowledgements}
This project has received support from the European Union's Horizon 2020 research and innovation programme under the Marie Sk\l{}odowska-Curie grant agreement No.~860881 (HIDDe$\nu$ network) and from the IN2P3 (CNRS) Master Project, ``Flavour probes: lepton sector and beyond'' (16-PH-169).

\appendix
\section{Constraints on HNL extensions of the SM}\label{app:constraints}
In this appendix we collect the most relevant constraints on heavy neutral fermion extensions of the SM, in which the additional states have non-negligible mixings with the light (mostly) active neutrinos. 
In addition to complying with oscillation data (which we summarise in the following subsection), numerous bounds arising from observation lead to constraints on the new degrees of freedom. The relevant ones for our study are subsequently summarised.

\subsection{Neutrino oscillation data}\label{app:nudata}
Below we collect the NuFIT 5.1 global fit results~\cite{Esteban:2020cvm} for neutrino mixing data as used in our analysis. Although the predictions for cLFV transitions are mostly independent of the ordering of the spectrum and of the lightest neutrino mass, 
for concreteness we notice that in the numerical analysis we have assumed a normal ordered light neutrino spectrum, with the lightest neutrino mass in the range $m_0 \in [10^{-5}, 10^{-3}]\:\mathrm{eV}$.

\renewcommand{\arraystretch}{1.3}
\begin{table}[h!]
    \centering
    \begin{tabular}{|c|c|c|}
        \hline
        & Normal ordering & Inverted ordering \\
        \hline
        \hline
      $\sin^2\theta_{12}$  & $0.304^{+0.013}_{-0.012}$ & $0.304^{+0.012}_{-0.012}$ \\
      \hline
      $\sin^2\theta_{23}$ & $0.573^{+0.018}_{-0.023}$ & $0.578^{+0.017}_{-0.021}$\\
      \hline
      $\sin^2\theta_{13}$ & $0.02220^{+0.00068}_{-0.00062}$ & $0.02238^{+0.00064}_{-0.00062}$\\
      \hline
      $\Delta m_{21}^2/10^{-5}\,\mathrm{eV}$ & $7.42^{+0.21}_{-0.20}$ & $7.42^{+0.21}_{-0.20}$\\
      \hline
      $\Delta m_{3\ell}^2/10^{-3}\,\mathrm{eV}$ & $2.515^{+0.028}_{-0.028}$ & $-2.498^{+0.028}_{-0.029}$\\
      \hline
    \end{tabular}
    \caption{Global fit results obtained by NuFIT 5.1~\cite{Esteban:2020cvm} for neutrino mixing data, not including experimental data from oscillation experiments measuring atmospheric neutrinos. In our numerical analysis we assume normal ordering of the light neutrino spectrum ($\Delta m_{3\ell}=\Delta m_{31}>0$) and fix the neutrino mixing parameters to their central values.}
    \label{tab:nufit}
\end{table}
\renewcommand{\arraystretch}{1.}

\subsection{Further phenomenological constraints}\label{app:HNLconstraints}
The mixing angles $\theta_{\alpha 4(5)}$ between the active and heavy states can be constrained from precision observables which are insensitive to the new CPV-violating phases, and which can be divided into high- and low-energy ones. 

At high energies, one can construct the lepton universality ratios of $W$ boson decays~\cite{Abada:2013aba}
\begin{equation}
    R_W^{\ell_\alpha \ell_\beta}\,=\,\frac{\Gamma(W\rightarrow \ell_\alpha\,\nu)}{\Gamma(W\rightarrow \ell_\beta\,\nu)}\,,
\end{equation}
with $\alpha, \beta \in\lbrace e,\mu,\tau\rbrace$. Moreover, the invisible decay width of the $Z$ boson, $\Gamma(Z\rightarrow\mathrm{inv})$, plays a crucial role in constraining scenarios with heavy neutral leptons, as it is typically reduced when compared to the SM prediction.

At low energies, several ratios from (semi-)leptonic decays of the $\tau$-lepton and leptonic decays of light mesons can be constructed; these are sensitive to the modification of the $W\ell \nu$ vertex in the presence of sterile fermions. In particular, one has~\cite{Abada:2013aba,Abada:2012mc}
\begin{eqnarray}
    R_{\tau} &\equiv& \frac{\Gamma(\tau\rightarrow \mu\,\nu\,\nu)}{\Gamma(\tau\rightarrow e\,\nu\,\nu)}\,,\\
    \Delta r_P &\equiv& \frac{R_P}{R_P^{\mathrm{SM}}}-1\,,\\
    \mathrm{where \, \, }\; R_P &\equiv& \frac{\Gamma(P^+\rightarrow e\,\nu)}{\Gamma(P^+\rightarrow \mu\,\nu)}\quad (P=\pi,K)\,,\\
    R_e &\equiv& \frac{\Gamma(\pi^+\rightarrow e\,\nu)}{\Gamma(K^+\rightarrow e\,\nu)}\,,\quad R_\mu \equiv \frac{\Gamma(\pi^+\rightarrow \mu\,\nu)}{\Gamma(K^+\rightarrow \mu\,\nu)}\,.
\end{eqnarray}
Additionally, we take into account upper bounds on the entries of $\eta$ (see Eq.~(\ref{eq:defPMNSeta})) as derived in~\cite{Fernandez-Martinez:2016lgt}, taking into account indirect modifications of the Fermi constant $G_F$, the weak mixing angle $\sin^2\theta_w$ and the mass of the $W$ boson, among others.

For masses of the HNL at the TeV scale (as we are interested in our study), constraints from direct searches at colliders or from cosmology (such as big bang nucleosynthesis) are not competitive and will not be taken into account; nonetheless, we can derive theoretical constraints by imposing that decays of the new heavy states comply with perturbative unitarity~\cite{Chanowitz:1978mv,Durand:1989zs,Bernabeu:1993up,Fajfer:1998px,Ilakovac:1999md}, posing a direct bound on their decay width $\Gamma_{\nu_i}/m_i<1/2$ for $i\geq 4$. Since the dominant contribution to their decay stems from the $W$ exchange, the bound can be written as
\begin{equation}
    m_{\nu_i}^2\,C_{ii}\,<2\,\frac{M_W^2}{\alpha_w}\quad (i\geq4)\,,
\end{equation}
where $\alpha_w=g_w^2/(4\pi)$.
Finally, the inclusion of sterile states leads to the modification of the prediction for the Majorana effective mass, $m_{ee}$, to which the amplitude for neutrinoless double beta decay is proportional. In the presence of $n_S$ new states, $m_{ee}$ can be written as
\begin{equation}
    m_{ee}\simeq\sum_{i=1}^3 \mathcal{U}_{ei}^2 \,m_i+\sum_{i=4}^{3+n_S}\mathcal{U}_{ei}^2\,p^2\, \frac{m_i}{p^2-m_i^2}\,,
\end{equation}
in which $p^2$ corresponds to the typical virtual momentum, $p^2\simeq -(100\,\mathrm{MeV})^2$. We take into account the KamLAND-ZEN upper limit~\cite{KamLAND-Zen:2016pfg} $m_{ee}\leq (61\div 165)$~MeV. 

\section{Feynman rules for SM extensions via Majorana sterile fermions}\label{app:feynrules}
In Table~\ref{table:feynrules}, we summarise in a compact way the relevant interaction vertices, as inferred from the Lagrangian terms in Eq.~(\ref{eq:lagrangian:WGHZ}). In the $Z$ and Higgs vertices, the arrows denote the momentum flow. 
We note here that diagrams including at least one $Z n_i n_j$ or $H n_i n_j$ vertex have to be symmetrised (factor 2) due to the Majorana nature of the physical neutrinos.
\hspace*{-2mm}
\begin{table}[hb!]
\begin{tabular}{m{2.4cm}m{5.35cm}m{2.4cm}m{5.35cm}}
        \begin{tikzpicture}
    \begin{feynman}
    \vertex (a) at (0,0) {\(Z_\mu\)};
    \vertex (b) at (1,0);
    \vertex (c) at (2,1.){\(n_i\)};
    \vertex (d) at (2,-1.){\(n_j\)};
    \diagram* {
    (a) -- [boson] (b),
    (c) -- [momentum'=\( \)] (b),
    (b) -- [momentum'=\( \)] (d),
    };
    \end{feynman}
    \end{tikzpicture} 
    & $= \, -i \dfrac{g_w}{4 c_w} \gamma_\mu \left(C_{ij}^* \,P_L -C_{ij} \,P_R \right)$ 
    & & 
\\
\\
        \begin{tikzpicture}
    \begin{feynman}
    \vertex (a) at (0,0) {\(H\)};
    \vertex (b) at (1,0);
    \vertex (c) at (2,1.){\(n_i\)};
    \vertex (d) at (2,-1.){\(n_j\)};
    \diagram* {
    (a) -- [scalar] (b),
    (c) -- [momentum'=\( \)] (b),
    (b) -- [momentum'=\( \)] (d),
    };
    \end{feynman}
    \end{tikzpicture}
    & $=\, -i\dfrac{g_w}{4 M_W} \left[C_{ij} \left( m_i\, P_L + m_j \,P_R \right)
    + C_{ij}^* \left( m_i\, P_R + m_j \,P_L \right) \right]$
    \\
    \\
    \begin{tikzpicture}
    \begin{feynman}
    \vertex (a) at (0,0) {\(W_\mu^-\)};
    \vertex (b) at (1,0);
    \vertex (c) at (2,1.){\(n_i\)};
    \vertex (d) at (2,-1.){\(\ell_\alpha^-\)};
    \diagram* {
    (a) -- [boson] (b),
    (c) -- [fermion] (b),
    (b) -- [fermion] (d),
    };
    \end{feynman}
    \end{tikzpicture}
    & $=\, -i\dfrac{g_w}{\sqrt{2}} \,\mathcal{U}_{\alpha i}\, \gamma_\mu\, P_L$
    &
    %
    \begin{tikzpicture}
    \begin{feynman}
    \vertex (a) at (0,0) {\(W_\mu^+\)};
    \vertex (b) at (1,0);
    \vertex (c) at (2,1.){\(\ell_\alpha^+\)};
    \vertex (d) at (2,-1.){\(n_i\)};
    \diagram* {
    (a) -- [boson] (b),
    (c) -- [fermion] (b),
    (b) -- [fermion] (d),
    };
    \end{feynman}
    \end{tikzpicture}
    &
    $=\, -i\dfrac{g_w}{\sqrt{2}} \,\mathcal{U}_{\alpha i}^*\, \gamma_\mu\, P_L$
    \\
    \\
    \begin{tikzpicture}
    \begin{feynman}
    \vertex (a) at (0,0) {\(G^-\)};
    \vertex (b) at (1,0);
    \vertex (c) at (2,1.){\(n_i\)};
    \vertex (d) at (2,-1.){\(\ell_\alpha^-\)};
    \diagram* {
    (a) -- [scalar] (b),
    (c) -- [fermion] (b),
    (b) -- [fermion] (d),
    };
    \end{feynman}
    \end{tikzpicture}  
    &
    $=\, i\dfrac{g_w}{\sqrt{2}M_W} \,\mathcal{U}_{\alpha i} \,\left(m_i\, P_R - m_\alpha\, P_L\right)$
&
    %
    \begin{tikzpicture}
    \begin{feynman}
    \vertex (a) at (0,0) {\(G^+\)};
    \vertex (b) at (1,0);
    \vertex (c) at (2,1.){\(\ell_\alpha^+\)};
    \vertex (d) at (2,-1.){\(n_i\)};
    \diagram* {
    (a) -- [scalar] (b),
    (c) -- [fermion] (b),
    (b) -- [fermion] (d),
    };
    \end{feynman}
    \end{tikzpicture}
    &
    $=\, i\dfrac{g_w}{\sqrt{2}M_W} \,\mathcal{U}_{\alpha i}^* \,\left(m_i \,P_L - m_\alpha \,P_R\right)$
\end{tabular}
\caption{Feynman rules for $W$, $Z$ and Higgs interactions (and associated Goldstone bosons) in SM extensions via Majorana sterile fermions: $n_i$ denotes neutrino mass eigenstates (with $i=1\dots5$), while $\ell_\alpha$ corresponds to charged leptons. }\label{table:feynrules}
\end{table}

\mathversion{bold}
\section{Additional information on leptonic $Z$ and  Higgs decays}\label{app:ZHdecays:formulae}
\mathversion{normal}
In this appendix we discuss in detail different aspects of the computation of the cLFV $Z$ and  Higgs decays. 
We present sub-dominant terms not included in the main text, and we also include the results derived  in the Feynman-’t Hooft gauge. 

\medskip
For the purpose of renormalisation, and in order to achieve finite and gauge-invariant results for the analytical calculation of the loop amplitudes, we work in the framework of low-scale type-I seesaw models~\cite{Minkowski:1977sc,Yanagida:1979as,Glashow:1979nm,Gell-Mann:1979vob,Mohapatra:1979ia}.
We thus hypothesise a neutrino Majorana mass matrix (complex and symmetric) of the form
\begin{equation}
    \mathcal M_\nu = \begin{pmatrix}\mathbb{0}_{3\times3} & m_D\\
    m_D^T & m_M        
        \end{pmatrix}\,.
\end{equation}
Notice that in the interaction basis the active neutrino mass vanishes due to gauge invariance. Light neutrino masses are generated by  Majorana mass terms from the heavy sector (with $n_S$ HNL) and Yukawa couplings.
The full (Majorana) neutrino mass matrix $\mathcal M_\nu$ can be diagonalised by a unitary rotation
\begin{equation}
    \mathcal U^\dagger \,\mathcal M_\nu\, \mathcal U^* = \mathrm{diag}(m_{\nu_1}, ..., m_{\nu_{3+n}})\,.
\end{equation}
Due to vanishing active neutrino masses in the interaction basis, the mixing matrix $\mathcal U$ has certain properties pertinent for the renormalisability of the model (which will be subsequently detailed).
The charged lepton masses are assumed to be diagonal, such that the leptonic mixing matrix is then given by $\mathcal U$.

In what follows, we provide detailed expressions for the  cLFV $Z$-boson decay rates, also discussing several points related with the renormalisation of the contributions.

\mathversion{bold}
\subsection{Full cLFV $Z$-boson decay rate}
\mathversion{normal}
The full decay rate (adding the sub-leading contributions to the 
width given in Eq.~(\ref{eq:Zwidth:compact})) is
\begin{eqnarray}
    \Gamma(Z\to\ell_\alpha^-\ell_\beta^+) &=& \frac{1}{48\pi^2\, M_Z}\sqrt{1 - \frac{(m_\alpha + m_\beta)^2}{M_Z^2}}\sqrt{1 - \frac{(m_\alpha - m_\beta)^2}{M_Z^2}}\times\Bigg\{\nonumber\\
    &\phantom{=}& \left(|F_V^L|^2 + |F_V^R|^2\right)\left(2 M_Z^2 - (m_\alpha^2 + m_\beta^2) - \frac{(m_\alpha^2 - m_\beta^2)^2}{M_Z^2}\right)\nonumber\\
    &\phantom{=}& +\, \left(|F_T^L|^2 + |F_T^R|^2\right)\left(M_Z^4 + (m_\alpha^2 + m_\beta^2)\,M_Z^2 - (m_\alpha^2  - m_\beta^2)^2\right)\nonumber\\
    &\phantom{=}& +\, 12\,\mathrm{Re}\left(F_V^L \,F_V^{R\,\ast} - F_T^L \,F_T^{R\,\ast} \,M_Z^2\right)\,m_\alpha\, m_\beta \nonumber\\
    &\phantom{=}& +\, 6\,\mathrm{Im}\left(F_T^L\, F_V^{L\,\ast} - F_T^R \,F_V^{R\,\ast}\right)(m_\alpha^3 - m_\alpha\, m_\beta^2 - m_\alpha\, M_Z^2)\nonumber\\
    &\phantom{=}& +\, 6\,\mathrm{Im}\left(F_T^L \,F_V^{R\,\ast} - F_T^R \,F_V^{L\,\ast}\right)(m_\beta^3 - m_\alpha^2 \,m_\beta - m_\beta\, M_Z^2)\Bigg\}\,.
\end{eqnarray}
\mathversion{bold}
\subsection{Divergences in cLFV $Z$-boson decays}
\mathversion{normal}
\label{app:divZ}
As mentioned in Section~\ref{sec:cLFVZUG}, terms proportional to the $B_{0,1}$ and $C_{00}$ functions contain UV-divergences which are regulated dimensionally (via an integration in $D=4-2\varepsilon$ dimensions).
The UV-divergent pieces of the different contributions are given by (the superscripts associated with the distinct terms refer to the 
diagrams of Fig.~\ref{fig:cLFVZdecays:UG}):
\begin{eqnarray}
    \mathrm{div}(F_V^{L\:(a)}) &=& \frac{g_w^3}{128 \pi^2 \,c_w\, M_W^2}\sum_{i,j}\,\mathcal U_{\alpha i}\,\mathcal U_{\beta j}^\ast\left[-C_{ij}\,(m_\alpha^2 + m_\beta^2 - 2(m_i^2 + m_j^2)) + C_{ij}^\ast \,m_i \,m_j\right]\,\Delta_\varepsilon\,,\nonumber\\
    \mathrm{div}(F_V^{L\:(b)}) &=& \frac{g_w^3 c_w^2}{128 \pi^2 \,c_w\, M_W^2}\sum_{i}\,\mathcal U_{\alpha i}\,\mathcal U_{\beta i}^\ast\left[(2 m_\alpha^2 + 2 m_\beta^2 - 6 m_i^2 - \frac{m_i^2 q^2}{M_W^2}) + \frac{16 M_W^2\, q^2 + q^4}{3 M_W^2}\right]\, \Delta_\varepsilon\,,\nonumber\\
    \mathrm{div}(F_V^{L\:(c+d)}) &=& \frac{g_w^3(1 - 2 c_w^2)}{128 \pi^2 \,c_w\, M_W^2}\sum_i\,\mathcal U_{\alpha i}\,\mathcal U_{\beta i}^\ast\left[m_\alpha^2 + m_\beta^2 - 3 m_i^2 \right]\,\Delta_\varepsilon\,,\nonumber\\
    \mathrm{div}(F_V^{R\:(a)}) &=& - \frac{g_w^3}{128\pi^2 \,c_w\, M_W^2}\,m_\alpha\, m_\beta \sum_{i,j}\,\mathcal U_{\alpha i}\,\mathcal U_{\beta j}^\ast C_{ij}\,{\Delta_\varepsilon}\,,\nonumber\\
    \mathrm{div}(F_V^{R\:(b)}) &=& \frac{g_w^3 c_w}{128\pi^2 \,M_W^4}\,m_\alpha\, m_\beta \sum_i \,\mathcal U_{\alpha i}\,\mathcal U_{\beta i}^\ast(2 M_W^2 - q^2)\,{\Delta_\varepsilon}\,,\nonumber\\
    \mathrm{div}(F_V^{R\:(c+d)}) &=& \frac{g_w^3 (1 - c_w^2)}{64\pi^2\, c_w \,M_W^2}\,m_\alpha\, m_\beta  \sum_i \,\mathcal U_{\alpha i}\,\mathcal U_{\beta i}^\ast\,{\Delta_\varepsilon}\,,
\end{eqnarray}
with 
\begin{equation}
    \Delta_\varepsilon \,= \,\frac{1}{\varepsilon} - \gamma_E + \log(4\pi)\,,
\end{equation}
in which $\gamma_E$ is the Euler-Mascheroni constant (all other contributions are finite).
Being a (semi-)uni\-tary matrix, $\mathcal U$ fulfils the following identities:
\begin{eqnarray}
    \sum_i \,\mathcal U_{\alpha i}\,\mathcal U_{\beta i}^\ast &=& \delta_{\alpha\beta}\label{eqn:iden:start}\\
    \sum_{i,j} \,\mathcal U_{\alpha i}\,\mathcal U_{\beta j}^\ast\, C_{ij} &=& \sum_{i,j,\rho} \,\mathcal U_{\alpha i}\,\mathcal U_{\beta j}^\ast \,\mathcal U_{\rho i}^\ast\, \mathcal U_{\rho j} = \delta_{\alpha\beta}\\
    \sum_{i,j} \,\mathcal U_{\alpha i}\,\mathcal U_{\beta j}^\ast\, C_{ij} \,m_j^2 &=& \sum_{i,j,\rho} \,\mathcal U_{\alpha i}\,\mathcal U_{\beta j}^\ast \,\mathcal U_{\rho i}^\ast \,\mathcal U_{\rho j} \,m_j^2 = \sum_{j,\rho} \delta_{\rho \alpha} \,\mathcal U_{\beta j}^\ast\,\mathcal U_{\rho j} \,m_j^2 = \sum_{i}\, \mathcal U_{\alpha i}\,\mathcal U_{\beta i}^\ast \,m_i^2\\
     \sum_{i,j} \,\mathcal U_{\alpha i}\,\mathcal U_{\beta j}^\ast \,C_{ij} \,m_i^2 &=& \sum_{i,j,\rho}U_{\alpha i}\,\mathcal U_{\beta j}^\ast \,\mathcal U_{\rho i}^\ast \,\mathcal U_{\rho j} \,m_i^2 = \sum_{i,\rho} \delta_{\rho \beta} \,\mathcal U_{\alpha i}\,\mathcal U_{\rho i}^\ast \,m_i^2 = \sum_{i}\, \mathcal U_{\alpha i}\,\mathcal U_{\beta i}^\ast \,m_i^2\,.
\end{eqnarray}
In renormalisable extensions of the SM with fermion singlets, such as the type-I seesaw (and its variants), the following additional identities hold:
\begin{eqnarray}
    \sum_{i} \,\mathcal U_{\alpha i}^\ast\,\mathcal U_{\beta i}^\ast\, m_i &=& \hat M_{\alpha\beta} = 0\,,\\
    \sum_{i,j} \,\mathcal U_{\alpha i}\,\mathcal U_{\beta j}^\ast C_{ij}^\ast\, m_i \,m_j &=& \sum_{i,j\rho} \,\mathcal U_{\alpha i}\,\mathcal U_{\beta j}^\ast\,\mathcal U_{\rho i}\,\mathcal U_{\rho j}^\ast \,m_i \,m_j = \sum_{i,\rho} \,\mathcal U_{\alpha i}\,\mathcal U_{\rho i} \,m_i \,\hat M_{\beta\rho} = 0\,,\label{eqn:iden:end}
\end{eqnarray}
in which $\hat M_{\alpha\beta}$ (with $\alpha,\beta = 1, 2, 3$) denotes the mass matrix of the active neutrinos in the interaction basis, which generally contains only vanishing elements. Using the identities in Eqs.~(\ref{eqn:iden:start}--\ref{eqn:iden:end}) and  the tree-level definition of $\cos^2\theta_w = M_W^2/M_Z^2$, it can be easily shown that the sum of diagrams (a)-(d) is (only) finite if the $Z$ boson is on-shell (i.e. $q^2 = M_Z^2$).
Thus, for the purpose of on-shell $Z$-boson decays, we can safely consider the limit $D\to4$ in Eqs.~(\ref{eqn:FVLaU}--\ref{eqn:FVLcdU}), and the (partial) amplitude is manifestly finite and independent of the 't Hooft renormalisation scale\footnote{In the case of the 
considered SM extension
the identity of Eq.~\eqref{eqn:iden:end} is in general not fulfilled and thus the corresponding UV-divergent term in $F_V^{L\:(a)}$ does not cancel. Therefore, in our numerical analysis, we minimally subtract this remaining divergence in the $\overline{\mathrm{MS}}$-scheme and fix the 't Hooft scale to $M_Z$.}.
For the off-shell cLFV vertex, the corresponding box-diagrams (that contribute to the same process as the $Z$-penguin) have to be taken into account in order to yield a finite and gauge-independent amplitude.

\mathversion{bold}
\subsection{Additional amplitudes for the cLFV $Z$-vertex}\label{app:ampZ}
\mathversion{normal}
Here we present the additional ``scalar'' amplitudes that appear in the effective cLFV $Z$-vertex, but which do not contribute to the decay rates since they vanish due to the Ward identity.
The contributions are given by (see Eq.~\eqref{eqn:Zamplitude} for their definition)
\begin{eqnarray}
    F_S^{L\:(a)} &=& \frac{g_w^3}{64\pi^2 \,c_w\, M_W^2} m_\alpha\sum_{i,j}\,\mathcal U_{\alpha i}\,\mathcal U_{\beta j}^\ast\Bigg\{ C_{ij}\Big[m_j^2 \,C_2 + m_\beta^2\,(C_{12} - C_{22} - C_2)\nonumber\\
    &\phantom{=}& + M_W^2\,\Big((D-2)\,(C_{11} - C_{12}) + (D-4)\,C_1 - 2C_2 -2 C_0 \Big)\Big]\nonumber\\
    &\phantom{=}& + m_i \,m_j\, C_{ij}^\ast\,\Big[C_{12} - C_{11}\Big]\Bigg\}\,,\label{eqn:FSLaU}\\
    F_S^{R\:(a)} &=& -\frac{g_w^3}{64\pi^2 \,c_w \,M_W^2} m_\beta\,\sum_{i,j}\,\mathcal U_{\alpha i}\,\mathcal U_{\beta j}^\ast\,\Bigg\{ C_{ij} \,\Big[m_i^2 \,C_1 + m_\alpha^2\,(C_{12} - C_1 - C_{11})\nonumber\\
    &\phantom{=}& + M_W^2\,\Big((D-2)\,(C_{22} - C_{12}) + (D-4)\,C_2 - 2C_1 - 2C_0\Big)\Big]\nonumber\\
    &\phantom{=}& + m_i \,m_j\, C_{ij}^\ast\,\Big[C_{12} - C_{22}\Big]\Bigg\}\,,\label{eqn:FSRaU}
\end{eqnarray}
with the Passarino-Veltman functions defined as in Eqs.~(\ref{eqn:FVLaU}--\ref{eqn:FTRaU}).
The remaining (non-vanishing) contributions are given by
\begin{eqnarray}
    F_S^{L\:(b)} &=& \frac{g_w^3 c_w}{64\pi^2 \,M_W^4}m_\alpha \sum_i\,\mathcal U_{\alpha i}\, \mathcal U_{\beta i}^\ast\,\Bigg\{(m_i^2 - 2M_W^2)\,B_0^\beta + m_\beta^2\,B_1^\beta - m_i^2\,(B_0^\alpha + B_1^\alpha) \nonumber\\
    &\phantom{=}& + A_0 - 2M_W^2\,B_0^q + (2 M_W^2 - q^2)\,(B_1^q + 2 B_{11}^q) + 2M_W^2\,(m_\alpha^2 - m_i^2 - M_W^2)\,C_0\nonumber\\
    &\phantom{=}& + (2M_W^2\,(m_\alpha^2 - m_\beta^2 + 2M_W^2) + m_i^2\,(2M_W^2 - q^2))\,C_1 + (2M_W^2\,(2M_W^2 -  m_i^2) - m_i^2\,(2M_W^2 - q^2))\,C_2\nonumber\\
    &\phantom{=}& + (m_i^2\,(2M_W^2 - q^2) + 2(D-2)\,M_W^4)C_{11} - m_\beta^2\,(2M_W^2 - q^2)\,C_{22}\nonumber\\
    &\phantom{=}& - (m_i^2\,(2M_W^2 - q^2) - m_\beta^2\,(2M_W^2 - q^2) + 2(D-2)\,M_W^4)\,C_{12}\Bigg\}\,,\label{eqn:FSLbU}\\
    F_S^{R\:(b)} &=& -\frac{g_w^3 c_w}{64\pi^2\, M_W^4}m_\beta \sum_i\,\mathcal U_{\alpha i}\, \mathcal U_{\beta i}^\ast\,\Bigg\{(m_i^2 - 2M_W^2) \,B_0^\alpha + m_\alpha^2\, B_1^\alpha - m_i^2\,(B_0^\beta + B_1^\beta)\nonumber\\
    &\phantom{=}& + A_0 - q^2\,B_0^q + (2 M_W^2 - q^2)\,(3B_1^q + 2B_{11}^q) + 2M_W^2\,(m_\beta^2 - m_i^2 - M_W^2)\,C_0\nonumber\\
    &\phantom{=}& + (2M_W^2\,(2M_W^2 - q^2) - m_i^2\,(2M_W^2 - q^2))\,C_1 + (2M_W^2\,(m_\beta^2 - m_\alpha^2 + 2M_W^2) + m_i^2\,(2M_W^2 - q^2)\,C_2\nonumber\\
    &\phantom{=}& - m_\alpha^2\,(2M_W^2 - q^2)\,C_{11} + (m_i^2\,(2M_W^2 - q^2) + 2(D-2)\,M_W^4)\,C_{22}\nonumber\\
    &\phantom{=}& - (m_i^2\,(2M_W^2 - q^2) - m_\alpha^2\,(2M_W^2 - q^2) + 2(D-2)\,M_W^4)\,C_{12}\Bigg\}\,,\label{eqn:FSRbU}
\end{eqnarray}
with the Passarino-Veltman functions as defined in Eqs.~(\ref{eqn:FVLbU}--\ref{eqn:FVRcdU}).
Note that the ``scalar'' amplitudes are neither finite nor gauge invariant on their own; UV-divergences and gauge-dependent terms are however cancelled once the corresponding box-diagrams (that contribute to the same process as the off-shell $Z$-penguin) are taken into account.

\mathversion{bold}
\subsection{Amplitudes for cLFV $Z$-boson decays in the Feynman-'t Hooft gauge}
\mathversion{normal}
For completeness, we further include the amplitudes for the cLFV $Z$ decays in the Feynman-'t Hooft gauge; the labels (a)-(d) refer to the topologies presented in Fig.~\ref{fig:cLFVZdecays:UG}, which now also include additional diagrams with Goldstone bosons.
\begin{eqnarray}
F_V^{L\: (a)} &=&
- \frac{g_w^3 }{64 \pi^2 \,c_w \,M_W^2}\sum_{i,j}\mathcal{U}_{\alpha i} \,\mathcal{U}_{\beta j}^*
\left\{C_{ij} 
\left[2  M_W^2 \,(B_0^\alpha +B_0^\beta) + (D-6)\, M_W^2\, B_0^q \right.\right.\nonumber\\
&&
+\left[(D-6)\, M_W^4   -m_\alpha^2 \,m_\beta^2 +m_\beta^2\, m_i^2 +m_\alpha^2 \,m_j^2 -m_i^2 \,m_j^2 +\right. \nonumber \\
&& \quad \quad \left. +
(2 m_i^2 -2 m_\alpha^2 - 2 m_\beta^2 +2 m_j^2 -2 q^2 )\, M_W^2 \right]\,C_{0} 
\nonumber\\
&&
+\left[-2 m_\alpha^2\, m_\beta^2 +m_\alpha^2 \,m_j^2 +m_\beta^2 \,m_i^2 -6 m_\alpha^2 \,M_W^2-2 m_\beta^2 \, M_W^2    \right]\,C_1
\nonumber\\
&&
+\left[-2 m_\alpha^2 \,m_\beta^2 +m_\alpha^2 \,m_j^2  +m_\beta^2 \,m_i^2 -2 m_\alpha^2  \,M_W^2-6 m_\beta^2 \, M_W^2 \right]\,C_2
\nonumber\\
&&
+\left[-m_\alpha^2 \,m_\beta^2 -(D-2) \,m_\alpha^2 \, M_W^2   \right]\,C_{11}
\nonumber\\
&&
+\left[-m_\alpha^2 \,m_\beta^2 -(D-2)\, m_\beta^2 \, M_W^2  \right]\,C_{22}
\nonumber\\
&&
+\left[-2 m_\alpha^2 \,m_\beta^2  -(D-2)\, m_\alpha^2 \, M_W^2 -(D-2) \,m_\beta^2 \, M_W^2   \right]\,C_{12}
\nonumber\\
&&\left.-2(D-2)\,M_W^2\, C_{00} \right]
\nonumber\\
&& \left.+ C_{ij}^* \,m_i \,m_j \left[- B_0^q +  (D-3) \,M_W^2\, C_{0}
+ m_\alpha^2 \, C_{11} +m_\beta^2\, C_{22} \right. \right. \nonumber \\ 
&& \quad \left. \left.
+ (m_\alpha^2  + m_\beta^2) \,C_{12} + 2 C_{00}\right]
\right\}\,, \label{eqn:FVLaF}
\\
F_V^{R\: (a)} &=& \frac{g_w^3 m_\alpha \,m_\beta }{64 \pi^2 \,c_w \,M_W^2} \sum_{i,j}\mathcal{U}_{\alpha i} \,\mathcal{U}_{\beta j}^*
\left\{C_{ij} \left[- B_0^q + (D-3)\, M_W^2 \,C_{0} + 2 (D-2) \,M_W^2 \left[C_{1} + C_2\right] \right.\right.
\nonumber\\
&& 
+ \left[m_\alpha^2 + (D-2)\, M_W^2 \right]\,C_{11}
+\left[m_\beta^2 + (D-2) \,M_W^2 \right] \,C_{22} 
\nonumber\\
&& 
+\left. \left[m_\alpha^2 + m_\beta^2  + 2(D-2)\,M_W^2 \right] \,C_{12}
+ 2 C_{00} \right]
\nonumber\\
&& 
- \left. C_{ij}^*\, m_i\, m_j \left[C_{11} + C_{22}+ 2 C_{12}\right] \right\}\,,\label{eqn:FVRaF}
\\
F_T^{L\: (a)} &=& -\frac{i g_w^3 m_\alpha }{64 \pi^2 \,c_w \,M_W^2} \sum_{i,j}\mathcal{U}_{\alpha i} \,\mathcal{U}_{\beta j}^*
\left\{C_{ij} \left[2 M_W^2\, C_{0} + D \,M_W^2 \,C_{1} +\left[m_\beta^2-m_j^2+2 M_W^2\right]\, C_{2} \right.\right. \nonumber \\
&&
+\left.  (D-2)\, M_W^2 \,C_{11} + m_\beta^2 \,C_{22} + \left[m_\beta^2 + (D-2 )\, M_W^2 \right]\,C_{12} \right]
\nonumber \\
&&
- \left. C_{ij}^* \,m_i \,m_j \left[C_{11} + C_{12}\right] \right\}\,,\label{eqn:FTLaF}
\\
F_T^{R\: (a)} &=& -\frac{i g_w^3 m_\beta }{64 \pi^2 \,c_w \,M_W^2} \sum_{i,j}\mathcal{U}_{\alpha i} \,\mathcal{U}_{\beta j}^*
\left\{C_{ij} \left[ 2 M_W^2 \,C_{0} +\left[m_\alpha^2 - m_i^2 +2 M_W^2  \right]\,C_{1} + D \,M_W^2\, C_{2}\right.\right.
\nonumber \\
&&
+\left. m_\alpha^2 \,C_{11} + (D-2)\, M_W^2\, C_{22} + \left[ m_\alpha^2 + (D-2)\, M_W^2 \right]\,C_{12} \right]
\nonumber \\
&&
-\left.C_{ij}^*\, m_i \,m_j \left[C_{22} + C_{12}\right] \right\} \,,\label{eqn:FTRaF}
\end{eqnarray}
with $C_{rs}=C_{rs}\left(m_\alpha^2,q^2,m_\beta^2,M_W^2,m_i^2,m_j^2\right)$, $B_0^q=B_0\left(q^2,m_i^2,m_j^2\right)$, $B_0^{\alpha} = B_0\left(m_{\alpha}^2,M_W^2,m_i^2\right)$ and $B_0^{\beta} = B_0\left(m_{\beta}^2,M_W^2,m_j^2\right)$.
\begin{eqnarray}
F_V^{L\: (b)} &=&
\frac{g_w^3 }{64 \pi^2 \,c_w \,M_W^2} \sum_i \mathcal{U}_{\alpha i}\, \mathcal{U}_{\beta i}^* 
\Bigg\{ 2 c_w^2 \,M_W^2 \left[B_0^\alpha  + B_0^\beta\right] 
\nonumber \\
&\phantom{=}&
+ \left[4 c_w^2 \,M_W^4 -2 c_w^2 \,M_W^2 (m_\alpha^2 + m_\beta^2)+  m_i^2 \,(c_w^2-s_w^2) \,(m_\alpha^2 + m_\beta^2) +4 c_w \,m_i^2 \,M_Z\, s_w^2 \, M_W\right]\,C_{0}
\nonumber \\
&\phantom{=}&
+ \left[(c_w^2- s_w^2)\, m_\alpha^2\, m_\beta^2 +2 (D-5)\, c_w^2\, m_\alpha^2 \, M_W^2  
+2 c_w^2 \,m_\beta^2 \, M_W^2 
- 2 c_w^2\, q^2 \, M_W^2 
\right.\nonumber \\
&\phantom{=}&\left.
+ 2 c_w \,M_Z \,s_w^2 \, M_W \,(m_\alpha^2 + m_\beta^2) 
+ 2 (c_w^2 - s_w^2) \,m_\alpha^2 \,m_i^2  
+ (c_w^2 - s_w^2 ) \,m_\beta^2\, m_i^2  \right]\,C_{1}
\nonumber \\
&\phantom{=}&+ \left[(c_w^2 - s_w^2) \,m_\alpha^2 \,m_\beta^2 + 2(D-5) \,c_w^2 \,m_\beta^2\,  M_W^2 +2 c_w^2\, m_\alpha^2 \, M_W^2 -2 c_w^2\, q^2 \, M_W^2 
\right.\nonumber \\
&\phantom{=}&\left.
+2 c_w \,M_Z\, s_w^2 \, M_W \,(m_\alpha^2 + m_\beta^2)
+2 (c_w^2 - s_w^2 )\,m_\beta^2 \,m_i^2 
+ (c_w^2-s_w^2) \, m_\alpha^2 \,m_i^2 \right]\,C_{2}
\nonumber \\
&\phantom{=}&+ \left[(c_w^2 - s_w^2)\, m_\alpha^2\, m_\beta^2 +2 (D-2)\, c_w^2 \,m_\alpha^2 \, M_W^2  +(c_w^2  - s_w^2 )\, m_\alpha^2\, m_i^2 \right]\,C_{11}
\nonumber \\
&\phantom{=}&
+ \left[(c_w^2 - s_w^2)\, m_\alpha^2 \,m_\beta^2 + 2 (D-2) \,c_w^2\, m_\beta^2  \,M_W^2 +(c_w^2 - s_w^2)\, m_\beta^2\, m_i^2 \right]\,C_{22}
\nonumber \\
&\phantom{=}&
+ \left[2 (c_w^2 - s_w^2)\, m_\alpha^2 \,m_\beta^2 +2 (D-2)\, c_w^2\, m_\alpha^2  \,M_W^2 +2 (D-2) \,c_w^2 \,m_\beta^2 \,M_W^2
+
\right. \nonumber \\
&& \quad \quad \left. +
(c_w^2 - s_w^2) \,(m_\alpha^2 + m_\beta^2)\, m_i^2 \right]\,C_{12}
\nonumber \\
&\phantom{=}&
+ \left[ 4 (D-2) \,c_w^2\,  M_W^2 + 2 (c_w^2 - s_w^2)\, m_i^2  
\right]\,C_{00}
\Bigg\}\,, \label{eqn:FVLbF}
\\
F_V^{R\: (b)} &=&
\frac{g_w^3 m_\alpha m_\beta}{64 \pi^2 \,c_w \,M_W^2} \sum_i \mathcal{U}_{\alpha i} \,\mathcal{U}_{\beta i}^* 
\Bigg\{2 (c_w^2-s_w^2)\, m_i^2 \, C_{0}
\nonumber \\
&\phantom{=}&
+\left[(c_w^2 - s_w^2) \,(3 m_i^2  +  m_\alpha^2) + 2(D-2) \,c_w^2 \, M_W^2 \right] \,C_{1}
\nonumber \\
&\phantom{=}&
+\left[(c_w^2 - s_w^2) \,(3 m_i^2  +  m_\beta^2) + 2(D-2) \,c_w^2 \, M_W^2 \right] \,C_{2}
\nonumber \\
&\phantom{=}&
+\left[(c_w^2 - s_w^2) \,(m_i^2  +  m_\alpha^2) +2 (D-2) \,c_w^2\, M_W^2 \right]\, C_{11}
\nonumber \\
&\phantom{=}&
+\left[(c_w^2 - s_w^2)\, (m_i^2  +  m_\beta^2) +2 (D-2) \,c_w^2\, M_W^2 \right] \,C_{22}
\nonumber \\
&\phantom{=}&
+\left[(c_w^2 - s_w^2) \,(2 m_i^2 +  m_\alpha^2 +  m_\beta^2) +4 (D-2)\,c_w^2 \, M_W^2 \right]\, C_{12}
\nonumber \\
&\phantom{=}&
+2 (c_w^2 - s_w^2) \,C_{00}
\Bigg\}
\,, \label{eqn:FVRbF}
\\
F_T^{L\: (b)} &=&
-\frac{i g_w^3 m_\alpha }{64 \pi^2 \,c_w\, M_W^2} \sum_i \mathcal{U}_{\alpha i} \mathcal{U}_{\beta i}^* \,
\Bigg\{ (c_w^2-s_w^2) \,m_i^2  \,C_{0}
\nonumber \\
&\phantom{=}&
+2\left[ (c_w^2-s_w^2) \,m_i^2 + (D-5)\, c_w^2 \, M_W^2 \right]\,C_{1}
\nonumber \\
&\phantom{=}&
+\left[(c_w^2-s_w^2) \,(m_\beta^2+m_i^2) + 2 c_w \,M_W \,M_Z\, s_w^2\right] C_{2}
\nonumber \\
&\phantom{=}&
+\left[(c_w^2-s_w^2) \,m_i^2 + 2(D-2)\, c_w^2  \,M_W^2  \right]\,C_{11}
+ (c_w^2-s_w^2)\, m_\beta^2\, C_{22} 
\nonumber \\
&\phantom{=}&
+\left[(c_w^2-s_w^2)\,(m_\beta^2 + m_i^2 ) +2 (D-2) \,c_w^2 \, M_W^2 \right]\,C_{12}
\Bigg\}
\,, \label{eqn:FTLbF}
\\
F_T^{R\: (b)} &=&
-\frac{i g_w^3 m_\beta \sum_i \mathcal{U}_{\alpha i} \,\mathcal{U}_{\beta i}^* }{64 \pi^2 \,c_w \,M_W^2} 
\Bigg\{(c_w^2-s_w^2) \,m_i^2 \,C_{0}
\nonumber \\
&\phantom{=}&
+ \left[(c_w^2-s_w^2) \,(m_\alpha^2 + m_i^2 ) +2 c_w\, M_W\, M_Z \,s_w^2 \right] \,C_{1}
\nonumber \\
&\phantom{=}&
+ 2 \left[(c_w^2-s_w^2) \,m_i^2 + (D-5)\, c_w^2 \, M_W^2 \right] \,C_{2}
\nonumber \\
&\phantom{=}&
+(c_w^2 - s_w^2) \,m_\alpha^2 \,C_{11} + \left[(c_w^2 - s_w^2) \,m_i^2 +2 (D-2)\, c_w^2 \, M_W^2 \right] \,C_{22}
\nonumber \\
&\phantom{=}&
+ \left[(c_w^2-s_w^2) \,(m_\alpha^2 + m_i^2 )  +2 (D-2) \,c_w^2\, M_W^2 
\right] \,C_{12}
\Bigg\}
\,, \label{eqn:FTRbF}
\end{eqnarray}
in which $C_{rs}= C_{rs}\left(m_\alpha^2,q^2,m_\beta^2,m_i^2,M_W^2,M_W^2\right)$
$B_0^{\alpha,\beta} = B_0\left(m_\alpha^2,m_i^2,M_W^2\right)$.
\begin{eqnarray}
F_V^{L\: (c+d)} &=&
-\frac{g_w^3 ({\bf C}_{A}+{\bf C}_{V})}{32 \pi^2 \,c_w \,M_W^2 \left(m_\alpha^2-m_\beta^2\right)} \sum_i \mathcal{U}_{\alpha i}\, \mathcal{U}_{\beta i}^*   
 \nonumber \\
&\phantom{=}&
\Bigg\{
\left[(D-2)\,M_W^2\, (M_W^2-m_\alpha^2)+m_\beta^2 \,m_i^2 -m_i^4- (D-3)\,m_i^2 \,M_W^2\right]\,B_0^\alpha
 \nonumber \\
&\phantom{=}&
-\left[(D-2)\,M_W^2\,(M_W^2-m_\beta^2) +m_\alpha^2 \,m_i^2 - m_i^4 - (D-3)\,m_i^2\,M_W^2\right]\,B_0^\beta 
 \nonumber \\
&\phantom{=}&
+\left[m_\alpha^2 \,(m_\beta^2 - m_i^2 - (D-2) \,M_W^2)\right]\,B_1^\alpha
 \nonumber \\
&\phantom{=}&
-\left[ m_\beta^2\,(m_\alpha^2 - m_i^2 - (D-2)\, M_W^2 )\right]\,B_1^\beta
\Bigg\}
\,, \label{eqn:FVLcdF}
\\
F_V^{R\: (c+d)} &=&
-\frac{g_w^3 ({\bf C}_{A}-{\bf C}_{V}) m_\alpha \,m_\beta }{32 \pi^2 \,c_w \,M_W^2 \left(m_\alpha^2-m_\beta^2\right)} \sum_i \mathcal{U}_{\alpha i} \,\mathcal{U}_{\beta i}^*   
\Bigg\{\left[m_\alpha^2 - m_i^2 - M_W^2\right] \,B_0^\alpha
 \nonumber \\
&\phantom{=}&
-\left[m_\beta^2 - m_i^2 - M_W^2\right] \,B_0^\beta
 \nonumber \\
&\phantom{=}&
+ \left[m_\alpha^2  -m_i^2 - (D-2) \,M_W^2 \right]\, B_1^\alpha
 \nonumber \\
&\phantom{=}&
-\left[m_\beta^2 - m_i^2 - (D-2) \,M_W^2\right] \,B_1^\beta
\Bigg\}
\,, \label{eqn:FVRcdF}
\\
F_T^{L\: (c+d)} &=& 0\,, \label{eqn:FTLcdF}
\\
F_T^{R\: (c+d)} &=& 0 \,, \label{eqn:FTRcdF}
\end{eqnarray}
where $B_{0,1}^{\alpha, \beta}=B_{0,1}\left(m_{\alpha, \beta}^2,m_i^2,M_W^2\right)$.

\medskip
We can now compare the analytic expressions in both unitary (UG) and Feynman-'t Hooft (FG) gauges.
After a substantial amount of algebra one can show that all differences in $F_V^R, F_T^L$ and $F_T^R$ between the two gauges (diagram by diagram) are proportional to a common factor of $(D-4)$ and are thus trivially $0$ in $D=4$ dimensions.
For $F_V^L$ the situation is slightly more involved due to a permutation of arguments in the two-point Passarino-Veltman functions.
Neglecting terms independent of the internal fermion masses, and setting $q^2 = M_Z^2\,, c_w^2 = M_W^2/M_Z^2$, the total difference is given by
\begin{eqnarray}
    F_V^{L\,,\mathrm{UG}} - F_V^{L\,,\mathrm{FG}} &=& -\dfrac{g_w^3 M_Z}{64\pi^2 M_W^3}\sum_{i,j}\,\mathcal U_{\alpha i}\,\mathcal U_{\beta j}^*\,C_{ij}\Big[(m_\alpha^2 - m_i^2)\,B_0(m_\alpha^2, M_W^2, m_i^2)\nonumber\\
    &\phantom{=}& + (m_\beta^2 - m_j^2)B_0(m_\beta^2, M_W^2, m_j^2) + m_\alpha^2\, B_1(m_\alpha^2, M_W^2, m_i^2) + m_\beta^2\, B_1(m_\beta^2, M_W^2, m_j^2)\Big]\nonumber\\
    &\phantom{=}& -\dfrac{g_w^3 M_Z}{64\pi^2 M_W^3} \sum_i\,\mathcal U_{\alpha i}\,\mathcal U_{\beta i}^*\Big[m_i^2\left(B_0(m_\alpha^2, m_i^2, M_W^2) + B_0(m_\beta^2, m_i^2, M_W^2)\right)\nonumber\\
   &\phantom{=}& + m_\alpha^2\, B_1(m_\alpha^2, m_i^2, M_W^2) + m_\beta^2\, B_1(m_\beta^2, m_i^2, M_W^2)\Big]\,,
\end{eqnarray}
which can be shown to vanish relying on the identities in Eqs.~(\ref{eqn:iden:start}--\ref{eqn:iden:end}), and by using the following property of the two-point functions
\begin{equation}
\label{eq:PVBproperty}
    B_0(p^2, m_0^2, m_1^2) + B_1(p^2, m_0^2, m_1^2) + B_1(p^2, m_1^2, m_0^2) = 0\,.
\end{equation}

\subsection{Divergences in cLFV Higgs decays}
As done for the $Z$-boson decays, we now briefly discuss divergences emerging in association with the cLFV Higgs decays diagrams:
\begin{eqnarray}
    \mathrm{div}(F_L^{(a)}) &=& - \frac{g_w^3 m_\alpha }{128 \pi^2\, M_W^3}\sum_{i,j} \,\mathcal{U}_{\alpha i}\, \mathcal{U}_{\beta j}^* \left[ C_{ij} \left(m_i^2 + 2 m_j^2\right) + C_{ij}^*\, 3m_i \,m_j \right]\Delta_\varepsilon \,,
    \\
    \mathrm{div}(F_L^{(b)}) &=& - \frac{g_w^3 m_\alpha }{128 \pi^2\, M_W^3}\sum_{i} \,\mathcal{U}_{\alpha i}\, \mathcal{U}_{\beta i}^* \left[ m_\beta^2 - 3 m_i^2 + q^2 \right]\Delta_\varepsilon \,,
    \\
    \mathrm{div}(F_L^{(c+d)}) &=& \frac{g_w^3 m_\alpha }{128 \pi^2\, M_W^3} \,m_\beta^2\sum_{i}\, \mathcal{U}_{\alpha i}\, \mathcal{U}_{\beta i}^* \, \Delta_\varepsilon \,,
    \\
    \mathrm{div}(F_R^{(a)}) &=& - \frac{g_w^3 m_\beta }{128 \pi^2\, M_W^3}\sum_{i,j} \,\mathcal{U}_{\alpha i}\, \mathcal{U}_{\beta j}^* \left[ C_{ij} \left(2 m_i^2 + m_j^2\right) + C_{ij}^*\, 3m_i \,m_j \right]\Delta_\varepsilon \,,
    \\
    \mathrm{div}(F_R^{(b)}) &=& - \frac{g_w^3 m_\beta }{128 \pi^2\, M_W^3}\sum_{i} \,\mathcal{U}_{\alpha i}\, \mathcal{U}_{\beta i}^* \left[ m_\alpha^2 - 3 m_i^2 + q^2 \right]\Delta_\varepsilon \,,
    \\
    \mathrm{div}(F_R^{(c+d)}) &=& \frac{g_w^3  m_\beta}{128 \pi^2\, M_W^3} \,m_\alpha^2 \sum_{i}\, \mathcal{U}_{\alpha i}\, \mathcal{U}_{\beta i}^* \, \Delta_\varepsilon \,.
\end{eqnarray}
In a similar fashion to what was explained for the cancellation of the divergences in the $Z$-boson decay, using the identities in Eqs.~(\ref{eqn:iden:start}--\ref{eqn:iden:end}), it can be easily shown that the sum of diagrams (a)--(d) from Fig.~\ref{fig:cLFVHiggsdecays:UG} is finite.

\subsection{Amplitudes for cLFV Higgs decays in the Feynman-'t Hooft gauge}
As done before, and for completeness, below we collect the results for the cLFV Higgs decays in the Feynman-'t Hooft gauge (the labels (a)--(d) refer to the topologies presented in Fig.~\ref{fig:cLFVHiggsdecays:UG}, which now also include additional diagrams with Goldstone bosons): 
\begin{eqnarray}
F^{(a)}_L &=& \frac{g_w^3 m_\alpha }{64 \pi^2 \,M_W^3} \sum_{i,j}\mathcal{U}_{\alpha i}\, \mathcal{U}_{\beta j}^*
\left\{ C_{ij} \left[ - m_j^2 \, B_0^q + (D-3)\, m_j^2\, M_W^2\,C_0 \right.\right. \nonumber \\
&\phantom{=}& \left.- \left[m_\alpha^2 \,m_j^2 + m_i^2 \left(m_\beta^2-2 m_j^2\right)-(D-2)\, M_W^2 \left(m_i^2+m_j^2\right) \right]\,C_1 \right] \nonumber \\
&\phantom{=}& 
+ \, C_{ij}^* \,m_i \,m_j  \left[- B_0^q  +  (D-3) \,M_W^2\, C_0 \right. \nonumber\\
&\phantom{=}& \left. \left. 
-  \left[m_\alpha^2+m_\beta^2-m_i^2-m_j^2-2 (D-2)\, M_W^2\right]\,C_1
\right] \right\} \,,
\end{eqnarray}
\begin{eqnarray}
F^{(a)}_R &=& \frac{g_w^3 m_\beta }{64 \pi^2\, M_W^3} \sum_{i,j}\mathcal{U}_{\alpha i}\,\mathcal{U}_{\beta j}^*
\left\{C_{ij} \left[ -m_i^2\, B_0^q + (D-3)\, m_i^2 \,M_W^2 \,C_0 \right.\right. \nonumber\\
&\phantom{=}& \left.
- \left[m_\alpha^2 \,m_j^2+m_i^2 \left(m_\beta^2-2 m_j^2\right)-(D-2) \,M_W^2 \left(m_i^2+m_j^2\right) \right]\,C_2
\right]\nonumber\\
&\phantom{=}&
+ \, C_{ij}^*\, m_i \,m_j \left[ - B_0^q + (D-3)\, M_W^2  \,C_0 \right.\nonumber\\
&\phantom{=}&
\left.\left. - \left[m_\alpha^2+m_\beta^2-m_i^2-m_j^2-2 (D-2)\, M_W^2\right]\,C_2  \right] \right\} \,,
\end{eqnarray}
with $C_{rs} =C_{rs}\left(m_\alpha^2,q^2,m_\beta^2,M_W^2,m_i^2,m_j^2\right)$ and $B_0^q = B_0\left(q^2,m_i^2,m_j^2\right)$.
\begin{eqnarray}
F^{(b)}_L &=& -\frac{g_w^3 m_\alpha }{64 \pi^2\, M_W^3}\sum_{i}\mathcal{U}_{\alpha i} \,\mathcal{U}_{\beta i}^*
\left\{ M_W^2 \,( 2 B_0^\beta - B^q_0 ) \right. \nonumber\\
&\phantom{=}& + 
\left[ m_H^2 \,m_i^2 - 2 M_W^2 \,(m_\alpha^2 - M_W^2)\right]\,C_0\nonumber\\
&\phantom{=}& +
\left[ m_H^2 \,m_i^2 - 2 M_W^2\, \left(m_\alpha^2+m_\beta^2-m_i^2-(D-2) \,M_W^2\right) \right]\,C_1\nonumber\\
&\phantom{=}& +\left.
\left[ m_H^2 \,m_\beta^2 - 2 M_W^2 \,m_\beta^2 \right]\,C_2
\right\}\,,
\end{eqnarray}
\begin{eqnarray}
F^{(b)}_R &=& -\frac{g_w^3 m_\beta  }{64 \pi^2\, M_W^3}\sum_{i}\mathcal{U}_{\alpha i}\, \mathcal{U}_{\beta i}^* \left\{ M_W^2 \left(2 B_0^\alpha - B^q_0\right)\right.\nonumber\\
&\phantom{=}& +
\left[ m_H^2 \,m_i^2 - 2 M_W^2 \,(m_\beta^2 - M_W^2)\right]\,C_0\nonumber\\
&\phantom{=}& +
\left[ m_H^2 \,m_\alpha^2 - 2 M_W^2 \,m_\alpha^2 \right]\,C_1\nonumber\\
&\phantom{=}& +
\left.\left[ m_H^2 \,m_i^2 - 2 M_W^2 \left(m_\alpha^2+m_\beta^2-m_i^2-(D-2)\, M_W^2\right) \right]\,C_2
\right\}\,,
\end{eqnarray}
where $C_{rs} = C_{rs}\left(m_\alpha^2,q^2,m_\beta^2,m_i^2,M_W^2,M_W^2\right)$,
$B^q_0 = B_0\left(q^2,M_W^2,M_W^2\right) $ and $B_0^{\alpha, \beta} = B_0\left(m_{\alpha, \beta}^2,m_i^2,M_W^2\right)$.
\begin{eqnarray}
F^{(c+d)}_L &=& \frac{g_w^3 m_\alpha }{64 \pi^2 \,M_W^3 (m_\alpha^2-m_\beta^2)}\sum_{i}\mathcal{U}_{\alpha i}\, \mathcal{U}_{\beta i}^*
\left\{m_\beta^2 \left[m_\alpha^2  - m_i^2  -M_W^2 \right]\,B_0^{\alpha} \right. 
\nonumber \\
&& \quad \quad 
+ m_\beta^2 \left[m_\alpha^2  - m_i^2 -(D-2) \,M_W^2  \right]\,B_1^{\alpha}
\nonumber \\
&\phantom{=}&
+ \left[m_\beta^2 -m_i^2 -(D-2) \,M_W^2\right]\left[A_0(M_W^2) - A_0(m_i^2)\right]
\nonumber \\
&\phantom{=}&
+ \left[m_\alpha^2\, m_i^2 - m_i^4  -(D-2)\, m_\beta^2 \,M_W^2 -(D-3) \,m_i^2 \,M_W^2 +(D-2)\, M_W^4 \right]\,B_0^{\beta}
\nonumber \\
&\phantom{=}&
\left. +  \,m_\beta^2 \left[m_\alpha^2-m_i^2-(D-2) \,M_W^2\right] \,B_1^{\beta}
\right\}\,,
\end{eqnarray}
\begin{eqnarray}
F^{(c+d)}_R &=& \frac{g_w^3 m_\beta  }{64 \pi^2 \,M_W^3 (m_\alpha^2-m_\beta^2)}\sum_{i}\mathcal{U}_{\alpha i}\, \mathcal{U}_{\beta i}^*
\left\{- m_\alpha^2\left[m_\beta^2-m_i^2-M_W^2\right] \,B_0^{\beta} -\right. 
\nonumber \\
&& \quad \quad 
- m_\alpha^2 \left[m_\beta^2-m_i^2-(D-2)\, M_W^2\right] \,B_1^{\beta} \nonumber \\
&\phantom{=}&
-\left[m_\alpha^2-m_i^2-(D-2) \,M_W^2\right]\left[  A_0(M_W^2)-A_0(m_i^2)\right]
\nonumber \\
&\phantom{=}&
- \left[m_\beta^2 \,m_i^2-m_i^4 -(D-2)  \,m_\alpha^2\, M_W^2 - (D-3) \,m_i^2 \,M_W^2 +(D-2) \,M_W^4\right] \,B_0^{\alpha}
\nonumber \\
&\phantom{=}&
\left. - \, m_\alpha^2 \left[m_\beta^2-m_i^2-(D-2) \,M_W^2\right] \,B_1^{\alpha}
\right\}\,,
\end{eqnarray}
with $B_{0,1}^{\alpha,\beta} = B_{0,1}\left(m_{\alpha,\beta}^2,m_i^2,M_W^2\right)$.

Comparing the analytic expressions in unitary and Feynman-'t Hooft gauges, one has, after setting $q^2 = m_h^2$ and neglecting terms independent of the internal neutrino masses,
\begin{eqnarray}
\Delta F_L^{\mathrm{UG}} - \Delta F_L^{\mathrm{FG}} &=&
\frac{g^3 m_\alpha }{64 \pi^2 m_W^3} \,\sum_i \, \mathcal{U}_{\alpha i}\, \mathcal{U}_{\beta i}^*\,
m_i^2   \left[B_0\left(m_\alpha^2,m_i^2,m_W^2\right) + B_1\left(m_\alpha^2,m_i^2,m_W^2\right)\right]
\nonumber\\
&\phantom{=}&
+ \frac{g^3 m_\alpha }{64 \pi^2 m_W^3}\, \sum_{i,j} \, \mathcal{U}_{\alpha i} \,\mathcal{U}_{\beta j}^* \, m_i  (C_{ij} m_i+C_{ij}^* m_j) B_1\left(m_\alpha^2,m_W^2,m_i^2\right)
\,,
\\
\Delta F_R^{\mathrm{UG}} - \Delta F_R^{\mathrm{FG}}
&=&
\frac{g^3 m_\beta }{64 \pi^2 m_W^3} \, \sum_i \, \mathcal{U}_{\alpha i} \, \mathcal{U}_{\beta i}^*\,
m_i^2   \left[B_{0}\left(m_\beta^2,m_i,m_W\right)+ B_{1}\left(m_\beta^2,m_i,m_W\right)\right]
\nonumber\\
&\phantom{=}&
+ \frac{g^3 m_\beta }{64 \pi^2 m_W^3}\, \sum_{i,j} \, \mathcal{U}_{\alpha i} \, \mathcal{U}_{\beta j}^*\, 
m_j (C_{ij} m_j+C_{ij}^* m_i) B_{1}\left(m_\beta^2,m_W,m_j\right)
\, ,
\end{eqnarray}
which can be shown to vanish using the identities in Eqs.~(\ref{eqn:iden:start}--\ref{eqn:iden:end}, \ref{eq:PVBproperty}).

\section{Leptonic cLFV observables}\label{app:cLFVobservables}
In this Appendix we collect several expressions concerning the decay widths of numerous cLFV leptonic decays and rare transitions, which are relevant for our numerical studies (constraints and implications for probing the model). Likewise, in 
Table~\ref{tab:cLFV_lep}, we summarise the bounds and
future sensitivities for these cLFV processes. 
\renewcommand{\arraystretch}{1.3}
\begin{table}[h!]
    \centering
    \hspace*{-2mm}{\small\begin{tabular}{|c|c|c|}
    \hline
    Observable & Current bound & Future sensitivity  \\
    \hline\hline
    $\text{BR}(\mu\to e \gamma)$    &
    \quad $<4.2\times 10^{-13}$ \quad (MEG~\cite{TheMEG:2016wtm})   &
    \quad $6\times 10^{-14}$ \quad (MEG II~\cite{Baldini:2018nnn}) \\
    $\text{BR}(\tau \to e \gamma)$  &
    \quad $<3.3\times 10^{-8}$ \quad (BaBar~\cite{Aubert:2009ag})    &
    \quad $3\times10^{-9}$ \quad (Belle II~\cite{Kou:2018nap})      \\
    $\text{BR}(\tau \to \mu \gamma)$    &
     \quad $ <4.4\times 10^{-8}$ \quad (BaBar~\cite{Aubert:2009ag})  &
    \quad $10^{-9}$ \quad (Belle II~\cite{Kou:2018nap})     \\
    \hline
    $\text{BR}(\mu \to 3 e)$    &
     \quad $<1.0\times 10^{-12}$ \quad (SINDRUM~\cite{Bellgardt:1987du})    &
     \quad $10^{-15(-16)}$ \quad (Mu3e~\cite{Blondel:2013ia})   \\
    $\text{BR}(\tau \to 3 e)$   &
    \quad $<2.7\times 10^{-8}$ \quad (Belle~\cite{Hayasaka:2010np})&
    \quad $5\times10^{-10}$ \quad (Belle II~\cite{Kou:2018nap})     \\
    $\text{BR}(\tau \to 3 \mu )$    &
    \quad $<3.3\times 10^{-8}$ \quad (Belle~\cite{Hayasaka:2010np})  &
    \quad $5\times10^{-10}$ \quad (Belle II~\cite{Kou:2018nap})     \\
    & & \quad$5\times 10^{-11}$\quad (FCC-ee~\cite{Abada:2019lih})\\
        $\text{BR}(\tau^- \to e^-\mu^+\mu^-)$   &
    \quad $<2.7\times 10^{-8}$ \quad (Belle~\cite{Hayasaka:2010np})&
    \quad $5\times10^{-10}$ \quad (Belle II~\cite{Kou:2018nap})     \\
    $\text{BR}(\tau^- \to \mu^-e^+e^-)$ &
    \quad $<1.8\times 10^{-8}$ \quad (Belle~\cite{Hayasaka:2010np})&
    \quad $5\times10^{-10}$ \quad (Belle II~\cite{Kou:2018nap})     \\
    $\text{BR}(\tau^- \to e^-\mu^+e^-)$ &
    \quad $<1.5\times 10^{-8}$ \quad (Belle~\cite{Hayasaka:2010np})&
    \quad $3\times10^{-10}$ \quad (Belle II~\cite{Kou:2018nap})     \\
    $\text{BR}(\tau^- \to \mu^-e^+\mu^-)$   &
    \quad $<1.7\times 10^{-8}$ \quad (Belle~\cite{Hayasaka:2010np})&
    \quad $4\times10^{-10}$ \quad (Belle II~\cite{Kou:2018nap})     \\
    \hline
    $\text{CR}(\mu- e, \text{N})$ &
     \quad $<7 \times 10^{-13}$ \quad  (Au, SINDRUM~\cite{Bertl:2006up}) &
    \quad $10^{-14}$  \quad (SiC, DeeMe~\cite{Nguyen:2015vkk})    \\
    & &  \quad $2.6\times 10^{-17}$  \quad (Al, COMET~\cite{Krikler:2015msn,Adamov:2018vin,KunoESPP19})  \\
    & &  \quad $8 \times 10^{-17}$  \quad (Al, Mu2e~\cite{Bartoszek:2014mya})\\
    \hline
    \end{tabular}}
    \caption{Current experimental bounds and future sensitivities on cLFV observables here considered. The quoted limits are given at $90\%\:\mathrm{C.L.}$ (Belle II sensitivities correspond to an integrated luminosity of $50\:\mathrm{ab}^{-1}$).}
    \label{tab:cLFV_lep}
\end{table}
\renewcommand{\arraystretch}{1.}

\subsection{Radiative and three-body decays}
Below we summarise the expressions for the radiative and $\ell_\beta \to 3 \ell_\alpha$ decays (the full expression for the most general case of the 3-body decay
can be found in~\cite{Ilakovac:1994kj}), closely following the notation of~\cite{Abada:2021zcm}. 
The rates for the radiative and three-body decays in the SM extended via
$n_S$ heavy sterile fermions, are given by~\cite{Alonso:2012ji}  \begin{equation}
    \mathrm{BR}(\ell_\beta\to \ell_\alpha \gamma) \,=
    \frac{\alpha_w^3\,
      s_w^2}{256\,\pi^2}\,\frac{m_{\beta}^4}{M_W^4}\,
\frac{m_{\beta}}{\Gamma_{\beta}}\, 
    \left|G_\gamma^{\beta \alpha} \right|^2\:, 
\end{equation}
\begin{eqnarray}
    \mathrm{BR}(\ell_\beta\to 3\ell_\alpha) &=&
    \frac{\alpha_w^4}{24576\,\pi^3}\,\frac{m_{\beta}^4}{M_W^4}\,
\frac{m_{\beta}}{\Gamma_{\beta}}\times\left\{2\left|\frac{1}{2}F_\text{box}^{\beta
      3\alpha} +F_Z^{\beta\alpha} - 2 s_w^2\,(F_Z^{\beta\alpha} -
    F_\gamma^{\beta\alpha})\right|^2 \right.  \nonumber\\ 
     &+& \left. 4 s_w^4\, |F_Z^{\beta\alpha} -
    F_\gamma^{\beta\alpha}|^2 + 16
    s_w^2\,\mathrm{Re}\left[(F_Z^{\beta\alpha} - \frac{1}{2}F_\text{box}^{\beta
        3\alpha})\,G_\gamma^{\beta \alpha
        \ast}\right]\right.\nonumber\\ 
     &-&\left. 48 s_w^4\,\mathrm{Re}\left[(F_Z^{\beta\alpha} -
      F_\gamma^{\beta\alpha})\,G_\gamma^{\beta\alpha \ast}\right] + 32
    s_w^4\,|G_\gamma^{\beta\alpha}|^2\left[\log\frac{m_{\beta}^2}{m_{\alpha}^2}
      - \frac{11}{4}\right] \right\}\,.  
\end{eqnarray}
As before, $M_W$ is the $W$ boson mass, $\alpha_w = g_w^2/4\pi$ denotes the weak coupling, $s_w$ the sine of the weak mixing angle, and $m_{\beta}$ ($\Gamma_\beta$) the mass (total width)
of the decaying charged lepton of flavour $\beta$. 
The form factors $G_\gamma^{\beta \alpha}$, $F_\gamma^{\beta \alpha}$, $F_\text{box}^{\beta 3 \alpha}$ are given by~\cite{Alonso:2012ji, Ilakovac:1994kj} 
\begin{eqnarray}
    G_\gamma^{\beta \alpha} &=& \sum_{i =1}^{3 + n_S}
    \mathcal{U}_{\alpha i}^{\phantom{\ast}}\,\mathcal{U}_{\beta i}^\ast\,
    G_\gamma(x_i)\:,\label{eq:cLFV:FF:Ggamma} \\
     F_\gamma^{\beta \alpha} &=& \sum_{i =1}^{3 + n_S}
    \mathcal{U}_{\alpha i}^{\phantom{\ast}}\,\mathcal{U}_{\beta i}^\ast
    \,F_\gamma(x_i)\:,
   \\ 
    F_Z^{\beta \alpha} &=& \sum_{i,j =1}^{3 + n_S}
    \mathcal{U}_{\alpha i}^{\phantom{\ast}}\,\mathcal{U}_{\beta j}^\ast
    \left[\delta_{ij} \,F_Z(x_j) + 
    C_{ij}\, G_Z(x_i, x_j) + C_{ij}^\ast \,H_Z(x_i,
    x_j)\right]\:, 
    \label{eq:cLFV:FF:FZ}
    \\  
    F_\text{box}^{\beta 3 \alpha} &=&\sum_{i,j = 1}^{3+n_S}
    \mathcal{U}_{\alpha i}^{\phantom{\ast}}\,\mathcal{U}_{\beta
      j}^\ast\left[\mathcal{U}_{\alpha i}^{\phantom{\ast}} \,\mathcal{U}_{\alpha
        j}^\ast\, G_\text{box}(x_i, x_j) - 2 \,\mathcal{U}_{\alpha
        i}^\ast \,\mathcal{U}_{\alpha j}^{\phantom{\ast}}\, F_\text{Xbox}(x_i, x_j)
      \right]\:.\label{eq:cLFV:FF:Fbox}
    \end{eqnarray}
In the above expressions, the sums are understood to be taken over all neutral mass eigenstates ($i,j=1,...,3+n_S$). 
We recall that the $C_{ij}$ function is defined as 
\begin{equation}
 C_{ij} = \sum_{\rho = 1}^3
  \mathcal{U}_{i\rho}^\dagger \,\mathcal{U}_{\rho j}^{\phantom{\dagger}}\:. \nonumber
\end{equation}
The distinct loop functions (with arguments defined as $x_i ={m_{i}^2}/{M_W^2}$) are summarised in Appendix~\ref{app:loopfunctions-cLFVlepton},
with the corresponding arguments defined as $x_i ={m_{i}^2}/{M_W^2}$. 

\subsection{cLFV in muonic atoms}
Although other cLFV transitions and decays can also occur in the presence of muonic atoms (as for example Muonium oscillations and decays, or the Coulomb enhanced decay $\mu e \to e e$), here we collect the expressions for the coherent conversion rate in the presence of a nuclei (N), as the latter turns out to be among the most relevant cLFV observables due to its constraining power. 
The neutrinoless conversion rate is given by~\cite{Alonso:2012ji}   
\begin{equation}\label{eq:def:CRfull}
    \mathrm{CR}(\mu - e,\,\mathrm{N}) = \frac{2 G_F^2\,\alpha_w^2\,
      m_\mu^5}{(4\pi)^2\,\Gamma_\text{capt.}}\left|4 V^{(p)}\left(2
    \widetilde F_u^{\mu e} + \widetilde F_d^{\mu e}\right) + 4 V^{(n)}\left(
    \widetilde F_u^{\mu e} + 2\widetilde F_d^{\mu e}\right)  + s_w^2
    \frac{G_\gamma^{\mu e}D}{2 e}\right|^2\,,  
\end{equation}
in which $\Gamma_\text{capt.}$ denotes the capture rate for the nucleus N, with  
$D$, $V^{(p)}$ and $V^{(n)}$ corresponding to nuclear form factors (see~\cite{Kitano:2002mt}), and $e$ is 
the unit electric charge. 
The above form factors are given by~\cite{Alonso:2012ji, Ilakovac:1994kj} 
\begin{eqnarray}
    \widetilde F^{\mu e}_d &=& -\frac{1}{3}s_w^2 F_\gamma^{\mu e} - F_Z^{\mu e}\left(\frac{1}{4} - \frac{1}{3}s_w^2 \right) + \frac{1}{4}F^{\mu e dd}_\text{box}\ ,\\
    \widetilde F^{\mu e}_u &=& \frac{2}{3}s_w^2 F_\gamma^{\mu e} + F_Z^{\mu e}\left(\frac{1}{4} - \frac{2}{3}s_w^2 \right) + \frac{1}{4}F^{\mu e uu}_\text{box}\,,
\end{eqnarray}
to which one must add 
\begin{eqnarray}
     F_\text{box}^{\mu e uu} &=& \sum_{i = 1}^{3 + n_S}\sum_{q_d = d, s,
      b} \mathcal{U}_{e i}^{\phantom{\ast}}\,\mathcal{U}_{\mu i}^\ast\, V_{u q_d}^{\phantom{\ast}}\,V_{u
      q_d}^\ast \:F_\text{box}(x_i, x_{q_d})\,, 
    \label{eq:cLFV:FF:mueuu}\\
    F_\text{box}^{\mu e dd} &=& \sum_{i = 1}^{3 + n_S}\sum_{q_u = u, c,
      t} \mathcal{U}_{e i}^{\phantom{\ast}}\,\mathcal{U}_{\mu i}^\ast\, V_{q_u
      d}^{\phantom{\ast}}\,V_{q_u d}^\ast \:F_\text{Xbox}(x_i, x_{q_u})\,.
    \label{eq:cLFV:FF:muedd}    
\end{eqnarray}
Here, $x_{q} ={m_{q}^2}/{M_W^2}$ and $V$ is the Cabibbo-Kobayashi-Maskawa (CKM) quark mixing matrix. All relevant loop functions can be found in Appendix~\ref{app:loopfunctions-cLFVlepton}.

\subsection{Loop functions}\label{app:loopfunctions-cLFVlepton}
For completeness we collect here the loop functions for the purely leptonic cLFV decays and transitions discussed above, as well as some relevant limits (as presented in~\cite{Alonso:2012ji,Ilakovac:1994kj}).

\paragraph{Photon dipole and anapole functions} 
\begin{eqnarray}
    F_\gamma(x) &=& \frac{7 x^3 - x^2 - 12x}{12(1-x)^3} - \frac{x^4 -
      10x^3 + 12x^2}{6(1-x)^4}\log x\,,\nonumber\\ 
    F_\gamma(x) &\xrightarrow[x\gg1]{}& -\frac{7}{12} -
    \frac{1}{6}\log x\,,\nonumber\\ 
    F_\gamma(0) &=& 0\,,\label{eqn:lfun:fgamma}\\
    G_\gamma(x) &=& -\frac{x(2x^2 + 5x - 1)}{4(1-x)^3} -
    \frac{3x^3}{2(1-x)^4}\log x\,,\nonumber\\ 
    G_\gamma(x) &\xrightarrow[x\gg1]{}& \frac{1}{2}\,,\nonumber\\
    G_\gamma(0) &=& 0\,.\label{eqn:lfun:ggamma}
\end{eqnarray}

\paragraph{$Z$-penguin: two- and three-point functions}
\begin{eqnarray}
    F_Z(x) &=& -\frac{5 x}{2(1 - x)} - \frac{5x^2}{2(1-x)^2}\log x\,,\nonumber\\
    F_Z(x) &\xrightarrow[x\gg 1]{}& \frac{5}{2} - \frac{5}{2}\log
    x\,,\nonumber\\ 
    F_Z(0) &=& 0\,,\label{eqn:lfun:fz}
\end{eqnarray}

\begin{eqnarray}
    G_Z(x,y) &=& -\frac{1}{2(x-y)}\left[\frac{x^2(1-y)}{1-x}\log x -
      \frac{y^2(1-x)}{1-y}\log y \right]\,,\nonumber\\ 
    G_Z(x, x) &=& -\frac{x}{2} - \frac{x\log x}{1-x}\,,\nonumber\\
    G_Z(0,x) &=& -\frac{x\log x}{2(1-x)}\,,\nonumber\\
    G_Z(0,x)&\xrightarrow[x\gg 1]{}& \frac{1}{2}\log x\,,\nonumber\\
    G_Z(0,0) &=& 0\,,\label{eqn:lfun:gz}\\
    H_Z(x,y) &=& \frac{\sqrt{xy}}{4(x-y)}\left[\frac{x^2 - 4x}{1 -
        x}\log x - \frac{y^2 - 4y}{1 - y}\log
      y\ \right]\,,\nonumber\\ 
    H_Z(x,x) &=& \frac{(3 - x)(1-x) - 3}{4(1-x)} - \frac{x^3 - 2x^2 +
      4x}{4(1-x)^2}\log x\,,\nonumber\\ 
    H_Z(0,x) &=& 0\,.\label{eqn:lfun:hz}
\end{eqnarray}

\paragraph{Box loop-functions}
\begin{eqnarray}
    F_\text{box}(x,y) &=& \frac{1}{x-y}\left\{\left(4 +
    \frac{xy}{4}\right)\left[\frac{1}{1-x} + \frac{x^2}{(1-x)^2} \log
      x - \frac{1}{1-y} - \frac{y^2}{(1-y)^2}\log
      y\right]\right.\nonumber\\  
    &\phantom{=}& \left. -2xy\left[\frac{1}{1-x} + \frac{x}{(1-x)^2}
      \log x - \frac{1}{1-y} - \frac{y}{(1-y)^2}\log y
      \right]\right\}\,,\nonumber\\ 
    F_\text{box}(x,x) &=& -\frac{1}{4(1-x)^3}\left[x^4 - 16x^3 + 31x^2
      - 16 + 2x\left(3x^2 + 4x - 16\right)\log x\right]\,,\nonumber\\ 
    F_\text{box}(0,x) &=& \frac{4}{1 - x} + \frac{4x}{(1-x)^2}\log
    x\,,\nonumber\\ 
    F_\text{box}(0,x)&\xrightarrow[x\gg 1]{}& 0\,,\nonumber\\
    F_\text{box}(0,0) &=& 4\,,\label{eqn:lfun:fbox}\\
    F_\text{Xbox}(x,y) &=& -\frac{1}{x-y}\left\{\left(1 + \frac{xy}{4}
    \right)\left[\frac{1}{1-x} + \frac{x^2}{(1-x)^2} \log x -
      \frac{1}{1-y} - \frac{y^2}{(1-y)^2}\log
      y\right]\right.\nonumber\\  
    &\phantom{=}& \left. -2xy\left[\frac{1}{1-x} + \frac{x}{(1-x)^2}
      \log x - \frac{1}{1-y} - \frac{y}{(1-y)^2}\log y
      \right]\right\}\,,\nonumber\\ 
    F_\text{Xbox}(x,x) &=& \frac{x^4 - 16x^3 + 19x^2 - 4}{4(1-x)^3} +
    \frac{3x^3 + 4x^2 - 4x}{2(1-x)^3}\log x\,,\nonumber\\ 
    F_\text{Xbox}(0,x) &=& -\frac{1}{1-x} - \frac{x}{(1 - x)^2}\log
    x\,,\nonumber\\ 
    F_\text{Xbox}(0,x)&\xrightarrow[x\gg 1]{}& 0\,,\nonumber\\
    F_\text{Xbox}(0,0) &=& -1\,,\label{eqn:lfun:fxbox}\\
    G_\text{box}(x,y) &=& -\frac{\sqrt{xy}}{x-y}\left\{(4 +
    xy)\left[\frac{1}{1-x} + \frac{x}{(1-x)^2} \log x - \frac{1}{1-y}
      - \frac{y}{(1-y)^2}\log y\right]\right.\nonumber\\  
    &\phantom{=}& \left. -2\left[\frac{1}{1-x} + \frac{x^2}{(1-x)^2}
      \log x - \frac{1}{1-y} - \frac{y^2}{(1-y)^2}\log y
      \right]\right\}\,,\nonumber\\ 
    G_\text{box}(x,x) &=& \frac{2x^4 - 4x^3 + 8x^2 - 6x}{(1-x)^3} -
    \frac{x^4 + x^3 + 4x}{(1-x)^3}\log x\,,\nonumber\\ 
    G_\text{box}(0,x) &=& 0\,.\label{eqn:lfun:gbox}
\end{eqnarray}
In the above, we highlight that $F_\text{Xbox}$ corresponds to $F_\text{box}$ in~\cite{Ilakovac:1994kj}; moreover it has an opposite global sign in comparison to~\cite{Alonso:2012ji}, 
  which further impacts 
  $F_\text{box}^{\beta 3\alpha}$ (see~\cite{Alonso:2012ji,Ilakovac:1994kj}).

\end{document}